\documentclass[useAMS,usenatbib,usegraphicx]{mn2e}

\usepackage{amssymb}
\usepackage{lscape}
\usepackage{graphicx}
\usepackage{rotating}
\usepackage{times}
\usepackage{color}
\usepackage{longtable}
\usepackage{supertabular}
\bibpunct{(}{)}{;}{a}{}{,}

\newcommand{\snia}{SN~Ia}
\newcommand{\sneia}{SNe~Ia}

\newcommand{\one}{\,{\sc i}}
\newcommand{\two}{\,{\sc ii}}
\newcommand{\three}{\,{\sc iii}}

\newcommand{\nifs}{\ensuremath{^{56}\rm{Ni}}}

\newcommand{\fefs}{\ensuremath{^{56}\rm{Fe}}}
\newcommand{\msun}{\ensuremath{\rm{M}_{\odot}}}

\newcommand{\kms}{\ensuremath{\rm{km\,s}^{-1}}}

\newcommand{\gcc}{\ensuremath{\rm{g\,cm}^{-3}}}
\newcommand{\ergs}{\ensuremath{\rm{erg\,s}^{-1}}}
\newcommand{\dmft}{\ensuremath{\Delta m_{15}(B)}}

\def\cmfgen{{\sc cmfgen}}

\title[Non-LTE spectra of SN~2002bo]{A one-dimensional
  Chandrasekhar-mass delayed-detonation model for the broad-lined
  Type Ia supernova 2002bo}

\author[S. Blondin et al.]
{
St\'ephane Blondin,$^{1}$\thanks{E-mail: stephane.blondin@lam.fr}
Luc Dessart,$^{2}$
and D.~John Hillier$^{3}$\\
$^{1}$Aix Marseille Universit\'e, CNRS, LAM (Laboratoire
d'Astrophysique de Marseille), UMR 7326, 13388 Marseille, France.\\ 
$^{2}$Laboratoire Lagrange, UMR 7293, Universit\'e Nice
Sophia-Antipolis, CNRS, Observatoire de la C\^ote d'Azur, 06300 Nice,
France.\\ 
$^{3}$Department of Physics and Astronomy \& Pittsburgh
  Particle Physics, Astrophysics, and Cosmology Center (PITT PACC),\\
  University of Pittsburgh, Pittsburgh, PA 15260, USA.
}

\begin{document}

\date{Accepted 2015 January 26.  Received 2015 January 15; in original form 2014 November 25}

\pagerange{\pageref{firstpage}--\pageref{lastpage}} \pubyear{2015}

\maketitle

\label{firstpage}

%%%%%%%%%%%%%%%%%%%%%%%%%%%%%%%%%%%%%%%%%%%%%%%%%%%%%%%%%%%%%%%%%%%%%%
%%%%%%%%%%%%%%%%%%%%%%%%%%%%%%%%%%%%%%%%%%%%%%%%%%%%%%%%%%%%%%%%%%%%%%

\begin{abstract}
We present 1D non-local thermodynamic equilibrium (non-LTE)
time-dependent radiative-transfer simulations of a Chandrasekhar-mass
delayed-detonation model which synthesizes 0.51~\msun\ of \nifs, and
confront our results to the Type Ia supernova (\snia) 2002bo over the
first 100 days of its evolution.  Assuming only homologous expansion,
this same model reproduces the bolometric and multi-band light curves,
the secondary near-infrared (NIR) maxima, and the optical and NIR
spectra.  The chemical stratification of our model qualitatively
agrees with previous inferences by Stehle et al., but reveals
significant quantitative differences for both iron-group and
intermediate-mass elements.  We show that $\pm$0.1~\msun\ (i.e.,
$\pm$20 per cent) variations in \nifs\ mass have a modest impact on
the bolometric and colour evolution of our model.  One notable
exception is the $U$-band, where a larger abundance of iron-group
elements results in {\it less} opaque ejecta through ionization
effects, our model with more \nifs\ displaying a higher near-UV flux
level.  In the NIR range, such variations in \nifs\ mass affect the
timing of the secondary maxima but not their magnitude, in agreement
with observational results. Moreover, the variation in the $I$, $J$,
and $K_s$ magnitudes is less than 0.1~mag within $\sim$10 days from
bolometric maximum, confirming the potential of NIR photometry of
\sneia\ for cosmology.  Overall, the delayed-detonation mechanism in
single Chandrasekhar-mass white dwarf progenitors seems well suited
for SN~2002bo and similar \sneia\ displaying a broad
Si\two\ 6355~\AA\ line.  Whatever multidimensional processes are at
play during the explosion leading to these events, they must conspire
to produce an ejecta comparable to
our spherically-symmetric model.
\end{abstract}


%%%%%%%%%%%%%%%%%%%%%%%%%%%%%%%%%%%%%%%%%%%%%%%%%%%%%%%%%%%%%%%%%%%%%%
%%%%%%%%%%%%%%%%%%%%%%%%%%%%%%%%%%%%%%%%%%%%%%%%%%%%%%%%%%%%%%%%%%%%%%

\section{Introduction}\label{sect:intro}

The most widely-accepted model for Type Ia supernovae is the
thermonuclear disruption of a white dwarf (WD) star
\citep{Hoyle/Fowler:1960} in a binary system, although there is
ongoing discussion about the combustion mode (pure deflagration or
delayed detonation), the progenitor mass (Chandrasekhar mass or not),
and the nature of the binary companion (another WD or a non-degenerate
star). Observational evidence for diversity in the \snia\ population
\citep[e.g.,][]{vanKerkwijk/etal:2010,Badenes/Maoz:2012} seems to
require multiple progenitor channels or explosion mechanisms.

In a recent study, we found a promising agreement between a sequence
of Chandrasekhar-mass delayed-detonation models and observations of
\sneia\ at maximum light \citep{Blondin/etal:2013} [hereafter
  B13]. The good match between our model DDC15 ($\sim$0.5~\msun\ of
\nifs) and the standard SN~2002bo motivates an in-depth study of this
model at earlier and later times.

Supernova 2002bo was first studied by \cite{Benetti/etal:2004}, who
favour a delayed-detonation explosion for this event
based on the presence of intermediate-mass elements (IMEs) at high
velocities and the lack of spectral signatures of C\one/{\sc ii}
associated with unburnt carbon at early times. Subsequent
spectroscopic modeling by \cite{Stehle/etal:2005} revealed a chemical
stratification qualitatively similar to the fast-deflagration W7 model
of \cite{W7}, but with an offset to higher velocities and strong
mixing.

A striking feature of SN~2002bo is the large Doppler width of the
Si\two~6355~\AA\ line around maximum light, which places this SN in
the ``broad-line'' subclass of \cite{Branch/etal:2006}.\footnote{Given
  the associated large blueshift at maximum absorption
  ($>12000$~\kms\ at maximum light), this SN is also part of the
  ``high-velocity'' subclass of \cite{WangX/etal:2009b} and the
  ``high-velocity-gradient'' subclass of \cite{Benetti/etal:2005}.}
Such a broad Si\two\ line is systematically predicted in maximum-light
spectra of the delayed-detonation models presented in B13.

In the next section we briefly present the numerical setup
(hydrodynamics and radiative transfer). The observational data and our
method of analysis are presented in Section~\ref{sect:data}. We then
confront the predicted bolometric (Section~\ref{sect:bol}), colour
(Section~\ref{sect:phot}), and spectroscopic (Section~\ref{sect:spec})
evolution of our model to observations of SN~2002bo,
both in the optical and NIR.  We illustrate the sensitivity of our
model results to $\pm$0.1~\msun\ (i.e., $\pm$20 per cent) variations
in \nifs\ mass in Section~\ref{sect:sens}. We compare the
abundance distributions of various elements in our input
hydrodynamical model with those inferred by \cite{Stehle/etal:2005} in
Section~\ref{sect:comp}.  A discussion on the nature of broad-lined
\sneia\ and conclusions follow in Section~\ref{sect:ccl}.

%%%%%%%%%%%%%%%%%%%%%%%%%%%%%%%%%%%%%%%%%%%%%%%%%%%%%%%%%%%%%%%%%%%%%%
%%%%%%%%%%%%%%%%%%%%%%%%%%%%%%%%%%%%%%%%%%%%%%%%%%%%%%%%%%%%%%%%%%%%%%

\section{Numerical setup}\label{sect:models}

The hydrodynamics of the explosion and the radiative-transfer
treatment is analogous to that presented in B13 and \cite{D14_tech}
[hereafter D14c]. We refer the reader to those papers for an in-depth
discussion of our numerical setup. As with all our previous SN studies,
our models are iterated until convergence is achieved at all depths in
the ejecta (see Appendix~\ref{sect:conv}).

\begin{figure}
\centering
\includegraphics{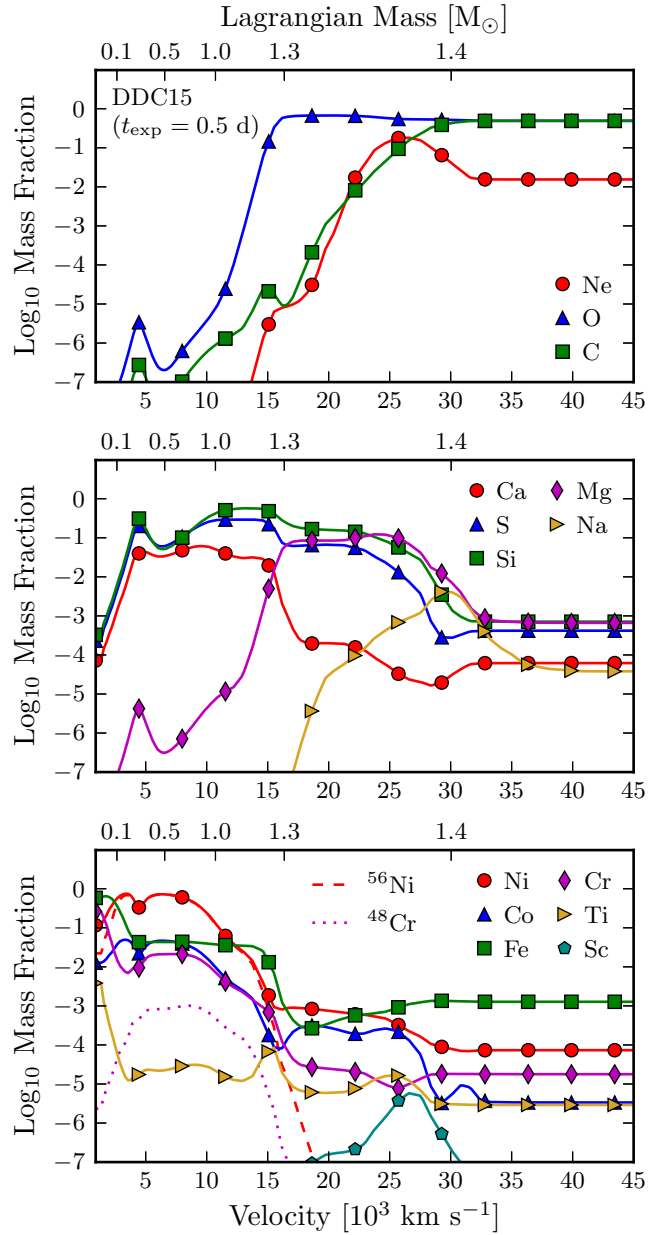}
\caption{\label{fig:elem_distrib} Abundance profiles in velocity space
  for C/O/Ne (top), and selected IMEs (middle), and IGEs (bottom; the
  profiles for \nifs\ and $^{48}$Cr at this time are shown as a dashed
  and dotted line, respectively) in model DDC15 at the start of our
  radiative-transfer simulations (0.5 days past explosion). The upper
  abscissa shows the Lagrangian mass coordinate.  }
\end{figure}

We use the Chandrasekhar-mass delayed-detonation model DDC15 presented
in B13, and summarize its basic properties in Table~\ref{tab:modinfo}.
Unlike B13, we apply a small radial mixing to the hydrodynamical input
with a characteristic velocity width $\Delta v_{\rm mix}=400$~\kms\ to
smooth sharp variations in composition (see \citealt{D14_PDD};
hereafter D14a). Moreover, we consider additional radioactive decay
chains (only \nifs\ decay was considered in B13), resulting in
non-negligible energy input beyond $\sim$15000~\kms\ where the ejecta
is deficient in \nifs\ (see model DDC10\_A4D1 in D14a). More
importantly, including the
$^{48}$Cr~$\rightarrow$~$^{48}$V~$\rightarrow$~$^{48}$Ti decay chain
leads to a $\sim$100-fold increase in the Ti mass fraction in the
inner ejecta ($v\lesssim 13000$~\kms; see
Fig.~\ref{fig:elem_distrib}), which influences sizeably the opacity in
the $B$-band at maximum light and beyond.

%%% FIGURE: composition profiles

The chemical stratification is typical of standard delayed-detonation
models. Iron-group elements (IGEs; Sc to Ni, $M_{\rm
  tot}\approx0.7$~\msun) dominate below $\sim$10000~\kms. Despite the
presence of \nifs\ at $\lesssim$2000~\kms\ due to our imposed radial
mixing, the innermost ejecta are dominated by {\it stable}
IGEs. Intermediate-mass elements (Na to Ca, $M_{\rm
  tot}\approx0.6$~\msun) are most abundant between
$\sim$10000~\kms\ and $\sim$30000~\kms. The outermost
$\sim0.01$~\msun\ of the ejecta is dominated by the composition of the
progenitor WD star (C/O, with traces of $^{22}$Ne, and solar abundances
for all other elements).

\begin{figure}
\centering
\includegraphics{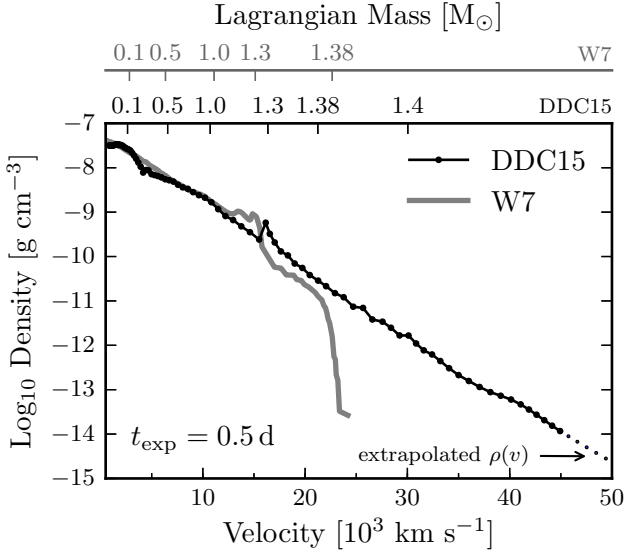}
\caption{\label{fig:dens_profile} Density profile at the start of our
  radiative transfer simulations (0.5 days past explosion; black
  line), compared to the W7 model (grey line) of
  {\protect\cite{W7}}. The solid dots correspond to the initial radial
  grid used in \cmfgen. The dotted curve shows the extrapolated
  density profile beyond $\sim$45000~\kms\ (see text). The upper axis
  gives the Lagrangian mass coordinate for both DDC15 (black) and W7
  (grey).  }
\end{figure}

The ejecta density profile at the start of our radiative-transfer
calculations (0.5 days past explosion) is shown in
Fig.~\ref{fig:dens_profile}. At these times, UV photons still interact
with the ejecta material in shells moving at a velocity $>40000$~\kms,
so we linearly extrapolate the 
hydrodynamical input beyond $\sim$45000~\kms\ up to $\sim$70000~\kms,
as done in D14a (dotted line in Fig.~\ref{fig:dens_profile}). The
density profile is reminiscent of standard delayed-detonation models
\citep[see, e.g.,][]{Khokhlov/etal:1993}, with a quasi-exponential
decrease with velocity.  Also shown is the density profile of the W7
model used as a reference in many \snia\ studies. It is nearly
indistinguishable from our model below $\sim$10000~\kms\ (or within a
Lagrangian mass coordinate of $\sim$1~\msun). The key difference
lies in the amount of mass contained at high velocities: It is a
factor of $\sim$2 larger beyond $\sim$15000~\kms\ in DDC15, resulting
in a $\sim$20 per cent larger kinetic energy
($\sim$1.5$\times$10$^{51}$~erg cf. $\sim$1.2$\times$10$^{51}$~erg for
W7; see Table~\ref{tab:modinfo}). The composition is also drastically
different, with IMEs confined to $\lesssim$15000~\kms, while they are
abundant beyond 20000~\kms\ in DDC15.

The radial mixing applied to the hydrodynamical input and the
  extended radial grid result in small differences in abundance
  between the DDC15 model presented in B13 and the
  version used in this paper. In particular, the initial \nifs\ mass
  for this model is now $\sim 0.51$~\msun, cf. $\sim 0.56$~\msun\ in
  B13 (see their Table~1). These differences 
%only lead to minor differences 
only have a modest impact at bolometric
maximum: A bolometric rise time of 17.57~d (cf. 17.76~d in B13), a
peak bolometric luminosity of
1.14$\times10^{43}$~\ergs\ (cf. 1.22$\times10^{43}$~\ergs), and a
$B-R$ colour of 0.04~mag (cf. $-0.02$~mag). 

In what follows, we use model DDC15 to discuss SN~2002bo and
broad-lined \sneia\ in general. However, we also study the
sensitivity of our results by comparing the predicted light curves and
spectra of delayed-detonation models endowed initially with $\pm$20
per cent of mass of \nifs\ (models DDC10 and DDC17 in
Table~\ref{tab:modinfo}). This comparison is particularly instructive
to understand the degeneracy (or lack thereof) of \snia\ properties.

%%% FIGURE: density profiles

%%% TABLE: basic model information
\begin{table*}
\footnotesize
\caption{Model parameters and nucleosynthetic yields for selected species at the start of our radiative-transfer calculations (0.5\,d past explosion). The \nifs\ mass is given at $t_{\rm exp}\approx0$. Models DDC10 and DDC17 will be discussed in Section~\ref{sect:sens}.}\label{tab:modinfo}
\begin{tabular}{l@{\hspace{1.6mm}}c@{\hspace{1.6mm}}c@{\hspace{1.6mm}}c@{\hspace{1.6mm}}c@{\hspace{1.6mm}}c@{\hspace{1.6mm}}c@{\hspace{1.6mm}}c@{\hspace{1.6mm}}c@{\hspace{1.6mm}}c@{\hspace{1.6mm}}c@{\hspace{1.6mm}}c@{\hspace{1.6mm}}c@{\hspace{1.6mm}}c@{\hspace{1.6mm}}c@{\hspace{1.6mm}}c@{\hspace{1.6mm}}c@{\hspace{1.6mm}}}
\hline
\multicolumn{1}{c}{Model} & $\rho_{\rm tr}$ & $E_{\rm kin}$ & $v(\nifs)$  & \nifs$_{t=0}$ & Ni & Co & Fe & Ti & Sc & Ca & S & Si & Mg & Na & O & C \\
 & [\gcc] & [\ergs] & [\kms]  & [M$_{\sun}$] & [M$_{\sun}$] & [M$_{\sun}$] & [M$_{\sun}$] & [M$_{\sun}$] & [M$_{\sun}$] & [M$_{\sun}$] & [M$_{\sun}$] & [M$_{\sun}$] & [M$_{\sun}$] & [M$_{\sun}$] & [M$_{\sun}$] & [M$_{\sun}$] \\
\hline
DDC15          &   1.8(7) &  1.465(51) &  1.12(4) &      0.511 &      0.516 &   3.44(-2) &      0.114 &   1.11(-4) &   4.09(-8) &   4.53(-2) &      0.197 &      0.306 &   1.14(-2) &   1.68(-5) &      0.105 &   2.73(-3) \\
\hline
DDC10          &   2.3(7) &  1.520(51) &  1.16(4) &      0.623 &      0.622 &   4.11(-2) &      0.115 &   1.10(-4) &   2.90(-8) &   4.10(-2) &      0.166 &      0.257 &   9.95(-3) &   1.25(-5) &      0.101 &   2.16(-3) \\
DDC17          &   1.6(7) &  1.459(51) &  1.08(4) &      0.412 &      0.421 &   2.84(-2) &      0.112 &   1.14(-4) &   6.16(-8) &   4.73(-2) &      0.222 &      0.353 &   1.79(-2) &   2.47(-5) &      0.152 &   3.80(-3) \\
\hline
\end{tabular}

\flushleft
{\bf Notes:}
Numbers in parenthesis correspond to powers of ten.
The deflagration velocity is set to 3\% of the local sound speed ahead of the flame for all models;
$\rho_{\rm tr}$ is the transition density at which the deflagration is artificially turned into a detonation;
$E_{\rm kin}$ is the asymptotic kinetic energy;
$v(\nifs)$ is the velocity of the ejecta shell that bounds 99\% of the total \nifs\ mass.
\end{table*}

%%%%%%%%%%%%%%%%%%%%%%%%%%%%%%%%%%%%%%%%%%%%%%%%%%%%%%%%%%%%%%%%%%%%%%
%%%%%%%%%%%%%%%%%%%%%%%%%%%%%%%%%%%%%%%%%%%%%%%%%%%%%%%%%%%%%%%%%%%%%%

\section{Observational data and Methodology}\label{sect:data}

Our data on SN~2002bo are mostly taken from \cite{Benetti/etal:2004},
with additional optical spectra from \cite{Blondin/etal:2012} and
\cite{Silverman/etal:2012a}. We use the near-infrared ($JHK_s$)
photometry published by \cite{Krisciunas/etal:2004c}. We select
spectra whose relative flux calibration agrees with photometry-based
colours to within 0.1~mag.  As in B13, we adopt a recession velocity
$cz=1293$~\kms\ for the host galaxy, a distance modulus
$\mu=31.90\pm0.20$~mag and a total reddening $E(B-V)=0.41\pm0.07$~mag.

The method for generating pseudo-bolometric light curves for both
models and data is analogous to that presented in B13. We specify each
time the bluest and the reddest band used in the integration, such
that $L_{U \rightarrow K_s}$ corresponds to the luminosity (in \ergs)
obtained by integrating $UBVRIJHK_s$ magnitudes.

Since there is no good definition of a photosphere through opacity
means in \sneia, we characterize the spectrum-formation region using
the spatial distribution of the optical (3000--10000~\AA) flux
instead. The fractional contribution to the total flux at wavelength
$\lambda$ from a given ejecta shell at radius $r$ and thickness $dr$ at
time $t$ since explosion is:

\begin{equation}
\label{eqn:dfr}
\delta F(r, \lambda, t) = \frac{2\pi}{D^2} \int \Delta
z\ \eta(p,z,\lambda,t)\ e^{-\tau(p,z,\lambda,t)}\ p\ dp,
\end{equation}

\noindent
where $D$ is the physical distance, $\Delta z$ is the projected shell
thickness for a ray with impact parameter $p$, $\eta$ is the
emissivity along the ray at $p$ and $z$, and $\tau$ is the ray optical
depth (at $\lambda$) at the ejecta location $(p,z)$. Because of homologous
expansion ($v=r/t$), surfaces of constant projected velocity are
planes perpendicular to the $z$ direction. Hence, the quantity $\delta
F(r,\lambda,t)$ mixes contributions from regions with the same radial
velocity but with very different projected velocities, hence Doppler
shift. The averaged contribution of a shell with velocity $v$
(corresponding to $r+dr/2$) over the 
wavelength interval [$\lambda_1,\lambda_2$] is given by:

\begin{equation}
\label{eqn:mean_df}
\overline{\delta F(v)} = \frac{ \int_{\lambda_1}^{\lambda_2} \delta F(r, \lambda,
  t)d\lambda} {|\lambda_1-\lambda_2|}.
\end{equation}

\noindent
A representative velocity for the spectrum-formation region in the
wavelength range $\lambda_1$ to $\lambda_2$ may be the velocity 
above which 50\% of the flux in that range is emitted, noted $v_{\rm
  1/2}$, such that:

\begin{equation}
\label{eqn:vhalf}
\frac{\sum\limits_{v \ge v_{\rm 1/2}}^{} \overline{\delta F(v)}}
{\sum\limits_{{\rm all}\ v}^{} \overline{\delta F(v)}} = \frac{1}{2},
\end{equation}

\noindent
where the sum in the denominator runs over the entire velocity grid
spanned by our simulation.

\begin{figure}
\centering
\includegraphics{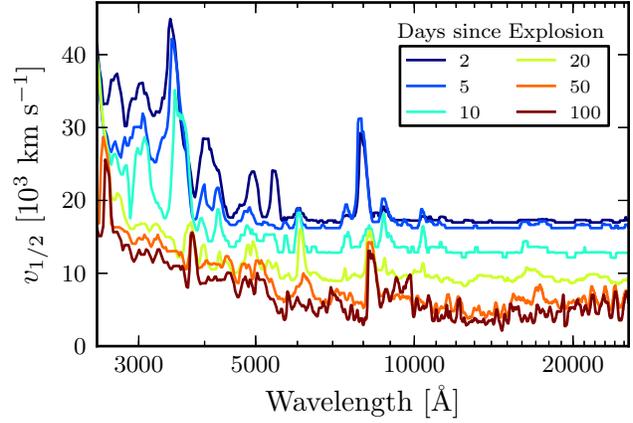}
\caption{\label{fig:vhalf}
  Illustration of the wavelength dependence
  of the spectrum-formation region, represented here using the
  velocity above which 50\% of the flux at a given wavelength is
  emitted. This velocity is computed based on equation~(\ref{eqn:vhalf}), but
replacing the wavelength-averaged flux contribution $\overline{\delta
  F(v)}$ of equation~(\ref{eqn:mean_df}) with the non-averaged $\delta F(r,
  \lambda, t)$ of equation~(\ref{eqn:dfr}).}
\end{figure}

In what follows this velocity is only used for illustration purposes,
as the averaging in equation~(\ref{eqn:mean_df}) occults the strong variation
of $v_{\rm 1/2}$ with wavelength (Fig.~\ref{fig:vhalf}). This
variation reflects the dominance of line opacity in \snia\ ejecta
\citep[see, e.g.,][]{Pinto/Eastman:2000b}.

%%% FIGURE: vhalf

%%%%%%%%%%%%%%%%%%%%%%%%%%%%%%%%%%%%%%%%%%%%%%%%%%%%%%%%%%%%%%%%%%%%%%
%%%%%%%%%%%%%%%%%%%%%%%%%%%%%%%%%%%%%%%%%%%%%%%%%%%%%%%%%%%%%%%%%%%%%%

\section{Bolometric Evolution}\label{sect:bol}

%%% FIGURE: UVOIR LC vs. 02bo
\begin{figure*}
\centering
\includegraphics{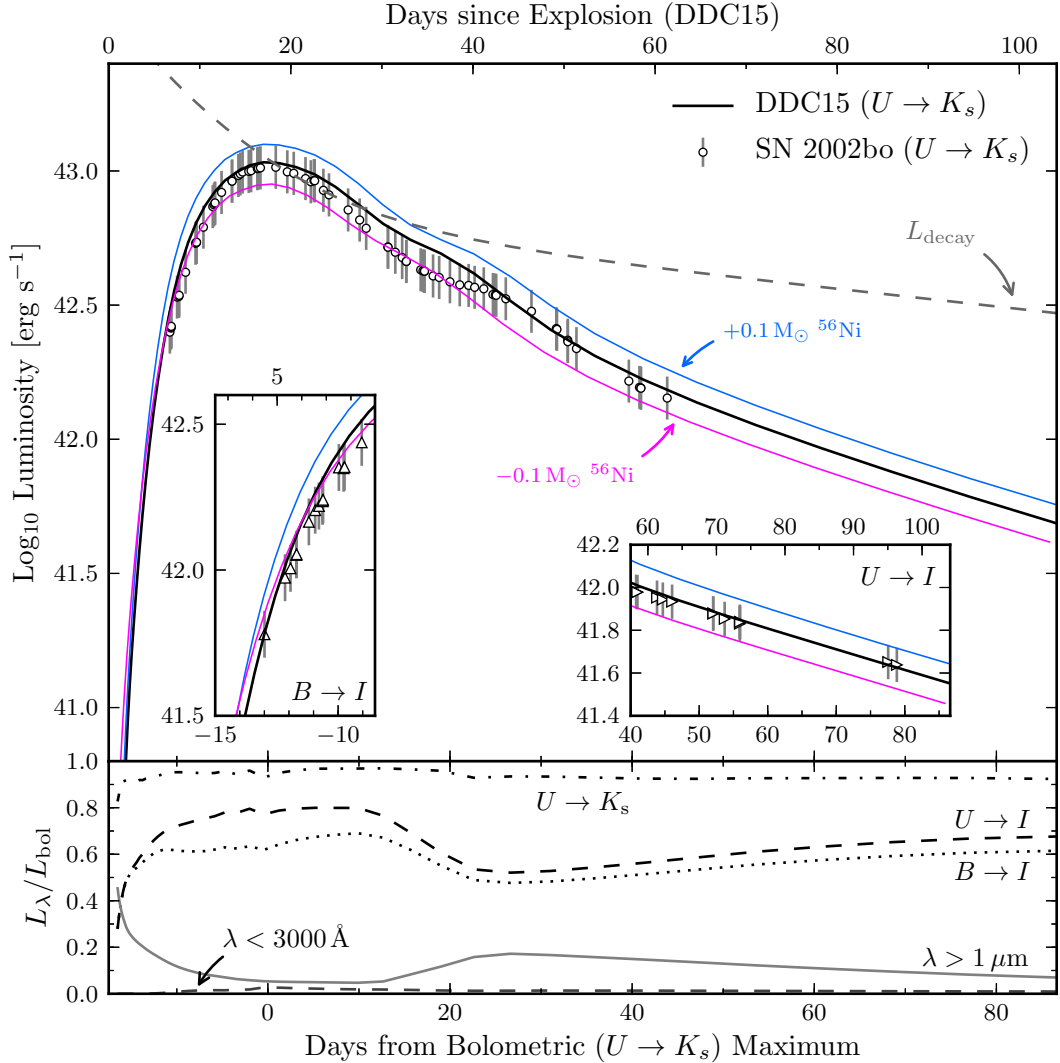}
\caption{\label{fig:comp_lcbol} {\bf Top:} Pseudo-bolometric
  ($U\rightarrow K_s$) light curve for model DDC15 (black) and
  SN~2002bo (open circles). The error bars take into account
  measurement, extinction, and distance errors. The grey dashed line
  shows the instantaneous rate of decay energy. Coloured lines
  correspond to models with $\pm$0.1~\msun\ of \nifs\ compared to
  DDC15.  We show close-up views of the early-time $B\rightarrow I$
  (left inset) and late-time $U\rightarrow I$ (right inset)
  pseudo-bolometric light curves, due to the lack of $U$-band or NIR
  data at these times.  The upper abscissa gives the time since
  explosion for model DDC15.  {\bf Bottom:} Contribution of the
  $U\rightarrow K_s$ (dash-dotted line), $U\rightarrow I$ (dashed
  line) and $B\rightarrow I$ (dotted line) pseudo-bolometric
  luminosities to the total bolometric luminosity. Also shown are the
  contributions of the UV ($\lambda<3000$~\AA; grey solid line) and IR
  ($\lambda>1$~$\mu$m; grey dashed line) ranges.  }
\end{figure*}

The pseudo-bolometric light curves for both our DDC15 model and
SN~2002bo are shown in Fig.~\ref{fig:comp_lcbol}.  Our model predicts
a rise to bolometric maximum of $\sim$17.6~d, while the first
observations of SN~2002bo are $\sim$14 days before peak. At these
early times, the bolometric evolution is well matched by our model,
albeit with a slightly larger luminosity (but still within the
errors). This suggests the \nifs\ distribution is adequate, in
particular its presence at $\gtrsim$15000~\kms. In W7, the \nifs\ mass
fraction drops to $10^{-5}$ at $\sim$12000~\kms, and the early-time
luminosity is significantly lower than observed in SN~2002bo
\citep[see also][]{Stehle/etal:2005}.

%%% TABLE: bolometric properties
\begin{table}
\footnotesize
\caption{Bolometric properties. Models DDC10 and DDC17 will be discussed in Section~\ref{sect:sens}.}\label{tab:bolprop}
\begin{tabular}{l@{\hspace{2.8mm}}c@{\hspace{2.8mm}}c@{\hspace{2.8mm}}c@{\hspace{2.8mm}}c@{\hspace{2.8mm}}c@{\hspace{2.8mm}}c@{\hspace{2.8mm}}c}
\hline
\multicolumn{1}{c}{Model} & \nifs   & $t_{\rm rise}$ & $\alpha$ & $L_{\rm peak}$ & $\Delta M_{15}$ & $Q_\gamma$ & $\mathcal{F}_{\rm dep}$ \\
                          & [\msun] & [d]            &          & [\ergs]        & [mag]           &            &                         \\
\hline
DDC15          & 0.511 &  17.57 & 3.4 &   1.14(43) & 0.72 & 1.04 &  0.95 \\
\hline
DDC10          & 0.623 &  17.13 & 3.3 &   1.38(43) & 0.74 & 1.00 &  0.95 \\
DDC17          & 0.412 &  18.60 & 3.3 &   9.10(42) & 0.67 & 0.99 &  0.97 \\
\hline
\end{tabular}

\flushleft
{\bf Notes:}
Numbers in parenthesis correspond to powers of ten;
$\alpha$ is the exponent of the power-law fit to the early-time ($t_{\rm exp}\lesssim3$\,d) luminosity, $L(t)\propto t^\alpha$;
$\Delta M_{15}$ is the bolometric magnitude decline between bolometric maximum and 15\,d after;
$Q_{\gamma}$ is the ratio of the peak bolometric luminosity ($L_{\rm peak}$) to the instantaneous decay luminosity;
$\mathcal{F}_{\rm dep}$ is the fraction of decay energy at peak actually deposited in the ejecta.
\end{table}

A power-law fit $L(t)\propto t^{\alpha}$ to the early ($t_{\rm
  exp}\lesssim3$~d) bolometric light curve of model DDC15 yields
$\alpha=3.4$ (Table~\ref{tab:bolprop}).  Deviations from a
``fireball'' model (for which $\alpha=2$; \citealt{Nugent/etal:2011})
are expected theoretically and confirmed observationally, as in
the recent SN~2013dy \citep{SN2013dy}.  Variations in the early-time
luminosity can result from outward mixing of
\nifs\ \citep[e.g.,][]{Piro/Nakar:2014}, or from structural changes in
the outer ejecta, as in the pulsational-delayed-detonation models of
D14a.

The lower panel of Fig.~\ref{fig:comp_lcbol} gives the fractional
contribution to the total bolometric luminosity of various filter
sets, as well as in the UV ($<3000$~\AA) and IR ($>1$~$\mu$m)
ranges. At maximum light, we recover 94 per cent of the true
bolometric luminosity with our $U\rightarrow K_s$ pseudo-bolometric
light curve (see also Table~\ref{tab:lc}). The remaining 6 per cent is
mostly emitted blueward of the $U$-band.

From bolometric maximum until the onset of the radioactive tail
$\sim$35 days later the $\gamma$-ray escape fraction increases from
$\sim$5 to $\sim$50 per cent. The decline in bolometric magnitude
between maximum light and 15 days later (noted $\Delta M_{15}$ in
Table~\ref{tab:bolprop}) is 0.72~mag, comparable to that inferred for
SN~2002bo based on its $U\rightarrow K_s$ pseudo-bolometric light
curve ($\sim$0.85~mag).

Past maximum, the transition to an optically-thin ejecta is
concomitant with a shift in ionization and a greater release of
radiation energy, causing a small bump in the bolometric light curve
around $\sim$20 days past maximum light ($\sim$30 days past maximum in
SN~2002bo; see Fig.~\ref{fig:comp_lcbol}).

Past $\sim$30 days from maximum light ($\sim$50 days since explosion),
the bolometric luminosity follows the instantaneous rate of energy
deposition.  The $\gamma$-ray escape fraction continues to increase,
reaching $\sim$80 per cent at $\sim$100 days past explosion. The
late-time ($>50$ days past maximum) pseudo-bolometric ($U\rightarrow
I$) decline is compatible with SN~2002bo ($\sim$0.024~mag\,day$^{-1}$
cf. $\sim$0.022$\pm$0.007~mag\,day$^{-1}$).

The kinetic energy, \nifs\ mass, and ejecta mass of our model yield a
bolometric light curve consistent with observations of SN~2002bo at
all times.  This represents a compelling check for the standard
Chandrasekhar-mass delayed-detonation model for this class of
\snia\ light-curve morphology. Moreover, the good match with a 1D
model suggests that asymmetries, if present, may only have a minor
role in shaping the bolometric light curve of SN~2002bo.

%%%%%%%%%%%%%%%%%%%%%%%%%%%%%%%%%%%%%%%%%%%%%%%%%%%%%%%%%%%%%%%%%%%%%%
%%%%%%%%%%%%%%%%%%%%%%%%%%%%%%%%%%%%%%%%%%%%%%%%%%%%%%%%%%%%%%%%%%%%%%

\section{Photometric Evolution}\label{sect:phot}

%%% FIGURE: time-evolution of temperature profile
\begin{figure}
\centering
\includegraphics{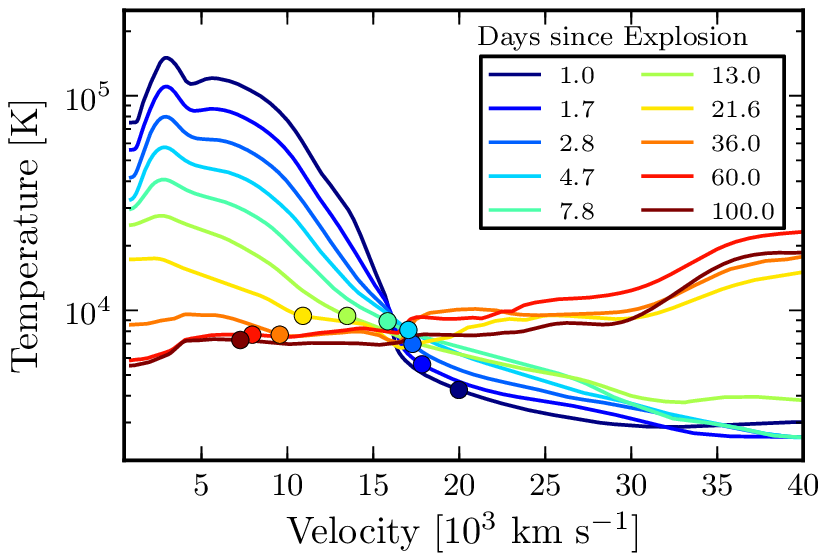}\vspace{.25cm}
\includegraphics{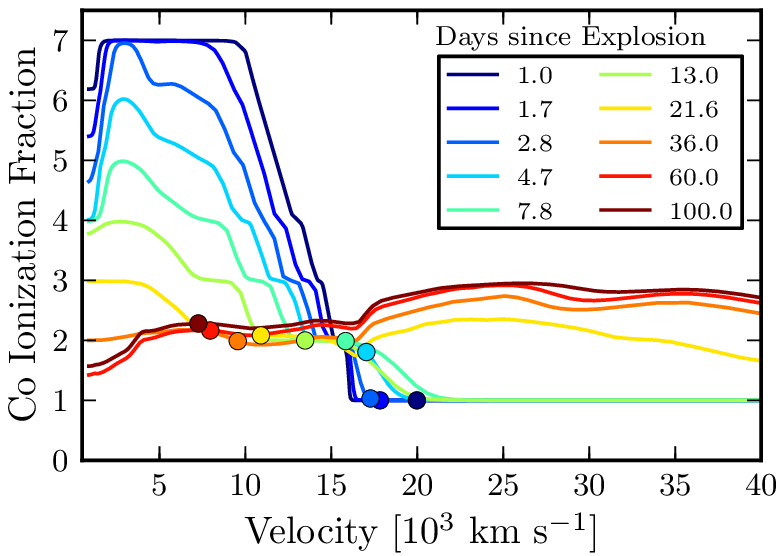}
\caption{\label{fig:temp_evol} Evolution of the gas temperature
  profile (top) and of the cobalt ionization fraction (bottom) in
  model DDC15, between 1 and 100 days since explosion (note the
    logarithmic time steps). Solid dots mark the velocity
    location above which 50\% of the optical flux is emitted, which we
  use to characterize the spectrum-formation region 
  (see Section~\ref{sect:data}).}
\end{figure}

%%% FIGURE: colour curves
\begin{figure}
\centering
\includegraphics{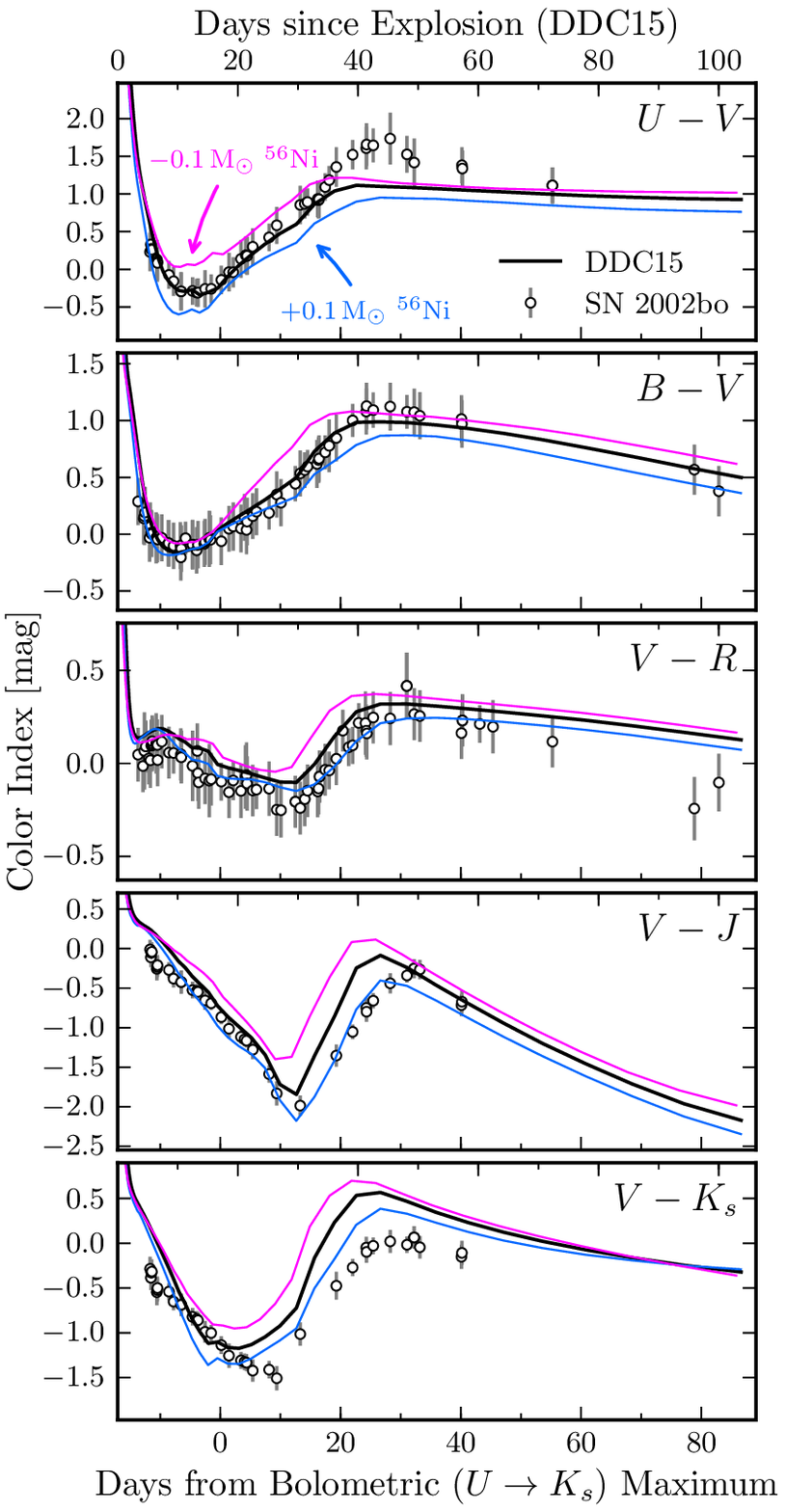}
\caption{\label{fig:comp_colors} Comparison of colour
  curves for model DDC15 (black) and SN~2002bo (dereddened; open circles).
  The error bars take into account measurement and extinction errors.
  Coloured lines correspond to models with $\pm$0.1~\msun\ of
  \nifs\ compared to DDC15. Note that the redder colour curve
  corresponds to
  the model with {\it less} \nifs\ (DDC17).  The upper abscissa gives
  the time since explosion for model DDC15.}
\end{figure}

The colour evolution of \sneia\ reflects both the global evolution of
ejecta properties (temperature, mean ionization) and the
wavelength-dependent opacity, and is thus very difficult to
accurately model.

\begin{figure*}
\centering
\includegraphics{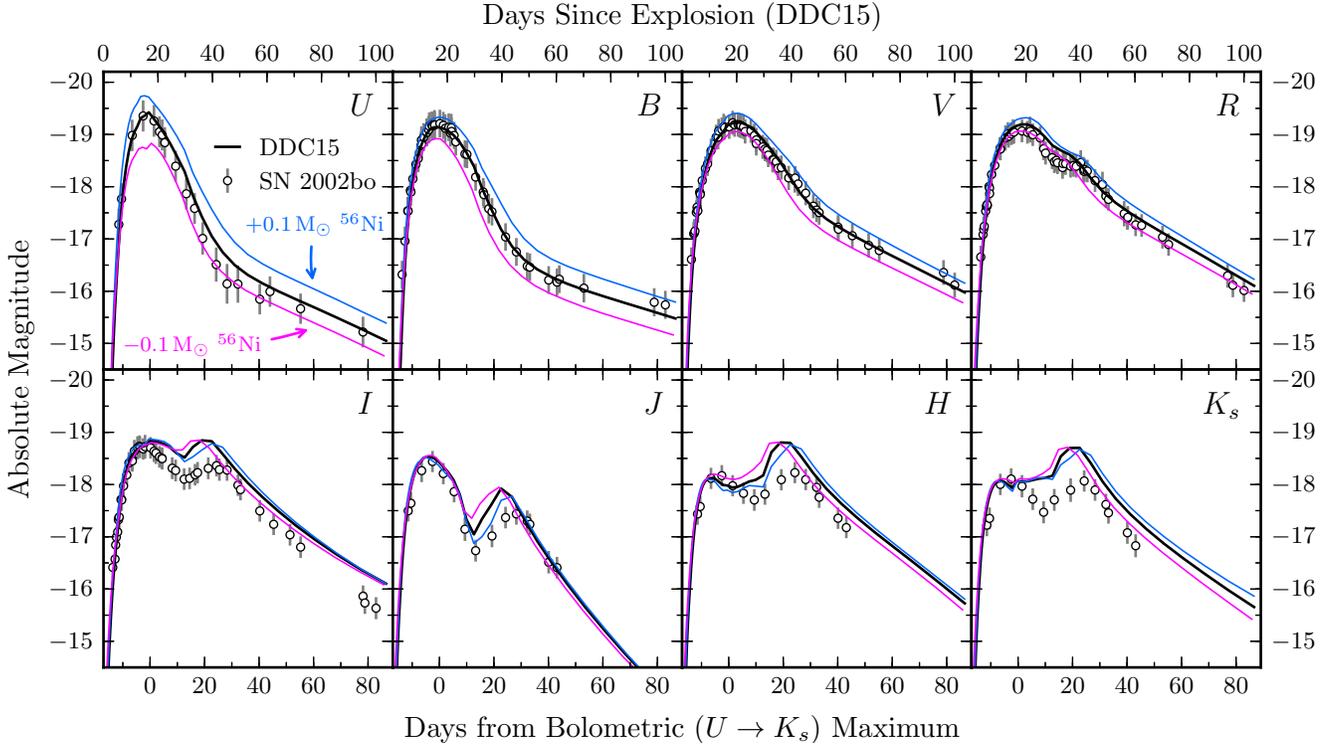}
\caption{\label{fig:comp_lc} Optical ($UBVRI$) and NIR ($JHK_s$) light
  curves for model DDC15 (solid lines) compared to SN~2002bo
  (dereddened; open circles). 
  The error bars take into account measurement, extinction, and
  distance errors.
Coloured lines correspond to models with
  $\pm$0.1~\msun\ of \nifs\ compared to DDC15.  The upper abscissa
  gives the time since explosion for model DDC15.  }
\end{figure*}

Initially, all bands experience a steep brightening due to the
increase in radius. The brightening is steeper for the bluer bands
(see the $\alpha$ exponent in Table~\ref{tab:lcprop}) due to the
temperature increase in the spectrum-formation layers (solid dots in
Fig.~\ref{fig:temp_evol}), which shifts the optical and NIR colours
shown in Fig.~\ref{fig:comp_colors} to the blue.  Because of the lower
opacity/emissivity compared to the optical, the NIR flux starts to
{\it decrease} as the spectral energy distribution (SED) hardens, as
early as 7 days prior to bolometric maximum for the $H$ and $K_s$
bands in model DDC15 (Fig.~\ref{fig:comp_lc}).  In SN~2002bo, the
initial peak in these bands occurs $\sim$3 days later than in our
models, similar to the evolution in the $J$ band (which is well
matched by our model).

%%% FIGURE: optical/NIR LC

%%% TABLE: lcprop
\begin{table*}
\footnotesize
\caption{Light-curve properties. For each parameter, the upper line corresponds to model DDC15 (the upper and lower values, when added to the DDC15 value, correspond to models DDC10 and DDC17, respectively), and the lower line to SN~2002bo (1$\sigma$ errors are given in between parentheses).}\label{tab:lcprop}
\begin{tabular}{l@{\hspace{2.8mm}}l@{\hspace{2.8mm}}l@{\hspace{2.8mm}}l@{\hspace{2.8mm}}l@{\hspace{2.8mm}}l@{\hspace{2.8mm}}l@{\hspace{2.8mm}}l@{\hspace{2.8mm}}l}
\hline
\multicolumn{1}{c}{Parameter} & \multicolumn{1}{c}{$U$} & \multicolumn{1}{c}{$B$} & \multicolumn{1}{c}{$V$} & \multicolumn{1}{c}{$R$} & \multicolumn{1}{c}{$I$} & \multicolumn{1}{c}{$J$} & \multicolumn{1}{c}{$H$} & \multicolumn{1}{c}{$K_s$} \\
\hline
$\alpha$ in $L(t)\propto t^\alpha$            &       \multicolumn{1}{c}{$4.0^{+1.2}_{+1.4}$} &       \multicolumn{1}{c}{$5.9^{-0.1}_{-0.3}$} &       \multicolumn{1}{c}{$5.3^{-0.6}_{-0.4}$} &       \multicolumn{1}{c}{$4.1^{-0.6}_{-0.3}$} &       \multicolumn{1}{c}{$3.7^{-0.5}_{-0.2}$} &       \multicolumn{1}{c}{$2.6^{-0.1}_{-0.1}$} &       \multicolumn{1}{c}{$2.6^{-0.1}_{-0.1}$} &       \multicolumn{1}{c}{$2.5^{-0.0}_{-0.0}$} \\
($t_{\rm exp} < 3$~d)                         &        \multicolumn{1}{c}{$\cdots$} &        \multicolumn{1}{c}{$\cdots$} &        \multicolumn{1}{c}{$\cdots$} &        \multicolumn{1}{c}{$\cdots$} &        \multicolumn{1}{c}{$\cdots$} &        \multicolumn{1}{c}{$\cdots$} &        \multicolumn{1}{c}{$\cdots$} &        \multicolumn{1}{c}{$\cdots$} \\ \hline
$\Delta t_{\rm peak}$                         &             $-1.10^{-0.65}_{+0.57}$ &             $-0.39^{+0.57}_{-0.65}$ &             $+3.25^{-0.24}_{-0.07}$ &             $+2.19^{+0.39}_{-0.99}$ &             $+1.55^{-0.53}_{-0.29}$ &             $-4.05^{-1.23}_{+0.69}$ &             $-6.55^{-1.19}_{+1.30}$ &             $-6.62^{-0.89}_{+0.68}$ \\
$[{\rm day}]$                                 &                      $-$0.68 (0.70) &                      $+$0.61 (0.60) &                      $+$2.44 (0.67) &                      $+$2.15 (0.54) &                      $-$1.47 (0.34) &                      $-$1.39 (0.14) &                      $-$1.10 (0.14) &                      $-$0.84 (0.28) \\ \hline
$\Delta t_{\rm 2^{\rm nd}\ max}$              &        \multicolumn{1}{c}{$\cdots$} &        \multicolumn{1}{c}{$\cdots$} &        \multicolumn{1}{c}{$\cdots$} &        \multicolumn{1}{c}{$\cdots$} &            $+20.42^{+2.91}_{-3.50}$ &            $+23.42^{+1.99}_{-2.08}$ &            $+20.66^{+2.91}_{-3.63}$ &            $+20.84^{+2.35}_{-3.51}$ \\
$[{\rm day}]$                                 &        \multicolumn{1}{c}{$\cdots$} &        \multicolumn{1}{c}{$\cdots$} &        \multicolumn{1}{c}{$\cdots$} &        \multicolumn{1}{c}{$\cdots$} &                     $+$24.45 (1.52) &                     $+$28.36 (0.39) &                     $+$24.93 (0.42) &                     $+$24.81 (0.98) \\ \hline
$M_{\rm peak}$                                &            $-19.40^{-0.34}_{+0.59}$ &            $-19.15^{-0.19}_{+0.22}$ &            $-19.23^{-0.17}_{+0.17}$ &            $-19.20^{-0.13}_{+0.13}$ &            $-18.83^{-0.04}_{+0.03}$ &            $-18.53^{+0.02}_{-0.03}$ &            $-18.13^{+0.02}_{-0.07}$ &            $-18.07^{+0.00}_{-0.04}$ \\
$[{\rm mag}]$                                 &                     $-19.38$ (0.22) &                     $-19.20$ (0.11) &                     $-19.20$ (0.10) &                     $-19.10$ (0.09) &                     $-18.75$ (0.10) &                     $-18.43$ (0.14) &                     $-18.18$ (0.14) &                     $-18.13$ (0.14) \\ \hline
$M_{2^{\rm nd}{\rm max}}$                     &        \multicolumn{1}{c}{$\cdots$} &        \multicolumn{1}{c}{$\cdots$} &        \multicolumn{1}{c}{$\cdots$} &        \multicolumn{1}{c}{$\cdots$} &            $-18.86^{+0.07}_{-0.00}$ &            $-17.93^{+0.14}_{-0.02}$ &            $-18.83^{+0.06}_{+0.01}$ &            $-18.73^{+0.06}_{+0.01}$ \\
$[{\rm mag}]$                                 &        \multicolumn{1}{c}{$\cdots$} &        \multicolumn{1}{c}{$\cdots$} &        \multicolumn{1}{c}{$\cdots$} &        \multicolumn{1}{c}{$\cdots$} &                     $-18.33$ (0.13) &                     $-17.44$ (0.13) &                     $-18.21$ (0.14) &                     $-18.05$ (0.15) \\ \hline
$\Delta M_{15}$                               & \phantom{$-$}$1.30^{-0.17}_{+0.10}$ & \phantom{$-$}$1.03^{-0.10}_{+0.24}$ & \phantom{$-$}$0.71^{-0.05}_{+0.17}$ & \phantom{$-$}$0.54^{+0.08}_{-0.03}$ & \phantom{$-$}$0.08^{+0.24}_{-0.12}$ & \phantom{$-$}$1.31^{-0.12}_{-0.11}$ & \phantom{$-$}$0.06^{+0.07}_{-0.19}$ &             $-0.04^{-0.02}_{-0.06}$ \\
$[{\rm mag}]$                                 &            \phantom{$-$}1.53 (0.13) &            \phantom{$-$}1.18 (0.07) &            \phantom{$-$}0.70 (0.03) &            \phantom{$-$}0.71 (0.04) &            \phantom{$-$}0.64 (0.03) &            \phantom{$-$}1.65 (0.06) &            \phantom{$-$}0.37 (0.03) &            \phantom{$-$}0.42 (0.07) \\ \hline
$\Delta M/\Delta t_{50+}$                     &           $0.024^{+0.001}_{+0.000}$ &           $0.016^{+0.001}_{+0.000}$ &           $0.026^{+0.001}_{-0.001}$ &           $0.030^{+0.001}_{-0.001}$ &           $0.034^{+0.002}_{-0.002}$ &           $0.056^{+0.001}_{-0.002}$ &           $0.041^{-0.000}_{+0.000}$ &           $0.037^{-0.002}_{+0.002}$ \\
$[{\rm mag\ day}^{-1}]$                       &                       0.019 (0.013) &                       0.011 (0.005) &                       0.023 (0.003) &                       0.033 (0.002) &                       0.041 (0.001) &        \multicolumn{1}{c}{$\cdots$} &        \multicolumn{1}{c}{$\cdots$} &        \multicolumn{1}{c}{$\cdots$} \\ \hline
\end{tabular}

\flushleft
{\bf Notes:}
the $I$-band light curves for models DDC10 and DDC15 have several glitches around maximum light, hence the parameter values are less reliable;
$\Delta t_{\rm peak}$ is the time difference between maximum light in a given band ({\it first} maximum for $IJHK_s$) and pseudo-bolometric ($U\rightarrow K_s$) maximum;
$\Delta t_{2^{\rm nd}{\rm max}}$ is the time difference between the secondary maximum ($IJHK_s$ only) and pseudo-bolometric ($U\rightarrow K_s$) maximum;
$\Delta M/\Delta t_{50+}$ is the average magnitude decline rate past +50~d from maximum light.
\end{table*}

In the optical, the shorter rise times in the $U$ and $B$ bands with
respect to $V$ and $R$ reflects the temperature decline in the
spectrum-formation region after about one week past explosion. By
maximum light, most of the flux originates from layers
$\lesssim$15000~\kms\ where the temperature is decreasing
(Fig.~\ref{fig:temp_evol}).

The $U-V$ and $B-V$ colours shift to the red at $\sim$10 days past
explosion (one week before $B$-band maximum), i.e. 10--20 days before
the other colours shown in Fig.~\ref{fig:comp_colors}. This earlier
shift in colour results from the increase in opacity from once-ionized
IGEs that deplete the flux in $U$ and $B$ relative to redder bands.

The \dmft\ parameter for our model is 1.03~mag, consistent with that
inferred for SN~2002bo ($1.18\pm0.07$~mag; see
Table~\ref{tab:lcprop}).  In the redder optical bands and in the NIR
range, the post-maximum decline rate is sensitive to the location of
line emission.  In the $IHK_s$ bands, line emission is sufficient to
maintain a near-constant flux level despite the overall decline in
bolometric luminosity ($\Delta M_{15}(IHK_s)\approx 0$~mag), while the
weaker line emission within the $J$ band causes a more pronounced
decline ($\Delta M_{15}(J)\approx 1.3$~mag).

At late times, the colours shown in Fig.~\ref{fig:comp_colors} become
progressively bluer despite the near-constant temperature
(7000--8000~K in the ejecta layers between 5000~\kms\ and 15000~\kms;
see Fig.~\ref{fig:temp_evol}): The low ejecta densities favour the
formation of forbidden lines, whose wavelength location impacts the
colour evolution independent of temperature.

Our model reproduces the complex evolution of the optical and NIR
colours of SN~2002bo out to $\sim$80 days past maximum light.  This
gives us confidence that the apparent red colours of SN~2002bo are
indeed due to extinction in the host galaxy. Moreover, the observed
colour curves are corrected for reddening using the classical
\cite{CCM89} extinction law with a standard value of the
total-to-selective extinction ratio $R_V=3.1$.

However,
the model is very red within the first $\sim$5 days since
explosion, where we lack SN~2002bo data. Earlier observations of
several other \sneia\ (e.g., SN~2011fe, \citealt{Nugent/etal:2011};
SN~2013dy, \citealt{SN2013dy}), all of which display a narrower
Si\two\ 6355~\AA\ line than SN~2002bo around maximum, reveal
significantly bluer colours than predicted by our model. Such blue
colours at early times are difficult to reproduce with standard 1D
delayed-detonation models, even with strong outward mixing of \nifs,
and instead appear to require some form of hydrodynamical interaction
to heat up the outer ejecta layers, as in the pulsating
delayed-detonation models of D14a.\footnote{Although such an
    interaction serves to decelerate these outer ejecta layers,
    inhibiting the formation of broad spectral lines as in
    SN~2002bo.} Earlier 
observations of a 
broad-lined \snia\ such as SN~2002bo would enable to check the
prediction of red early-time optical colours in our model.

%%%%%%%%%%%%%%%%%%%%%%%%%%%%%%%%%%%%%%%%%%%%%%%%%%%%%%%%%%%%%%%%%%%%%%
%%%%%%%%%%%%%%%%%%%%%%%%%%%%%%%%%%%%%%%%%%%%%%%%%%%%%%%%%%%%%%%%%%%%%%

\section{Spectroscopic Evolution}\label{sect:spec}

\begin{figure*}
\centering
\includegraphics{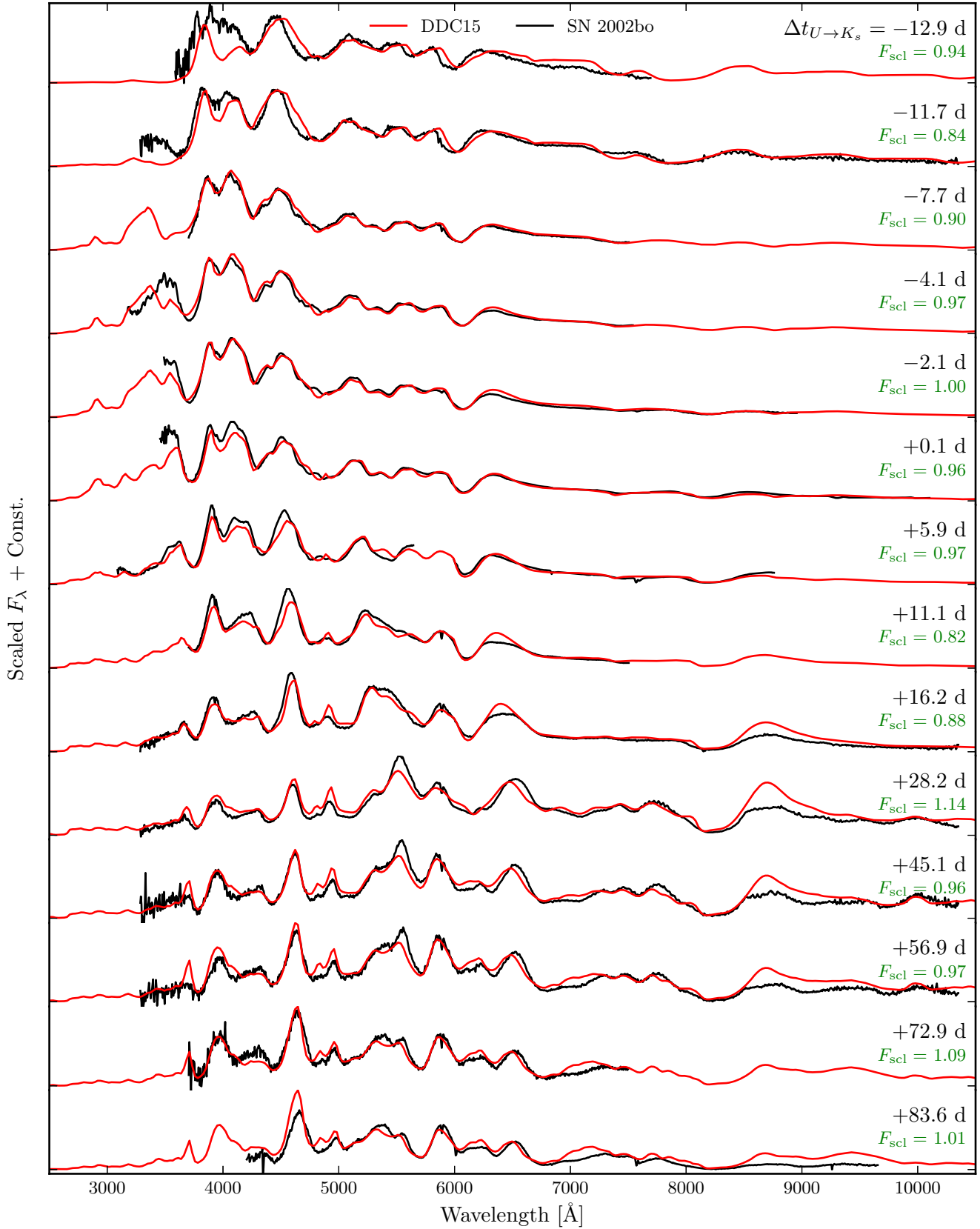}
\caption{\label{fig:spec_opt} Optical spectroscopic evolution of model
  DDC15 (red) compared to SN~2002bo (black), between $-12.9$~d and
  +83.6~d from pseudo-bolometric ($U\rightarrow K_s$) maximum.  The
  tickmarks on the ordinate give the zero-flux level.  The observed
  spectra have been de-redshifted, de-reddened, and scaled to match
  the absolute $V$-band magnitude inferred from the corresponding
  photometry. An additional scaling ($F_{\rm scl}$; green label) has
  been applied to the synthetic spectra to reproduce the mean
  observed flux in the range 5000--6500~\AA.}
\end{figure*}

We describe below the evolution of the optical
(Fig.~\ref{fig:spec_opt}) and NIR (Fig.~\ref{fig:spec_nir}) spectra of
our model and compare them to SN~2002bo observations.  Tables with
detailed line identifications, and a collection of plots
illustrating the influence of individual ions on the synthetic
spectra, are available in Appendices~\ref{sect:line_ids} and
\ref{sect:ladder}, respectively.

%%% FIGURE: comparison of DDC15 vs. 02bo optical spectra at all times

%%%%%%%%%%%%%%%%%%%%%%%%%%%%%%%%%%%%%%%%%%%%%%%%%%%%%%%%%%%%%%%%%%%%%%

\subsection{Early evolution}\label{sect:spec_early}

During the first $\sim$10 days past explosion, the majority of the
optical flux emerges from the IME-rich regions of the ejecta at
$\gtrsim$15000~\kms. The optical spectrum is thus dominated by broad
lines of Mg\two, Si\two, S\two, and Ca\two. We also note the presence
of O\one\,7773~\AA, which causes a deep absorption blueward of the
Ca\two\ 8500~\AA\ triplet. 
Iron-group elements, most notably Sc\two, Ti\two, Cr\two, Fe\two,
Co\two, and Ni\two, also contribute to the flux shortward of
$\sim$5000~\AA. 

We obtain high-velocity absorption ($>25000$~\kms\ blueshifts) in both
the Ca\two\ H \& K and 8500~\AA\ triplet features.  These remain
optically thick well beyond 30000~\kms\ where the Ca abundance is at
its primordial solar value ($X\approx6\times10^{-5}$), but their blue
wings are affected by absorption from other lines and thus cannot
easily serve as probes of the progenitor metallicity (see
 also \citealt{Lentz/etal:2000} in the context of the W7 model). The
Si\two\ 3858~\AA\ line has negligible impact on our synthetic spectra,
and the predicted double-absorption feature at $\sim$3500~\AA\ seen,
e.g., in the $-7.7$~d spectrum results from Ca\two\ alone. 
Multiple absorption components in both Ca\two\ features are
  routinely observed in early-time \snia\ spectra
  \citep[e.g.,][]{Childress/etal:2014}.

Unburnt material is located in the outermost ($\gtrsim$20000~\kms)
low-density layers of the ejecta. This configuration makes the
associated spectral signatures of carbon (e.g., C\two\ 6578~\AA) too
weak to be visible. The O\one\ 7773~\AA\ line can in principle serve
to constrain the amount of unburnt oxygen, but it forms below
$\sim$25000~\kms\ in our model and hence results in part from
C-burning ash.

In the NIR range, our model reproduces the overall shape of
the earliest spectrum of SN~2002bo at $-8.5$~d
($\sim$9 days past explosion based on model DDC15), but
  predicts a higher flux level than observed ($F_{\rm scl}=0.64$).
Three ions are responsible for the spectral signatures
seen: O\one, Mg\two, and Si\two\ (see Fig.~C2, upper-right
  panel). The two most notable 
features at these early times are a P-Cygni profile due to the
Mg\two\ 10927~\AA\ triplet and a broad emission feature at
$\sim$1.7~$\mu$m due to the Si\two\ 16931~\AA\ doublet.

Lines from IGEs are completely absent from the NIR range at these
early times, although lines of Sc\two, Ti\two, Cr\two, and Fe\two--{\sc
  iii} (with weaker absorptions from Co\two\ and Ni\two\ in the
$U$-band) contribute to the optical flux. The majority of the NIR flux
emerges from ejecta layers $\lesssim$20000~\kms\ where most IGEs are
at least twice-ionized, and contribute opacity at much shorter
wavelengths.

The earliest spectrum at $\sim$13 days before bolometric maximum,
however, shows that we predict too little flux around 4000~\AA, due to
excess absorption from Sc\two\ 4247~\AA\ and several Ti\two\ lines,
and that the blue extent of certain features is not well matched, in
particular the Fe\two\ absorption complex around 4800~\AA, and the two
Si\two\ 5972~\AA\ and 6355~\AA\ features. The
O\one\ 7773~\AA\ absorption is also too strong in our model. These
discrepancies are related to the low ejecta ionization fraction at
$\sim$20000~\kms\ (Fig.~\ref{fig:temp_evol}), and suggest a stronger
outward mixing of \nifs\ than adopted here. Another possibility is
that our extrapolated density profile at large velocity is steeper in
reality (see Fig.~\ref{fig:dens_profile}).

%%%%%%%%%%%%%%%%%%%%%%%%%%%%%%%%%%%%%%%%%%%%%%%%%%%%%%%%%%%%%%%%%%%%%%

\subsection{Maximum-light phase}\label{sect:spec_max}

Around maximum light the optical spectra are still dominated by strong
lines of singly-ionized IMEs, with IGEs absorbing the flux at
$\lesssim$5000~\AA. The drop in the Mg and O mass fractions below
$\sim$15000~\kms\ results in a gradual weakening of the Mg\two\ and
O\one\ 7773~\AA\ lines and reveals the transition from the C-burning
to the O-burning regime.  We note the disappearance of the
high-velocity absorption associated with Ca\two\ H\&K and the
8500~\AA\ triplet, as these transitions turn thin beyond
$\sim$20000~\kms.

In the near-infrared, we note the appearance of emission features
associated with IGEs, most notably permitted Co\two\ and forbidden
Co\three\ transitions.  There are no published NIR spectra of
SN~2002bo around maximum light to compare against our model.

%%% FIGURE: comparison of DDC15 vs. 02bo NIR spectra at all times
\begin{figure}
\centering
\includegraphics{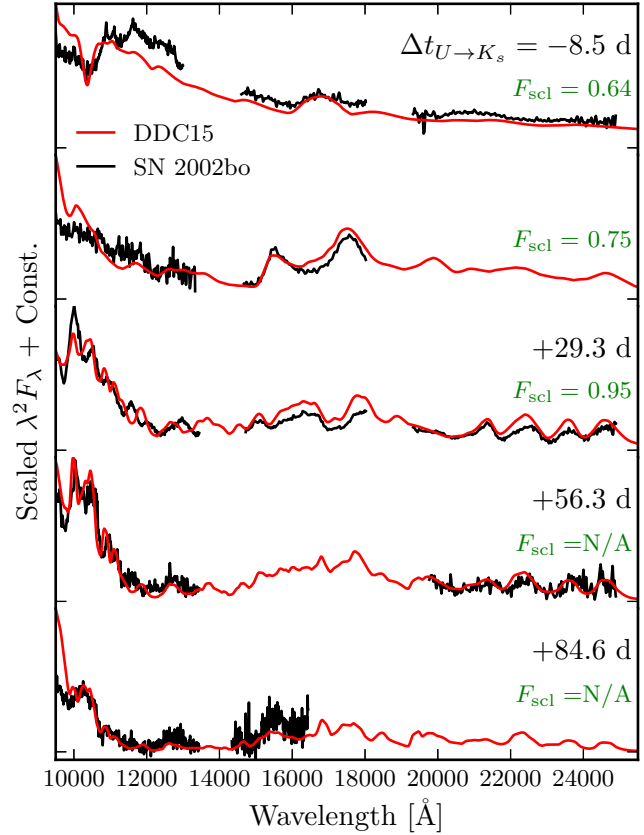}
\caption{\label{fig:spec_nir} Near-infrared spectroscopic evolution of
  model DDC15 (red) compared to SN~2002bo (black), between $-8.5$~d
  and +84.6~d from pseudo-bolometric ($U\rightarrow K_s$) maximum. The
  flux has been scaled by $\lambda^2$ for better visibility, and the
  tickmarks on the ordinate give the zero-flux level. The observed
  spectra have been de-redshifted, de-reddened, and scaled to match
  the absolute $J$-band magnitude inferred from the corresponding
  photometry, when available. An additional scaling ($F_{\rm scl}$;
  green label) has been applied to the synthetic spectra to reproduce
  the mean observed flux in the range 10000--11500~\AA.}
\end{figure}

%%%%%%%%%%%%%%%%%%%%%%%%%%%%%%%%%%%%%%%%%%%%%%%%%%%%%%%%%%%%%%%%%%%%%%

\subsection{Post-maximum and transition to nebular phase}\label{sect:spec_postmax}

From $\sim$10 until $\lesssim$40 days past maximum light the majority
of the flux emerges from the outer IGE-rich regions of the ejecta
($\lesssim$10000~\kms). Some strong lines of IMEs survive in the
optical range (Si\two, S\two, Ca\two), although they become weaker and
narrower with time. Most of the spectrum is now shaped by lines of
IGEs such as Ti\two, Cr\two, Fe\two--{\sc iii}, and Co\two--{\sc iii}.

The SED becomes redder due to the gradual drop in temperature in the
inner ejecta layers.  Radiation cooling dominates over expansion
cooling, with a growing importance played by forbidden-line
transitions.  A notable example is the 5900~\AA\ feature, well matched
by our model, and associated with [Co\three] 5888~\AA. This feature is
often mistaken for Na\one~D (see discussion in \citealt{D14_CoIII}).

This cooling further enhances absorption by Cr\two/Ti\two\ transitions
around 4000--4300~\AA. Several Cr\two\ lines are responsible for the
rapid change in relative flux around $\sim$5300~\AA\ between +16.2~d
and +28.2~d (Fig.~\ref{fig:spec_opt}), which reflects the rapid
evolution of the Cr\two/{\sc iii} ratio.

In the NIR, {\sc iii}$\rightarrow${\sc ii} recombination of IGEs
causes a gradual weakening of Co\three\ lines (primarily forbidden
transitions) and a strengthening of Co\two\ lines (primarily
permitted). In some bands ($H$ and $K_s$), emission by Co\two\ is
sufficient to maintain a near-constant flux level and causes an
earlier shift to the red of the $V-K_s$ colour curve with
respect to $V-J$ (Fig.~\ref{fig:comp_colors}).

The strengthening of the Co\two\ emission is partly responsible
for the mismatch in the Ca\two\ 8500~\AA\ triplet emission profile
between our model and SN~2002bo. Co\two\ lines blueward of
$\sim$8500~\AA\ emit photons in the inner ejecta
($\lesssim$10000~\kms) that are redshifted into resonance with the
Ca\two\ 8500~\AA\ triplet transitions in higher-velocity shells (see
also D14c). The resulting boost in emissivity explains the slower
decline in the $I$-band in our model compared to SN~2002bo
(cf. Fig.~\ref{fig:comp_lc} and Table~\ref{tab:lcprop}).

%%%%%%%%%%%%%%%%%%%%%%%%%%%%%%%%%%%%%%%%%%%%%%%%%%%%%%%%%%%%%%%%%%%%%%

\subsection{Late-time evolution out to $\sim$100 days past explosion}\label{sect:spec_late}

From $\sim$40~days past maximum onwards the spectrum-formation region
is almost exclusively located in the IGE-rich layers of the ejecta
below 10000~\kms. Apart from the two strong Ca\two\ H\&K and
8500~\AA\ triplet features and traces of Si\two\ 6355~\AA, which
retain their P-Cygni profile morphology, the only optical lines from
IMEs are optically-thin transitions of [S\two], [S\three],
and [Ar\three]. Beyond 1$\mu$m, the only emission from IMEs comes from
[S\two] and Ca\two\ transitions, although the impact on the observed
flux is negligible given the overlap with stronger emission from
Fe\two\ and Co\two.

Non-thermal ionization prevents a complete {\sc iii}$\rightarrow${\sc
  ii} recombination of IGEs in the inner ejecta, resulting in
increasing emission from lines of [Fe\three] and [Co\three], now
favoured due to the low ejecta density. Some of these lines represent
a fair fraction of the total flux in a given band: At $\sim$100 days
past explosion, the [Fe\three] 4658~\AA\ and [Co\three]
5888~\AA\ lines alone contribute $\sim$15 per cent of the total
flux in the $B$ and $R$ bands, respectively. The ratio of the total flux
emitted in both lines follows the increase in the Fe/Co abundance
ratio resulting from $^{56}$Co~$\rightarrow$~$^{56}$Fe decay
\citep[see also][]{Kuchner/etal:1994}.  The late-time colour evolution
is thus largely determined by the wavelength location of individual
strong transitions and no longer reflects the evolution of the ejecta
temperature, which remains almost constant in the spectrum-formation
region at these times (Fig.~\ref{fig:temp_evol}).

In the NIR, we predict the gradual emergence of [Ni\two] 1.94~$\mu$m,
due to {\it stable} Ni in the innermost ejecta ($\lesssim 2000$~\kms;
see Fig.~\ref{fig:elem_distrib}), resulting from burning at high
densities characteristic of a Chandrasekhar-mass WD progenitor
\citep[see, e.g.,][]{Hoeflich/etal:2002}.  However, most of the
emission just blueward of 2~$\mu$m is in fact due to lines of
[Co\three] in our model, not [Ni\two] as recently suggested by
\cite{Friesen/etal:2014} for SNe~2001fe, 2003du, and 2014J based on
the 1D delayed-detonation models of \cite{Dominguez/etal:2001}.
Unfortunately, the NIR spectrum of SN~2002bo at +56.3~d does not cover
this feature.

The match between the spectra of model DDC15 and SN~2002bo over the
first $\sim$100 days of its evolution suggests the composition of the
ejecta in our model is adequate at all depths. It also shows our
time-dependent radiative-transfer treatment consistently predicts the
evolution of the gas properties, in particular its ionization state.

%%%%%%%%%%%%%%%%%%%%%%%%%%%%%%%%%%%%%%%%%%%%%%%%%%%%%%%%%%%%%%%%%%%%%%
%%%%%%%%%%%%%%%%%%%%%%%%%%%%%%%%%%%%%%%%%%%%%%%%%%%%%%%%%%%%%%%%%%%%%%

\section{Sensitivity to \nifs\ mass}\label{sect:sens}

Our model DDC15 seems well suited to match the observed properties of
SN~2002bo. In this section, we investigate whether delayed-detonation
models with $\pm$0.1~\msun\ (i.e., $\pm$20 per cent) variations in
\nifs\ mass also provide a satisfactory match to this event (models
DDC10 and DDC17 of B13; see Table~\ref{tab:modinfo}). This is of
particular relevance to assess the actual accuracy of abundance
determinations in \sneia\ and to understand the degeneracy (or lack
thereof) of their properties.  We discuss in turn the impact on the
early brightening phase (Section~\ref{sect:sens_early}), on the
post-maximum decline rate and the width-luminosity relation
(\ref{sect:sens_wlr}), on the late-time evolution
(\ref{sect:sens_late}), and on the relative homogeneity of
\snia\ properties in the NIR (\ref{sect:sens_nir}). Finally, we comment
on the 
impact of the IGE abundance on the near-UV flux level
(\ref{sect:sens_abund}).

%%% FIGURE: sensitivity of optical/NIR spectra at selected times to 56Ni mass
\begin{figure*}
\centering
\includegraphics{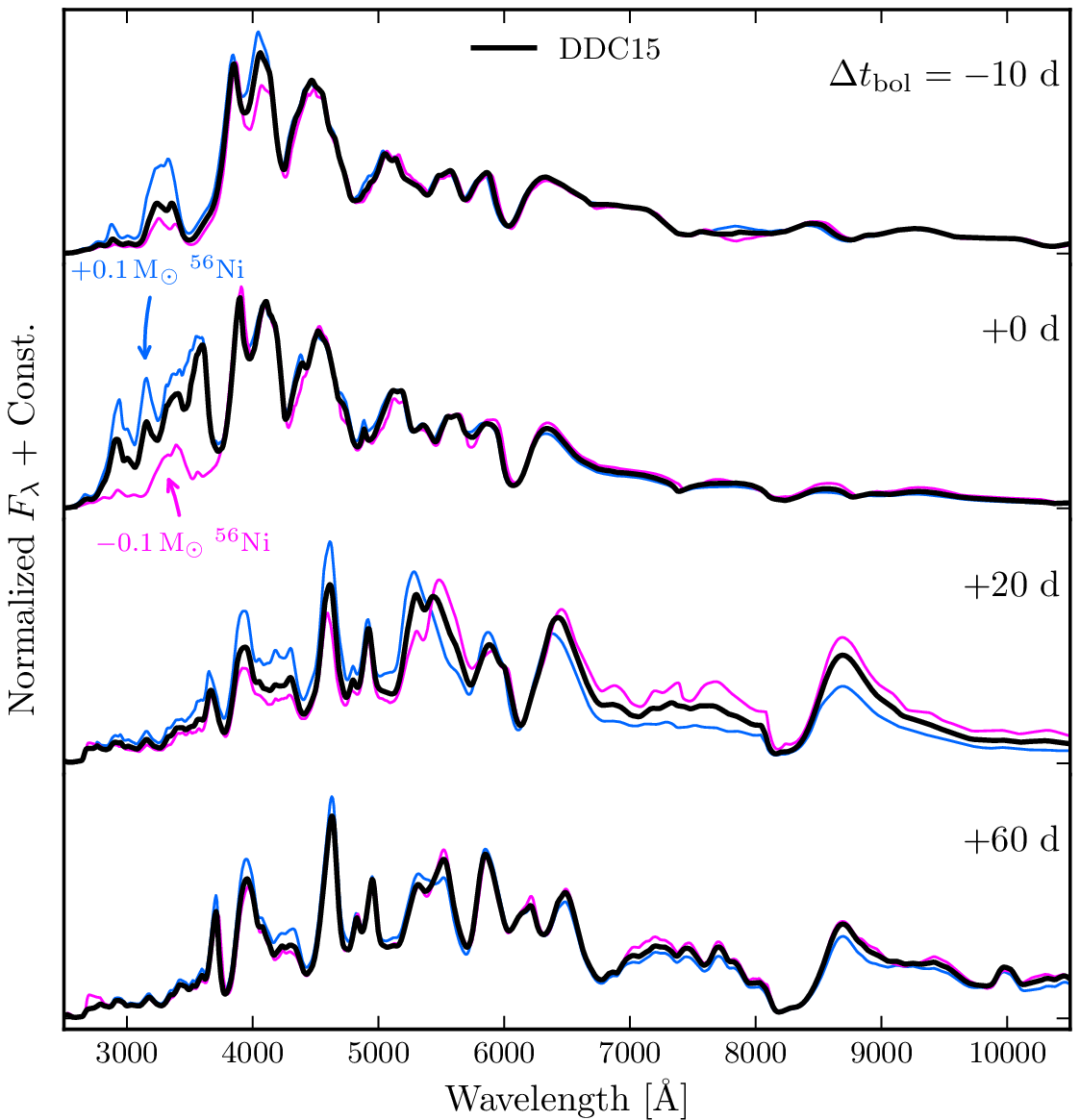}\hspace*{0cm}
\includegraphics{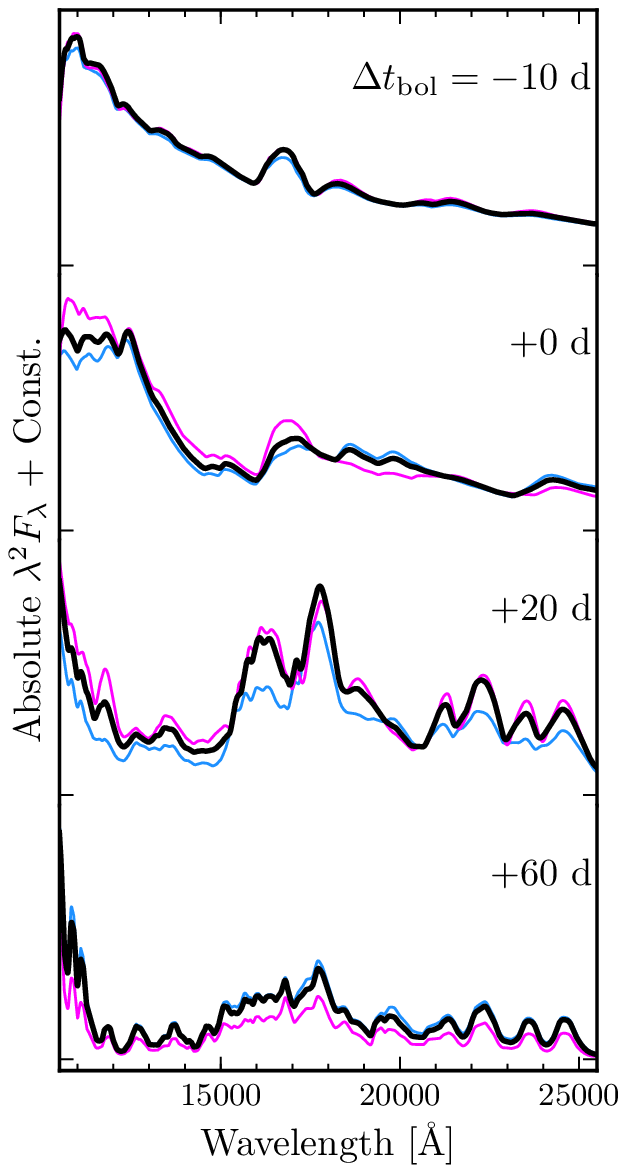}
\caption{\label{fig:spec_56ni} Illustration of the impact of a
  $\pm$0.1~\msun\ variation in \nifs\ mass (coloured curves) with respect to our
  reference model DDC15 (black) on the optical (left) and NIR (right)
  spectra at four selected times between $-10$ and +60~days from
  bolometric maximum.
  The optical spectra have been normalized to the same mean
    flux in the range 5000--6500~\AA\ to better visualize relative
    differences in the SED. The NIR spectra on the other hand are not
    normalized (but the flux has been scaled by $\lambda^2$ for better
    visibility), such that differences between the models correspond to
    real variations in absolute flux.
  The tickmarks on the ordinate give the zero-flux level.
  Notice the lower flux level in the near-UV relative to
  the optical in our model with {\it less} \nifs.}
\end{figure*}

%%%%%%%%%%%%%%%%%%%%%%%%%%%%%%%%%%%%%%%%%%%%%%%%%%%%%%%%%%%%%%%%%%%%%%

\subsection{Early brightening phase and high-velocity features}\label{sect:sens_early}

The early rate of increase in bolometric luminosity is largely
insensitive to the \nifs\ mass, as seen through the $<3$ per cent
variation in the power-law exponent $\alpha$ in
Table~\ref{tab:lcprop}. The same is true in individual bands, although
it is larger in $UBVR$ and is hardly noticeable in $I$ and the NIR
bands (Fig.~\ref{fig:comp_lc}).  This suggests that the early-time
luminosity, if powered by \nifs\ alone, is more sensitive to mixing
than slight variations in total \nifs\ mass (see D14a). Indeed, mixing
can increase the \nifs\ mass fraction by several orders of magnitude
in the outermost layers (see also Sect.~\ref{sect:comp}) where an
increase in \nifs\ mass only results in a modest variation of the
outer extent of the \nifs\ distribution (cf. $v(\nifs)$ column in
Table~\ref{tab:modinfo}).

Spectroscopically, the effect is most clearly seen in the UV range
(blueward of $\sim$3500~\AA; see Fig.~\ref{fig:spec_56ni}), through
variations in the absorption by once-ionized IGEs (Sc\two, Ti\two,
Cr\two, Fe\two, Co\two, and Ni\two).  The relative strength of the
high-velocity components to the Ca\two\ H\&K and 8500~\AA\ triplet
absorptions also varies significantly amongst our three models. They
are weaker in our model with more \nifs\ (DDC10) due to the higher Ca
ionization at 25000--35000~\kms. Conversely, the higher excitation in
the outer ejecta leads to enhanced absorption in the blue wings of
several lines, including the Fe\two\ absorption complex around
4800~\AA.  The increase in \nifs\ mass is not sufficient to match the
large Si\two~6355~\AA\ blueshift at early times, which might indicate
a greater outward mixing of \nifs\ (D14a; see also
\citealt{Baron/etal:2012}).

%%%%%%%%%%%%%%%%%%%%%%%%%%%%%%%%%%%%%%%%%%%%%%%%%%%%%%%%%%%%%%%%%%%%%%

\subsection{Post-maximum decline and the width-luminosity relation}\label{sect:sens_wlr}

The bolometric post-maximum decline rate (parametrized via the $\Delta
M_{15}$ parameter in Table~\ref{tab:bolprop}) changes by less than
0.1~mag from model DDC10 to DDC17. Part of this degeneracy with
\nifs\ mass stems from the increase in the $\gamma$-ray escape
fraction for models with more \nifs. By 20 days past explosion,
$\sim$10 per cent of the decay energy directly leaks out of the ejecta
in model DDC10, cf. only $\sim$5 per cent in DDC17.

However, an increase in the \nifs\ mass results in bluer colours at any
given time and a more gradual shift to redder colours
(Fig.~\ref{fig:comp_colors}), which affects the post-maximum decline
rate in the individual bands.  In the $B$-band, the $\Delta M_{15}$
parameter varies by $\sim0.4$~mag (0.92, 1.03, and 1.27~mag in order
of decreasing \nifs\ mass), or four times the bolometric
variation. Despite the relatively narrow range in peak $M_B$, we can
compute a width-luminosity relation (WLR) for these three models: $M_B
= 1.12[\dmft-1.1] - 19.11$.  This WLR is consistent with the observed
one for broad-lined \sneia\ in \cite{Blondin/etal:2012}.

Spectroscopically, the most notable impact of \nifs\ mass (other than
the absolute flux difference) is in the Cr\two-dominated region around
5300~\AA. As noted in Section~\ref{sect:spec_postmax}, this region is
particularly sensitive to the Cr ionization level.

%%%%%%%%%%%%%%%%%%%%%%%%%%%%%%%%%%%%%%%%%%%%%%%%%%%%%%%%%%%%%%%%%%%%%%

\subsection{Late-time evolution}\label{sect:sens_late}

The late-time bolometric magnitude decline is the same for our three
models ($\sim$0.028~mag\,day$^{-1}$), i.e., the rate of change of the
$\gamma$-ray escape fraction is nearly identical. This is a natural
consequence of the similar density structure and \nifs\ distribution
of the inner ejecta in our three Chandrasekhar-mass models.

The spectra are remarkably similar at these times, with minor
differences in SED confined to $\lesssim$6000~\AA, 
owing to the more uniform ionization state of IGEs. Variations in
Cr\two\ line opacity continue to affect the flux around 5300~\AA. 
Although not apparent in Fig.~\ref{fig:spec_56ni} due to the flux
normalization, the [Co\three] 5888~\AA\ emission strength correlates
with \nifs\ mass. This correlation is also
present in observations of \sneia\ and supports the association of
this feature with cobalt \citep[see][and references therein]{D14_CoIII}.

%%%%%%%%%%%%%%%%%%%%%%%%%%%%%%%%%%%%%%%%%%%%%%%%%%%%%%%%%%%%%%%%%%%%%%

\subsection{NIR homogeneity and secondary maxima}\label{sect:sens_nir}

An increase in \nifs\ mass causes an earlier hardening of the SED and
a delayed cooling of the ejecta. The result is an earlier decline (and
hence an earlier first maximum) in the NIR bands
(Table~\ref{tab:lcprop}).

\begin{figure}
\centering
\includegraphics{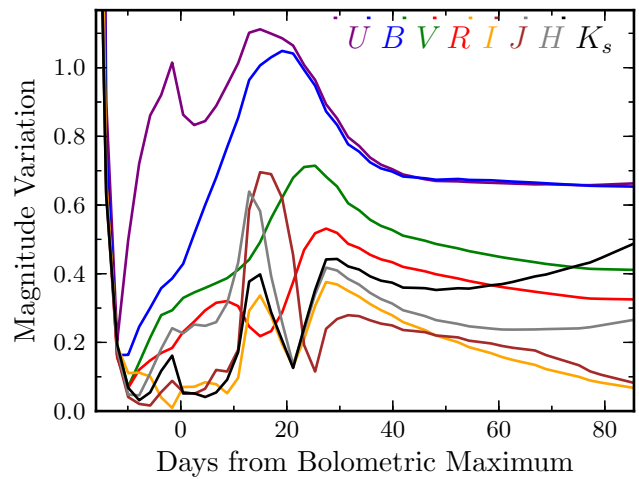}
\caption{\label{fig:mag_range} Evolution of the magnitude variation in
  various bands amongst our three models, spanning $\sim$0.2~\msun\ in
  \nifs\ mass.}
\end{figure}

The associated delayed {\sc iii}$\rightarrow${\sc ii} recombination of
IGEs has a stronger impact in the bluest bands, resulting in a larger
magnitude variation amongst our three models compared to redder
optical and NIR bands (Fig.~\ref{fig:mag_range}; see also B13,
Appendix~C).  The variation in the $I$, $J$, and $K_s$ magnitudes is
less than 0.1~mag (and occasionally less than 0.05~mag) between the
epoch of first maximum in those bands up until $\sim$10 days past
bolometric maximum. Such a small intrinsic dispersion (also reflected
in the uniformity of their spectra; see Fig.~\ref{fig:spec_56ni})
confirms the potential of NIR photometry of \sneia\ for accurate
distance determinations \cite[e.g.,][]{Krisciunas/etal:2004a}.

The same effect delays the time of NIR secondary maxima for larger
\nifs\ mass \citep[see also][]{Kasen:2006}, but leaves their magnitude
unchanged (Table~\ref{tab:lcprop}), in agreement with observations
\citep[][their Fig.~8]{Biscardi/etal:2012}. The result is a larger
magnitude variation in the NIR range at 15--20~d from bolometric
maximum, even exceeding that in the $V$ and $R$ bands.

%%% FIGURE: magnitude scatter in various bands vs. time

%%%%%%%%%%%%%%%%%%%%%%%%%%%%%%%%%%%%%%%%%%%%%%%%%%%%%%%%%%%%%%%%%%%%%%

\subsection{IGE abundance and near-UV flux level}\label{sect:sens_abund}

The $\pm$0.1~\msun\ variation in \nifs\ mass represents a
non-negligible variation in the relative abundances of IGEs and
IMEs\footnote{Since all but $\lesssim 0.01$~\msun\ of the progenitor
  WD is burnt in our delayed-detonation models, the
  $\pm$0.1~\msun\ variation in \nifs\ mass results in a comparable
  variation (of opposite sign) in IME mass.}  (see
Table~\ref{tab:modinfo}), which often result in a modest impact on the
predicted observables, illustrating the degeneracy of
\snia\ properties and the difficulties associated with abundance
determinations in these events (see next section).

Part of this degeneracy results from variations in ionization whose
impact often supersedes those associated with variations in abundance
in determining the ejecta opacity. In the near-UV range,
where most of the opacity is due to IGEs, it is our model with {\it
  less} \nifs\ (and hence less IGEs) that has the lowest near-UV flux
level relative to the optical up until maximum light
(Fig.~\ref{fig:spec_56ni}), due to the lower ionization level of
Ni/Co/Fe.
Our models thus show that an increase in the IGE abundance in
\sneia\ does not necessarily result in more opaque ejecta
  \cite[see also][]{Sauer/etal:2008}, which 
complicates the use of near-UV spectra to constrain the progenitor
metallicity \citep[see, e.g.,][]{Foley/Kirshner:2013}.

The larger sensitivity of the $U$-band flux to the mass of
\nifs\ results in a $\lesssim$1 mag variation in this band amongst our
three models around maximum light (Fig.~\ref{fig:mag_range}), and a
larger variation in $U-V$ compared to other colour indices
(Fig.~\ref{fig:comp_colors}), in line with observational findings
\citep[e.g.][]{Jha/etal:2006}. This large variation could significantly
affect reddening determinations of \sneia.

%%%%%%%%%%%%%%%%%%%%%%%%%%%%%%%%%%%%%%%%%%%%%%%%%%%%%%%%%%%%%%%%%%%%%%
%%%%%%%%%%%%%%%%%%%%%%%%%%%%%%%%%%%%%%%%%%%%%%%%%%%%%%%%%%%%%%%%%%%%%%

\section{Comparison to previous work}\label{sect:comp}

\begin{figure*}
\centering
\includegraphics{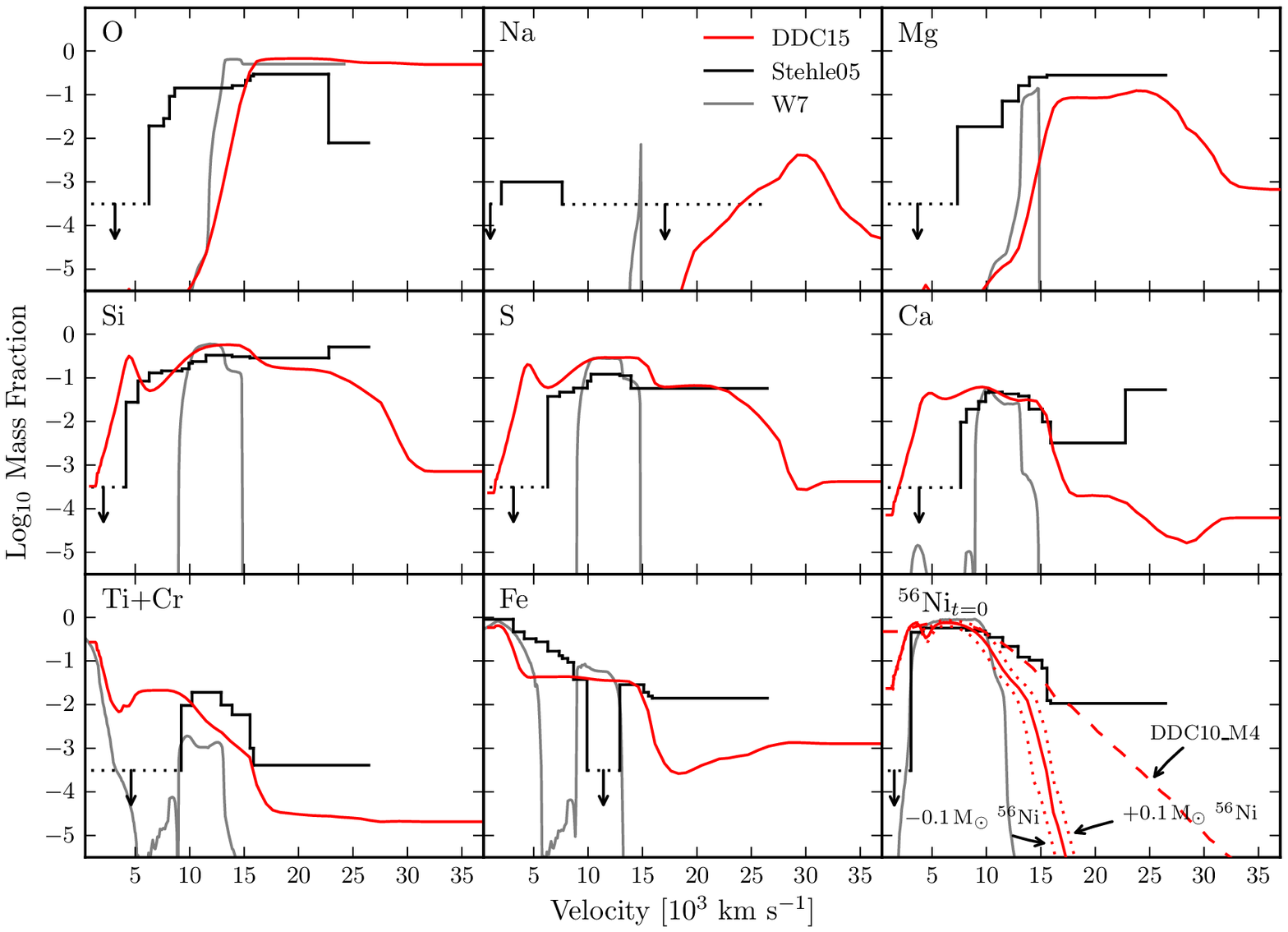}
\caption{\label{fig:comp_abund} Comparison of abundance profiles in
  our reference model DDC15 (red, solid line) 
 with those derived
  via spectroscopic modeling by {\protect\cite{Stehle/etal:2005}}
  [black; the data were manually scanned from their Fig.~5a] and the
  W7 model of {\protect\cite{W7}} [grey]. The \nifs\ distribution is
  shown at $t_{\rm exp}\approx 0$. Since the ordinate of Fig.~5a of
  {\protect\cite{Stehle/etal:2005}} does not extend below
  $\sim$3.1$\times$10$^{-4}$, we give upper limits as dotted lines
  plus a downward-pointing arrow. We also show the \nifs\ distribution
  for our models with $\pm$0.1~\msun\ of \nifs\ (red dotted
    lines), and
  for a mixed version of our model with +0.1~\msun\ of \nifs\ (noted
  DDC10\_M4 in D14a; red dashed line).  The abundances for Mg through
  Fe are solar beyond $\sim$35000~\kms\ in our model.}
\end{figure*}

In-depth studies of SN~2002bo have been presented by
\cite{Benetti/etal:2004} and \cite{Stehle/etal:2005}. The latter study
attempts to infer the chemical stratification of the SN~2002bo ejecta
based on steady-state spectroscopic modelling under the assumption of
a sharp photosphere emitting a pure thermal continuum. Stehle et al.
assume an ejecta structure (the density profile of W7 shown in
Fig.~\ref{fig:dens_profile}) and vary the abundances in order to
reproduce the observed spectra at selected epochs. Their abundance
profiles are shown in Fig.~\ref{fig:comp_abund}. For comparison, we
overplot the abundance profiles for our model DDC15 and those of
W7. The elemental yields are reported in
Table~\ref{tab:comp_abund}. We discuss in turn the inferred chemical
stratification and the validity of a diffusive inner boundary, which
is a fundamental ansatz of the method applied by
\cite{Stehle/etal:2005}.

\subsection{Chemical stratification}\label{sect:comp_strat}

The most striking difference of our results with those of
\cite{Stehle/etal:2005} is that they require a large overabundance of
IGEs and IMEs in the outer layers ($\gtrsim 15000$~\kms) in order to
reproduce the broad lines of SN~2002bo, including the blue wings of
Fe\two\ 4800~\AA\ at early times. Although our models possess
$\sim$1000 times less Ca (by mass) beyond $\sim$20000~\kms, the resulting
synthetic spectra exhibit high-velocity Ca\two\ features.  Such a
huge difference illustrates the difficulties associated with some
abundance determinations that are highly sensitive to the ionization
balance and outer density structure.

%%% FIGURE: comparison with Stehle05 abundance profiles

In the intermediate layers (10000--15000~\kms) the distribution of
IMEs is in good agreement with our model, although
\cite{Stehle/etal:2005} infer a larger Mg abundance
(0.08~\msun\ cf. $\sim 0.01$~\msun\ in our model) in order to
reproduce the Mg\two\ absorption feature around 4200~\AA\ in the
early-time spectra. Past maximum light, however, the deep absorption
at $\sim$4400~\AA\ is caused by several Fe\two\ lines in our model,
not by Mg\two, hence we can reproduce this feature despite the sharp
drop in Mg abundance below $\sim$15000~\kms. \cite{Stehle/etal:2005}
also argue for significant downward mixing of oxygen ($X_{\rm O}
\gtrsim 0.1$ above $\sim 8500$~\kms), when we satisfactorily fit the
O\one~7773~\AA\ feature up until shortly after maximum light with no
oxygen present below $\sim 10000$~\kms.

Contrary to our model, \cite{Stehle/etal:2005} infer the presence of
Na and the absence of Ti/Cr below $\sim 10000$~\kms. For Na, their
abundance constraint is most likely a misidentification of the
emission feature at $\sim 5900$~\AA\ past maximum light, which is due
to [Co\three] 5888~\AA, not Na\one\ \citep[see][]{D14_CoIII}.  As for
Ti/Cr, several Cr\two\ lines affect the flux level around
$\sim$5300~\AA\ past maximum light (Sect.~\ref{sect:spec_postmax}), so
that chromium must be present in the inner ejecta.\footnote{At the
  time their paper was written, \cite{Stehle/etal:2005} did not have
  access to any spectra of SN~2002bo between +6~d and +263~d past
  $B$-band maximum, and thus could provide only weak constraints on
  the abundance distribution in these intermediate ejecta layers.}

The \nifs\ distribution in our model is comparable to that inferred by
\cite{Stehle/etal:2005} in the range 5000--10000~\kms, but drops more
rapidly beyond 10000~\kms. With our distribution we reproduce the
early rise in bolometric luminosity (Fig.~\ref{fig:comp_lcbol}).
Nonetheless, we argued for some additional outward mixing of
\nifs\ based on the lack of absorption in the blue wings of several
features in the early-time spectra (Sect.~\ref{sect:spec_early}). In
fact, the \nifs\ distribution of the mixed model DDC10\_M4 of D14a
closely matches the \nifs\ distribution of
\cite{Stehle/etal:2005}. Note that we cannot really comment on the
distribution below $\sim$3000~\kms\ since we did not evolve our model
beyond $\sim$100 days past explosion, when most of the flux is
expected to emerge from the inner $\sim$0.1~\msun\ of ejecta.

%%% TABLE: comparison of abundances DDC15, S05, W7
\begin{table}
\footnotesize
\caption{Comparison of nucleosynthetic yields at $t_{\rm exp}\approx 0$.}\label{tab:comp_abund}
\begin{tabular}{lccc}
\hline
\multicolumn{1}{c}{Species} & DDC15 & Stehle05 & W7 \\
 & [\msun] & [\msun] & [\msun] \\
\hline
$^{56}$Ni            &           0.511 &           0.520 &           0.587 \\
Fe$^a$               &           0.100 &           0.360 &           0.145 \\
Ti$+$Cr              &           0.018 &           0.003 &           0.005 \\
Ca                   &           0.045 &           0.020 &           0.012 \\
S                    &           0.197 &           0.067 &           0.085 \\
Si                   &           0.306 &           0.220 &           0.154 \\
Mg                   &           0.011 &           0.080 &           0.013 \\
Na                   &        1.68(-5) &           0.001 &        6.32(-5) \\
O                    &           0.105 &           0.110 &           0.138 \\
C                    &           0.003 & $\lesssim$0.002 &           0.047 \\
\hline
\end{tabular}

\flushleft
{\bf Notes:}
Numbers in parenthesis correspond to powers of ten.\\
$^a$Stable isotopes only ($^{54}$Fe, $^{56}$Fe, $^{57}$Fe, and $^{58}$Fe).
Since these yields are given at $t_{\rm exp}\approx 0$, the stable iron yield does not include \fefs\ from \nifs\ decay.
\end{table}

\subsection{The validity of a diffusive inner boundary}\label{sect:nlte}

\begin{figure}
\centering
\includegraphics{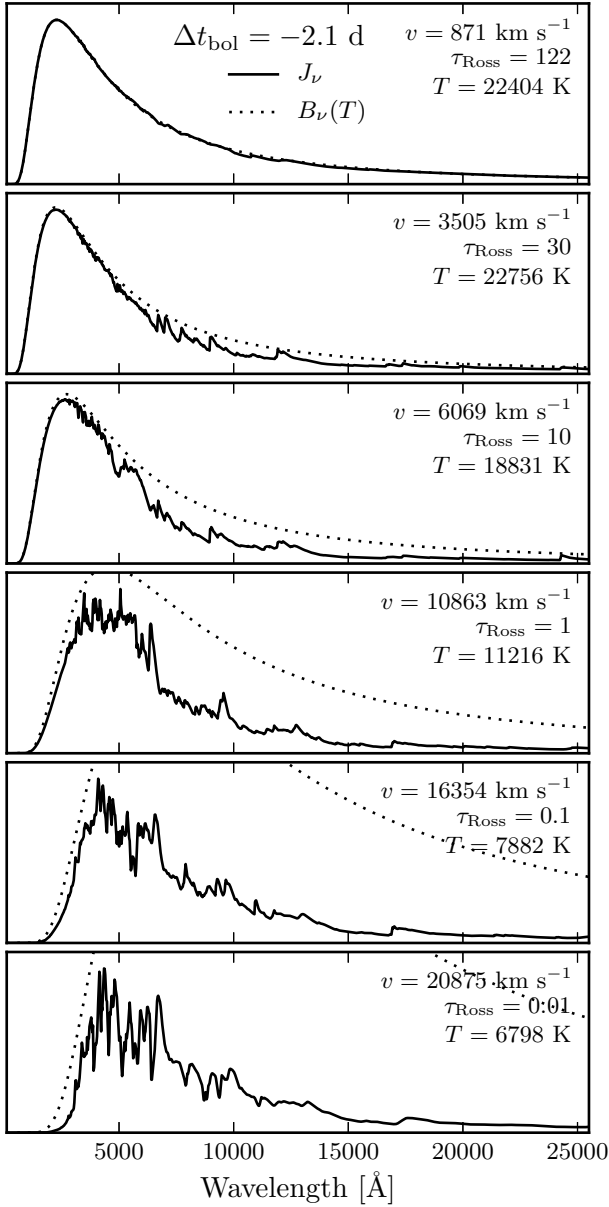}
\caption{\label{fig:jb} Illustration of the wavelength dependence of
  the mean intensity $J_\nu$ (solid lines) and the Planck function
  $B_\nu$ (dotted lines) at different depths in the ejecta for model
  DDC15 at $-2.1$~d from bolometric maximum. From top to bottom, the
  panels correspond to locations from the inner to the outer
  ejecta. For each location, we give the velocity, the Rosseland-mean
  optical depth, and the gas temperature. As the optical depth
  decreases, $J_\nu$ deviates more and more from $B_\nu(T)$.}
\end{figure}

When modelling the early-time spectra of SN~2002bo
\cite{Stehle/etal:2005} assume there exists a sharp photosphere at a
pre-defined velocity emitting a thermal continuum, yet the importance
of line scattering in \sneia\ leads to strong departures from local
thermodynamic equilibrium (LTE) in both the gas (level populations)
and the radiation field (mean intensity $J_\nu$). In our simulations,
we solve the non-LTE time-dependent radiative transfer for the whole
ejecta, with a zero-flux inner boundary condition at the base of the
innermost ejecta shell. We can thus check how $J_\nu$ and $B_\nu$
deviate from each other as function of time since explosion, and as a
function of depth in the ejecta.  In Fig.~\ref{fig:jb} we illustrate
the departures of the mean intensity from a Planckian distribution
based on our non-LTE time-dependent radiative transfer 
calculations with \cmfgen\ for model DDC15 around bolometric
maximum.

In the innermost regions, at a Rosseland-mean optical depth $\tau_{\rm
  Ross}\gtrsim 100$ ($v\lesssim 1500$~\kms), we have $J_\nu\approx
B_\nu(T)$ at all UV, optical, and NIR wavelengths. Hence, the non-LTE
solver recovers LTE conditions, i.e., the radiation and the gas are in
equilibrium at the same (local) temperature $T$. At $\tau_{\rm
  Ross}\approx 30$ ($v\approx 3500$~\kms), $J_\nu < B_\nu(T)$ in the
red part of the optical and in the NIR, but the radiation is still
Planckian below $\sim$5000~\AA, where the large number of lines
ensures an efficient thermalization. As we progress further out to
$\tau_{\rm Ross}$ of 10 and 1 ($\sim$6000 and $\sim$11000~\kms,
respectively), $J_\nu$ drops below $B_\nu(T)$ even in the UV, although
the SED remains smooth below
$\sim$2500~\AA. This persists out to $\tau_{\rm Ross} \approx 0.1$
($v\approx 16000$~\kms), although $B_\nu(T)$ is then no longer a good
representation of $J_\nu$.

Thus, already at an optical depth of several tens, injecting photons
with a Planckian distribution would lead to an overestimate of the
flux in the red part of the spectrum.  The situation is even worse
when only the outermost (optically thin) ejecta are treated and one
assumes a sharp blackbody emitting photosphere at $\tau\approx1$, as
done by \cite{Stehle/etal:2005}. As a result, they invoke a lower
reddening for SN~2002bo than that inferred from observational
constraints.  In contrast, our full non-LTE solution treats the entire
ejecta and yields a satisfactory match to the overall SED at all times
with a total reddening $E(B-V)=0.41$~mag
(Figs.~\ref{fig:spec_opt}--\ref{fig:spec_nir}).

%%% FIGURE: J vs. B as a function of depth, time

%%%%%%%%%%%%%%%%%%%%%%%%%%%%%%%%%%%%%%%%%%%%%%%%%%%%%%%%%%%%%%%%%%%%%%
%%%%%%%%%%%%%%%%%%%%%%%%%%%%%%%%%%%%%%%%%%%%%%%%%%%%%%%%%%%%%%%%%%%%%%

\section{Discussion and Conclusion}\label{sect:ccl}

We have presented a detailed analysis of the photometric and
spectroscopic evolution of a 1D Chandrasekhar-mass delayed-detonation
model for the Type Ia SN~2002bo. Beyond providing a suitable model for
this event, it provides a reference for understanding all
\sneia\ similar to that prototype.

The good match we obtain with the light curves and spectra over the
first $\sim100$~days since explosion gives some support to the
standard single-degenerate model for these objects. It remains to be
studied whether WD-WD merger models \citep[e.g.,][]{Pakmor/etal:2012}
can reproduce events like SN~2002bo as well as delayed-detonations of
single WDs at the Chandrasekhar mass. The ejected mass distribution of
\sneia\ found by \cite{Scalzo/Ruiter/Sim:2014} is strongly peaked at
$\sim$1.4~\msun, yet reveals a significant fraction of
sub-Chandrasekhar-mass progenitors, although these are associated with
the least luminous events.  While it is unclear to what extent a
direct confrontation to observations reveals the successes and
shortcomings either of the radiative-transfer treatment or of the
underlying explosion model (or both), a proper assessment requires
treating the entire SN ejecta in non-LTE at all times (see
section~\ref{sect:nlte}).

Our assumption of spherical symmetry and our somewhat artificial
treatment of the delayed-detonation mechanism are not detrimental to
reproducing the radiative properties of a proto-typical \snia.  The
mild disagreement between our model and SN~2002bo at the earliest
times may be resolved by invoking a greater outward mixing of \nifs,
but could also indicate some fundamental characteristic of the
  explosion not captured by our artificial, one-dimensional setup.

In B13 we noted a general overestimate of the width of the
characteristic Si\two\ 6355~\AA\ line in our models to account for the
bulk of normal \sneia.  While this was originally flagged as a
potential overestimate of the kinetic energy in our input 1D
hydrodynamical models, it has since then become clear that the
observed diversity in spectral line widths could reflect profound
physical differences in the explosion mechanism (in particular,
pulsating delayed detonations result in a narrower
Si\two\ 6355~\AA\ line; see D14a). The delayed-detonation models
discussed in B13 are thus well-suited to broad-lined \sneia.

The observed paucity of C\two\ lines in the early-time spectra of
broad-lined \sneia\ is then physical and not related, for example, to line
overlap. In the pulsating delayed-detonation models of D14a, carbon is
present down to $\sim$15000~\kms, close to the spectrum-formation
region at early times. C\two\ lines are thus predicted in these
models, along with a narrower Si\two\ 6355~\AA\ line profile. The
preferential association of C\two\ detections with narrow-lined
\sneia\ (termed ``Core Normal'' by \citealt{Branch/etal:2006})
is backed up by extensive data sets \citep[e.g.,][]{Parrent/etal:2011}.

We show that $\pm$0.1~\msun\ variations in \nifs\ mass have a modest
impact on the bolometric and colour evolution of our
model. Spectroscopically, most of the effect is seen blueward of
$\sim$3500~\AA\ up until maximum light, our model with less
\nifs\ displaying a lower near-UV flux level despite the lower
abundance of IGEs. In the NIR, an increase in \nifs\ mass delays the
secondary maxima but does not affect their magnitude, as found in
observational studies. More importantly, our models confirm the
homogeneous nature of \sneia\ in the NIR, with a $<0.1$~mag variation
in the $I$, $J$, and $K_s$ bands around maximum light.

Allowing for homologous expansion of the ejecta inherited from the
hydrodynamical modelling of the explosion, combined with a
self-consistent time-dependent radiative-transfer treatment, our
models yield a better match to SN~2002bo than previous parametrized
approaches. The complex nature of the radiative transfer problem (line
overlap, line saturation, uncertain ionization, scattering
vs. absorption, non-thermal processes, etc.) does not permit a direct
and accurate determination of elemental abundances. Instead, a global
modeling of the ejecta to predict simultaneously the bolometric and
multi-band light curves, as well as the optical and NIR spectra,
allows a better assessment of the adequacy of a given model.

%%%%%%%%%%%%%%%%%%%%%%%%%%%%%%%%%%%%%%%%%%%%%%%%%%%%%%%%%%%%%%%%%%%%%%
%%%%%%%%%%%%%%%%%%%%%%%%%%%%%%%%%%%%%%%%%%%%%%%%%%%%%%%%%%%%%%%%%%%%%%

\section*{acknowledgements}

SB acknowledges useful discussions with Suhail Dhawan and Bruno
Leibundgut during a one-month visit to ESO as part of the ESO
Scientific Visitor Programme. Thanks to Ken'ichi Nomoto for sending
before-decay abundances of the W7 model, and to Stefano Benetti for
sending us the NIR spectrum of SN~2002bo at +56.3 days from
pseudo-bolometric ($U\rightarrow K_s$) maximum. LD and SB acknowledge
financial support from the European Community through an International
Re-integration Grant, under grant number PIRG04-GA-2008-239184, 
from ``Agence Nationale de la Recherche'' grant
ANR-2011-Blanc-SIMI-5-6-007-01, and from the Programme National de 
Physique Stellaire (PNPS) of CNRS/INSU, France. DJH acknowledges support from STScI
theory grant HST-AR-12640.01, and NASA theory grant NNX14AB41G.  DJH
would also like to acknowledge the hospitality and support of the
Distinguished Visitor program at the Research School of Astronomy and
Astrophysics (RSAA) at the Australian National University (ANU).  This
work was granted access to the HPC resources of CINES under the
allocation c2013046608 and c2014046608 made by GENCI (Grand Equipement
National de Calcul Intensif).  This research has made use of the CfA
Supernova Archive, which is funded in part by the National Science
Foundation through grant AST 0907903, of the Online Supernova Spectrum
Archive (SUSPECT; http://www.nhn.ou.edu/~suspect/), of the UC Berkeley
Supernova Database
(http://hercules.berkeley.edu/database/index\_public.html), and of the
NASA/IPAC Extragalactic Database (NED) which is operated by the Jet
Propulsion Laboratory, California Institute of Technology, under
contract with the National Aeronautics and Space Administration.

%%%%%%%%%%%%%%%%%%%%%%%%%%%%%%%%%%%%%%%%%%%%%%%%%%%%%%%%%%%%%%%%%%%%%%
%%%%%%%%%%%%%%%%%%%%%%%%%%%%%%%%%%%%%%%%%%%%%%%%%%%%%%%%%%%%%%%%%%%%%%

\clearpage

\appendix

%%%%%%%%%%%%%%%%%%%%%%%%%%%%%%%%%%%%%%%%%%%%%%%%%%%%%%%%%%%%%%%%%%%%%%

\section{Convergence}\label{sect:conv}

\cmfgen\ employs an iterative method for solving the statistical and
radiative equilibrium equations \citep[see][]{Hillier:1990}. A model
is converged when the maximum correction to any of the variables
involved (temperature, electron density, and level populations) at all
depths falls below some pre-defined value set by the {\tt EPS\_TERM}
parameter (typically $<0.1$\%). The vast majority of
corrections are well below this limit.

\begin{figure*}
\centering
\includegraphics{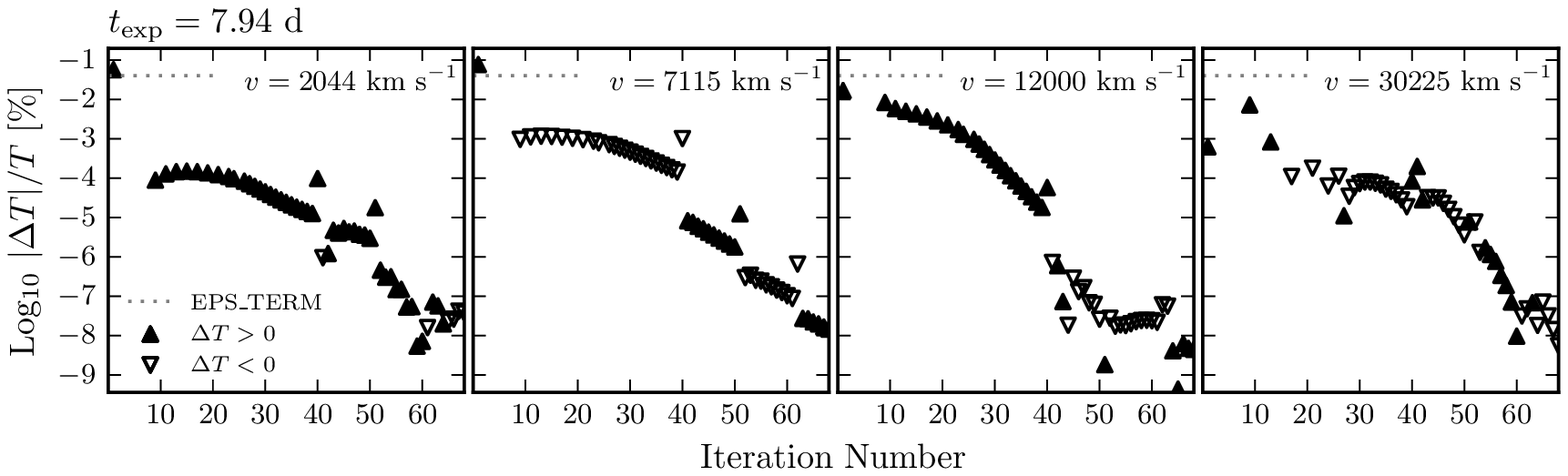}
\includegraphics{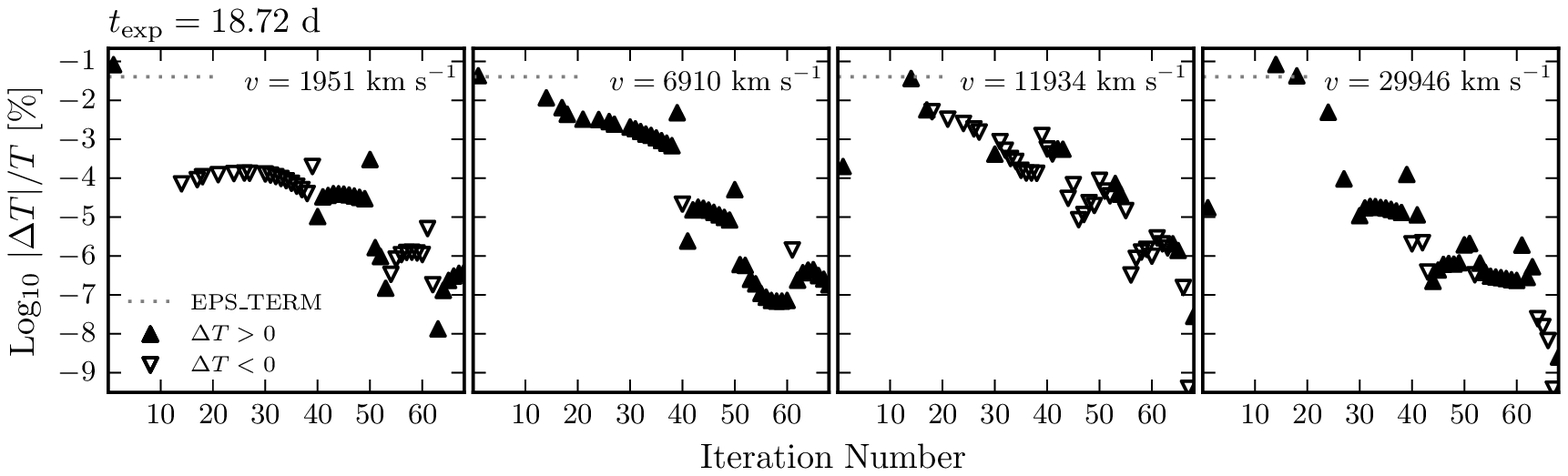}
\includegraphics{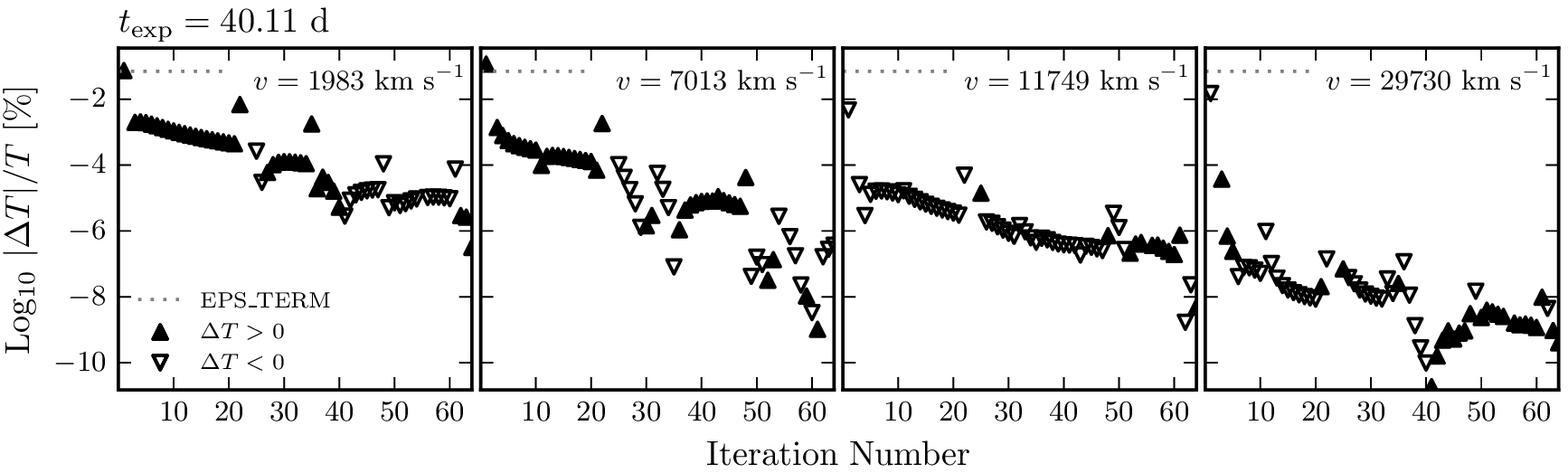}
\includegraphics{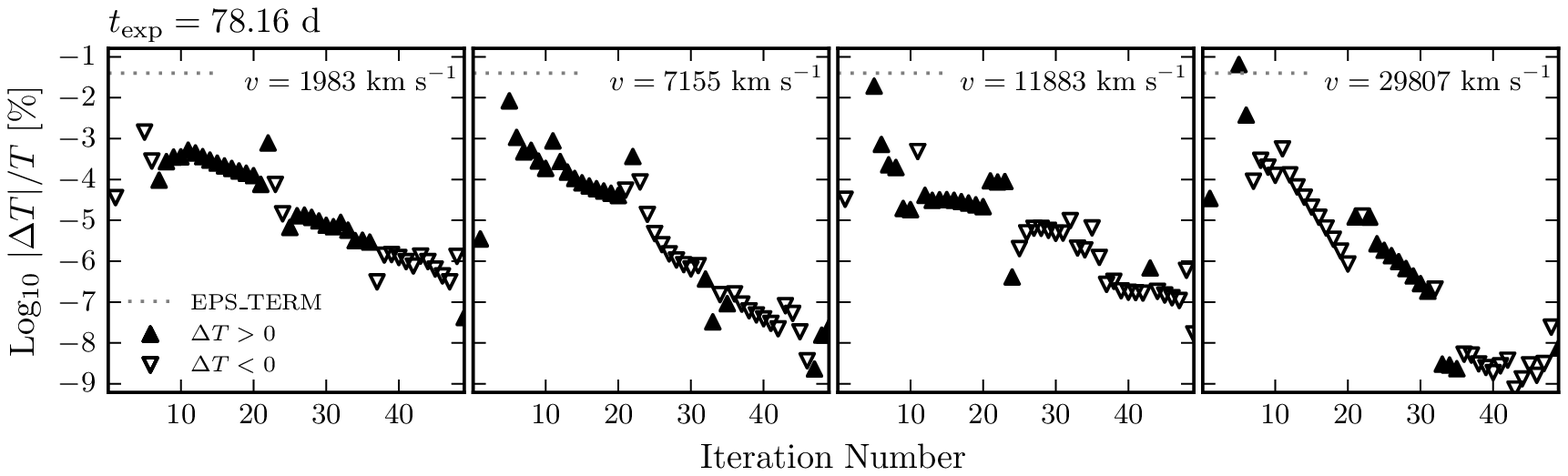}
\caption{\label{fig:temp_conv} Illustration of the temperature
  convergence at four selected time steps in our model sequence
  (increasing from top to bottom). At each time, we show the
  fractional change in temperature $|\Delta T| / T = 100 \times
  |T_{i-1}-T_i|/T_{i-1}$ versus iteration number $i$, at four
  different ejecta locations: In the inner, \nifs-deficient region
  ($v\approx 2000$~\kms); in the \nifs-rich layers ($v\approx
  7000$~\kms); in the IME-rich layers ($v\approx 12000$~\kms);
  in the outer C/O-dominated layers ($v\approx 30000$~\kms). The
  symbols indicate the sign of the correction (positive for filled
  upward-pointing triangles; negative for open downward-pointing
  triangles). The dotted line in each plot corresponds to the value of
  the convergence criterion, noted {\tt EPS\_TERM} in \cmfgen. It is
  set to 0.04\% at all time steps shown here, except at $t_{\rm
    exp}=40.11$~d where it is set to 0.07\%.}
\end{figure*}

We illustrate the convergence properties of our DDC15 model in
Fig.~\ref{fig:temp_conv}, at four selected time steps and for four
different ejecta locations. We only show the temperature, since its rate
of convergence largely determines the global convergence properties of
the model. Since small corrections to the temperature can result in
potentially large corrections of the level populations, the
temperature corrections are orders of magnitude less than {\tt
  EPS\_TERM} by the time convergence is achieved (typically in 40--70
iterations).

After the first iteration, the temperature is set to that obtained via
a grey calculation, and the fractional change is with respect to the
converged solution of the previous time step, hence the potentially
large correction ($\sim10$\%) associated with the first datum. The
temperature is then held fixed (i.e. the temperature correction is
zero) during a series of $\Lambda$ and full iterations in order to
stabilize the corrections to the level populations
\citep[see][]{Hillier/Dessart:2012}. Sharp variations in the
temperature correction result from Ng accelerations \citep{Ng:1974},
used periodically to increase the convergence rate.  Various
diagnostics, such as electron energy balance, global energy
conservation, and ionization equilibrium, are available as independent
checks of the convergence.

%%% FIGURE: 

%%%%%%%%%%%%%%%%%%%%%%%%%%%%%%%%%%%%%%%%%%%%%%%%%%%%%%%%%%%%%%%%%%%%%%

\clearpage

\section{Bolometric luminosity and absolute magnitudes}

Table~\ref{tab:lc} lists the bolometric and pseudo-bolometric
($U\rightarrow K_s$) luminosities, as well as the optical-NIR absolute
magnitudes of model DDC15 as a function of time since explosion.

%%% TABLE: LC
\begin{table*}
\caption{Bolometric luminosity, pseudo-bolometric ($U\rightarrow K_s$) luminosity, and optical-NIR absolute magnitudes for model DDC15.}\label{tab:lc}
\begin{tabular}{rrrrrrrrrrr}
\hline
\multicolumn{1}{c}{$t_{\rm exp}$} & \multicolumn{1}{c}{$L_{\rm bol}$} & \multicolumn{1}{c}{$L_{U\rightarrow K_s}$} & \multicolumn{1}{c}{$U$} & \multicolumn{1}{c}{$B$} & \multicolumn{1}{c}{$V$} & \multicolumn{1}{c}{$R$} & \multicolumn{1}{c}{$I$} & \multicolumn{1}{c}{$J$} & \multicolumn{1}{c}{$H$} & \multicolumn{1}{c}{$K_s$} \\
\multicolumn{1}{c}{[d]}           & \multicolumn{1}{c}{[erg s$^{-1}$]} & \multicolumn{1}{c}{[erg s$^{-1}$]} & [mag] & [mag] & [mag] & [mag] & [mag] & [mag] & [mag] & [mag] \\
\hline
  0.98 & 6.02(39) & 4.98(39) &  $-$5.64 &  $-$7.79 &  $-$9.85 & $-$11.26 & $-$11.22 & $-$12.43 & $-$12.47 & $-$12.65 \\
  1.07 & 8.71(39) & 7.37(39) &  $-$5.68 &  $-$8.40 & $-$10.45 & $-$11.77 & $-$11.67 & $-$12.74 & $-$12.80 & $-$12.93 \\
  1.18 & 1.27(40) & 1.10(40) &  $-$6.11 &  $-$9.00 & $-$11.05 & $-$12.27 & $-$12.14 & $-$13.05 & $-$13.12 & $-$13.22 \\
  1.30 & 1.84(40) & 1.63(40) &  $-$6.65 &  $-$9.61 & $-$11.62 & $-$12.74 & $-$12.61 & $-$13.36 & $-$13.44 & $-$13.52 \\
  1.43 & 2.62(40) & 2.37(40) &  $-$7.38 & $-$10.26 & $-$12.16 & $-$13.14 & $-$13.01 & $-$13.65 & $-$13.75 & $-$13.82 \\
  1.57 & 3.69(40) & 3.37(40) &  $-$8.28 & $-$10.91 & $-$12.68 & $-$13.52 & $-$13.37 & $-$13.95 & $-$14.05 & $-$14.11 \\
  1.73 & 5.15(40) & 4.72(40) &  $-$9.24 & $-$11.54 & $-$13.16 & $-$13.87 & $-$13.71 & $-$14.24 & $-$14.33 & $-$14.39 \\
  1.90 & 7.06(40) & 6.48(40) & $-$10.13 & $-$12.11 & $-$13.60 & $-$14.18 & $-$14.01 & $-$14.51 & $-$14.60 & $-$14.65 \\
  2.09 & 9.65(40) & 8.85(40) & $-$10.93 & $-$12.63 & $-$14.03 & $-$14.48 & $-$14.31 & $-$14.78 & $-$14.87 & $-$14.93 \\
  2.30 & 1.31(41) & 1.20(41) & $-$11.61 & $-$13.11 & $-$14.42 & $-$14.76 & $-$14.61 & $-$15.06 & $-$15.14 & $-$15.20 \\
  2.53 & 1.76(41) & 1.60(41) & $-$12.23 & $-$13.55 & $-$14.79 & $-$15.04 & $-$14.89 & $-$15.33 & $-$15.40 & $-$15.46 \\
  2.78 & 2.33(41) & 2.13(41) & $-$12.84 & $-$13.98 & $-$15.11 & $-$15.32 & $-$15.16 & $-$15.59 & $-$15.64 & $-$15.72 \\
  3.06 & 3.11(41) & 2.84(41) & $-$13.43 & $-$14.40 & $-$15.43 & $-$15.60 & $-$15.43 & $-$15.86 & $-$15.89 & $-$15.98 \\
  3.37 & 4.17(41) & 3.83(41) & $-$14.01 & $-$14.84 & $-$15.74 & $-$15.89 & $-$15.70 & $-$16.13 & $-$16.15 & $-$16.24 \\
  3.70 & 5.67(41) & 5.20(41) & $-$14.57 & $-$15.29 & $-$16.06 & $-$16.20 & $-$15.99 & $-$16.42 & $-$16.42 & $-$16.51 \\
  4.07 & 7.63(41) & 7.05(41) & $-$15.07 & $-$15.73 & $-$16.35 & $-$16.49 & $-$16.26 & $-$16.68 & $-$16.68 & $-$16.76 \\
  4.48 & 1.02(42) & 9.50(41) & $-$15.54 & $-$16.17 & $-$16.64 & $-$16.78 & $-$16.53 & $-$16.95 & $-$16.93 & $-$16.99 \\
  4.93 & 1.36(42) & 1.27(42) & $-$15.99 & $-$16.58 & $-$16.91 & $-$17.06 & $-$16.79 & $-$17.19 & $-$17.16 & $-$17.21 \\
  5.42 & 1.78(42) & 1.67(42) & $-$16.43 & $-$16.96 & $-$17.16 & $-$17.33 & $-$17.04 & $-$17.43 & $-$17.37 & $-$17.39 \\
  5.96 & 2.32(42) & 2.19(42) & $-$16.87 & $-$17.31 & $-$17.41 & $-$17.58 & $-$17.30 & $-$17.64 & $-$17.56 & $-$17.56 \\
  6.56 & 2.98(42) & 2.83(42) & $-$17.31 & $-$17.63 & $-$17.64 & $-$17.82 & $-$17.58 & $-$17.83 & $-$17.73 & $-$17.70 \\
  7.22 & 3.78(42) & 3.60(42) & $-$17.72 & $-$17.92 & $-$17.85 & $-$18.04 & $-$17.84 & $-$17.99 & $-$17.87 & $-$17.83 \\
  7.94 & 4.70(42) & 4.48(42) & $-$18.09 & $-$18.17 & $-$18.06 & $-$18.25 & $-$18.07 & $-$18.14 & $-$17.98 & $-$17.93 \\
  8.73 & 5.70(42) & 5.42(42) & $-$18.39 & $-$18.39 & $-$18.25 & $-$18.43 & $-$18.25 & $-$18.26 & $-$18.06 & $-$18.00 \\
  9.60 & 6.74(42) & 6.40(42) & $-$18.65 & $-$18.59 & $-$18.43 & $-$18.59 & $-$18.40 & $-$18.36 & $-$18.10 & $-$18.05 \\
 10.56 & 7.80(42) & 7.40(42) & $-$18.88 & $-$18.76 & $-$18.60 & $-$18.73 & $-$18.52 & $-$18.43 & $-$18.12 & $-$18.06 \\
 11.62 & 8.78(42) & 8.34(42) & $-$19.04 & $-$18.89 & $-$18.75 & $-$18.86 & $-$18.62 & $-$18.51 & $-$18.13 & $-$18.07 \\
 12.78 & 9.63(42) & 9.22(42) & $-$19.15 & $-$19.01 & $-$18.88 & $-$18.97 & $-$18.67 & $-$18.53 & $-$18.10 & $-$18.04 \\
 14.06 & 1.05(43) & 9.90(42) & $-$19.31 & $-$19.10 & $-$18.97 & $-$19.05 & $-$18.72 & $-$18.53 & $-$18.04 & $-$17.98 \\
 15.47 & 1.09(43) & 1.05(43) & $-$19.36 & $-$19.14 & $-$19.05 & $-$19.12 & $-$18.74 & $-$18.50 & $-$17.99 & $-$17.94 \\
 17.02 & 1.16(43) & 1.08(43) & $-$19.42 & $-$19.14 & $-$19.17 & $-$19.17 & $-$18.83 & $-$18.45 & $-$17.98 & $-$18.07 \\
 18.72 & 1.13(43) & 1.07(43) & $-$19.31 & $-$19.12 & $-$19.21 & $-$19.19 & $-$18.83 & $-$18.35 & $-$17.94 & $-$18.05 \\
 20.59 & 1.08(43) & 1.03(43) & $-$19.18 & $-$19.07 & $-$19.23 & $-$19.19 & $-$18.81 & $-$18.24 & $-$17.95 & $-$18.06 \\
 22.65 & 1.00(43) & 9.70(42) & $-$19.02 & $-$18.99 & $-$19.21 & $-$19.16 & $-$18.77 & $-$18.09 & $-$18.00 & $-$18.08 \\
 24.91 & 9.01(42) & 8.72(42) & $-$18.82 & $-$18.85 & $-$19.16 & $-$19.08 & $-$18.70 & $-$17.81 & $-$18.06 & $-$18.11 \\
 27.40 & 7.76(42) & 7.51(42) & $-$18.57 & $-$18.65 & $-$19.05 & $-$18.95 & $-$18.59 & $-$17.33 & $-$18.09 & $-$18.13 \\
 30.14 & 6.56(42) & 6.36(42) & $-$18.29 & $-$18.39 & $-$18.89 & $-$18.79 & $-$18.52 & $-$17.05 & $-$18.18 & $-$18.17 \\
 33.15 & 5.74(42) & 5.54(42) & $-$17.83 & $-$17.97 & $-$18.70 & $-$18.70 & $-$18.72 & $-$17.34 & $-$18.62 & $-$18.54 \\
 36.46 & 5.14(42) & 4.93(42) & $-$17.43 & $-$17.58 & $-$18.47 & $-$18.61 & $-$18.85 & $-$17.61 & $-$18.80 & $-$18.70 \\
 40.11 & 4.47(42) & 4.15(42) & $-$17.05 & $-$17.18 & $-$18.17 & $-$18.45 & $-$18.83 & $-$17.92 & $-$18.80 & $-$18.70 \\
 44.12 & 3.51(42) & 3.28(42) & $-$16.75 & $-$16.86 & $-$17.85 & $-$18.17 & $-$18.58 & $-$17.76 & $-$18.53 & $-$18.41 \\
 48.53 & 2.74(42) & 2.56(42) & $-$16.50 & $-$16.61 & $-$17.59 & $-$17.91 & $-$18.30 & $-$17.35 & $-$18.22 & $-$18.05 \\
 53.38 & 2.20(42) & 2.05(42) & $-$16.31 & $-$16.42 & $-$17.38 & $-$17.69 & $-$18.03 & $-$16.92 & $-$17.92 & $-$17.72 \\
 58.72 & 1.80(42) & 1.67(42) & $-$16.15 & $-$16.27 & $-$17.20 & $-$17.49 & $-$17.76 & $-$16.50 & $-$17.64 & $-$17.43 \\
 64.59 & 1.48(42) & 1.37(42) & $-$16.00 & $-$16.14 & $-$17.03 & $-$17.30 & $-$17.49 & $-$16.08 & $-$17.37 & $-$17.15 \\
 71.05 & 1.22(42) & 1.13(42) & $-$15.84 & $-$16.02 & $-$16.85 & $-$17.10 & $-$17.21 & $-$15.65 & $-$17.09 & $-$16.87 \\
 78.16 & 1.00(42) & 9.29(41) & $-$15.68 & $-$15.90 & $-$16.66 & $-$16.89 & $-$16.94 & $-$15.20 & $-$16.80 & $-$16.59 \\
 86.00 & 8.15(41) & 7.56(41) & $-$15.49 & $-$15.77 & $-$16.45 & $-$16.65 & $-$16.66 & $-$14.74 & $-$16.48 & $-$16.29 \\
 94.60 & 6.57(41) & 6.10(41) & $-$15.29 & $-$15.63 & $-$16.23 & $-$16.39 & $-$16.38 & $-$14.26 & $-$16.12 & $-$15.98 \\
104.10 & 5.26(41) & 4.86(41) & $-$15.05 & $-$15.48 & $-$15.98 & $-$16.10 & $-$16.10 & $-$13.81 & $-$15.73 & $-$15.66 \\
\hline
\end{tabular}

\flushleft
{\bf Note:} Numbers in parentheses correspond to powers of ten.
$UBVRI$ magnitudes are based on the passbands of \cite{Bessell:1990};
$JHK_s$ magnitudes are in the 2MASS system \citep{Cohen/etal:2003}.
\end{table*}

%%%%%%%%%%%%%%%%%%%%%%%%%%%%%%%%%%%%%%%%%%%%%%%%%%%%%%%%%%%%%%%%%%%%%%

\clearpage

\section{Strong lines}\label{sect:line_ids}

In tables~\ref{tab:stronglines_4.07}--~\ref{tab:stronglines_104.10} we
report the strongest lines in the optical (3000-10000~\AA) and NIR
(1--2.5~$\mu$m) ranges at a given time between 4.1~d and 104.1~d past
explosion ($-13.5$~d and +86.5~d from bolometric maximum). A line is
considered strong when its absolute Sobolev equivalent width (EW)
exceeds 10 per cent of the largest absolute EW within
$\pm10^4$~\kms\ of the line's central wavelength.  Forbidden and
semi-forbidden transitions are noted using the appropriate brackets
around the line wavelength (in air).  We highlight transitions
connected to the ground state with a ``$\dagger$'' symbol.  These
tables are meant to serve as a guide for line identification in
spectra of \sneia\ similar to SN~2002bo. However, we caution the
reader that the Sobolev EWs are only approximate and neglect important
effects such as line overlap. Moreover, lines with a P-Cygni profile
morphology may be important contributors to the resulting spectrum yet
yield a small absolute EW. This explains for instance why the strong
Ca\two\ K line (3934~\AA) does not show up in Table~C2 ($t_{\rm
  exp}=5.4$~d) where Fig.~D1 shows the strong impact of the
Ca\two\ H\&K feature on the optical spectra at these times, and the
Ca\two~K line indeed appears in Tables~C1 ($t_{\rm exp}=4.1$~d) and C3
($t_{\rm exp}=9.6$~d).

\begin{table}
\caption{Strong lines in the optical and NIR at 4.1\,d past explosion.}\label{tab:stronglines_4.07}
\begin{tabular}{lcc}
\hline
\multicolumn{3}{c}{$t_{\rm exp}=4.1$\,d ($\Delta t_{\rm bol}=-13.5$\,d)} \\
\hline
\multicolumn{1}{c}{} & \multicolumn{1}{c}{3000--10000\,\AA} & \multicolumn{1}{c}{1.0--2.5\,$\mu$m} \\
\multicolumn{1}{c}{Ion} & \multicolumn{1}{c}{$\lambda_{\rm air}$ [\AA]} & \multicolumn{1}{c}{$\lambda_{\rm air}$ [$\mu$m]} \\
\hline
O\,{\sc i}     & \phantom{[}9266.01\phantom{$^{\dagger}$}\phantom{]}          & \phantom{[}1.316\phantom{$^{\dagger}$}\phantom{]}            \\
\\[-1.8ex]
Mg\,{\sc ii}   & \phantom{[}7896.37\phantom{$^{\dagger}$}\phantom{]}          & \phantom{[}1.009\phantom{$^{\dagger}$}\phantom{]}            \\
               &                                                              & \phantom{[}1.091\phantom{$^{\dagger}$}\phantom{]}            \\
               &                                                              & \phantom{[}1.863\phantom{$^{\dagger}$}\phantom{]}            \\
\hline
\end{tabular}
\end{table}

\begin{table}
\caption{Strong lines in the optical and NIR at 5.4\,d past explosion.}\label{tab:stronglines_5.42}
\begin{tabular}{lcc}
\hline
\multicolumn{3}{c}{$t_{\rm exp}=5.4$\,d ($\Delta t_{\rm bol}=-12.1$\,d)} \\
\hline
\multicolumn{1}{c}{} & \multicolumn{1}{c}{3000--10000\,\AA} & \multicolumn{1}{c}{1.0--2.5\,$\mu$m} \\
\multicolumn{1}{c}{Ion} & \multicolumn{1}{c}{$\lambda_{\rm air}$ [\AA]} & \multicolumn{1}{c}{$\lambda_{\rm air}$ [$\mu$m]} \\
\hline
O\,{\sc i}     & \phantom{[}7771.94\phantom{$^{\dagger}$}\phantom{]}          & \phantom{[}1.316\phantom{$^{\dagger}$}\phantom{]}            \\
\\[-1.8ex]
Mg\,{\sc ii}   & \phantom{[}7896.37\phantom{$^{\dagger}$}\phantom{]}          & \phantom{[}1.091\phantom{$^{\dagger}$}\phantom{]}            \\
               &                                                              & \phantom{[}1.863\phantom{$^{\dagger}$}\phantom{]}            \\
               &                                                              & \phantom{[}2.137\phantom{$^{\dagger}$}\phantom{]}            \\
\hline
\end{tabular}
\end{table}

\begin{table}
\caption{Strong lines in the optical and NIR at 9.6\,d past explosion.}\label{tab:stronglines_9.60}
\begin{tabular}{lcc}
\hline
\multicolumn{3}{c}{$t_{\rm exp}=9.6$\,d ($\Delta t_{\rm bol}=-8.0$\,d)} \\
\hline
\multicolumn{1}{c}{} & \multicolumn{1}{c}{3000--10000\,\AA} & \multicolumn{1}{c}{1.0--2.5\,$\mu$m} \\
\multicolumn{1}{c}{Ion} & \multicolumn{1}{c}{$\lambda_{\rm air}$ [\AA]} & \multicolumn{1}{c}{$\lambda_{\rm air}$ [$\mu$m]} \\
\hline
O\,{\sc i}     & \phantom{[}7771.94\phantom{$^{\dagger}$}\phantom{]}          & \phantom{[}1.316\phantom{$^{\dagger}$}\phantom{]}            \\
\\[-1.8ex]
Mg\,{\sc ii}   & \phantom{[}9218.25\phantom{$^{\dagger}$}\phantom{]}          & \phantom{[}1.091\phantom{$^{\dagger}$}\phantom{]}            \\
\\[-1.8ex]
Si\,{\sc ii}   & \phantom{[}6347.11\phantom{$^{\dagger}$}\phantom{]}          & \phantom{[}1.370\phantom{$^{\dagger}$}\phantom{]}            \\
\hline
\end{tabular}
\end{table}

\begin{table}
\caption{Strong lines in the optical and NIR at 12.8\,d past explosion.}\label{tab:stronglines_12.78}
\begin{tabular}{lcc}
\hline
\multicolumn{3}{c}{$t_{\rm exp}=12.8$\,d ($\Delta t_{\rm bol}=-4.8$\,d)} \\
\hline
\multicolumn{1}{c}{} & \multicolumn{1}{c}{3000--10000\,\AA} & \multicolumn{1}{c}{1.0--2.5\,$\mu$m} \\
\multicolumn{1}{c}{Ion} & \multicolumn{1}{c}{$\lambda_{\rm air}$ [\AA]} & \multicolumn{1}{c}{$\lambda_{\rm air}$ [$\mu$m]} \\
\hline
O\,{\sc i}     & \phantom{[}7771.94\phantom{$^{\dagger}$}\phantom{]}          & $\cdots$                                                     \\
\\[-1.8ex]
Mg\,{\sc ii}   & \phantom{[}9218.25\phantom{$^{\dagger}$}\phantom{]}          & \phantom{[}1.091\phantom{$^{\dagger}$}\phantom{]}            \\
\\[-1.8ex]
Si\,{\sc ii}   & \phantom{[}5978.93\phantom{$^{\dagger}$}\phantom{]}          & \phantom{[}1.370\phantom{$^{\dagger}$}\phantom{]}            \\
\hline
\end{tabular}
\end{table}

\begin{table}
\caption{Strong lines in the optical and NIR at 15.5\,d past explosion.}\label{tab:stronglines_15.47}
\begin{tabular}{lcc}
\hline
\multicolumn{3}{c}{$t_{\rm exp}=15.5$\,d ($\Delta t_{\rm bol}=-2.1$\,d)} \\
\hline
\multicolumn{1}{c}{} & \multicolumn{1}{c}{3000--10000\,\AA} & \multicolumn{1}{c}{1.0--2.5\,$\mu$m} \\
\multicolumn{1}{c}{Ion} & \multicolumn{1}{c}{$\lambda_{\rm air}$ [\AA]} & \multicolumn{1}{c}{$\lambda_{\rm air}$ [$\mu$m]} \\
\hline
O\,{\sc i}     & \phantom{[}7771.94\phantom{$^{\dagger}$}\phantom{]}          & $\cdots$                                                     \\
\\[-1.8ex]
Mg\,{\sc ii}   & \phantom{[}9218.25\phantom{$^{\dagger}$}\phantom{]}          & \phantom{[}1.091\phantom{$^{\dagger}$}\phantom{]}            \\
\\[-1.8ex]
Si\,{\sc ii}   & \phantom{[}6347.11\phantom{$^{\dagger}$}\phantom{]}          & \phantom{[}1.691\phantom{$^{\dagger}$}\phantom{]}            \\
\hline
\end{tabular}
\end{table}

\begin{table}
\caption{Strong lines in the optical and NIR at 17.0\,d past explosion.}\label{tab:stronglines_17.02}
\begin{tabular}{lcc}
\hline
\multicolumn{3}{c}{$t_{\rm exp}=17.0$\,d ($\Delta t_{\rm bol}=-0.5$\,d)} \\
\hline
\multicolumn{1}{c}{} & \multicolumn{1}{c}{3000--10000\,\AA} & \multicolumn{1}{c}{1.0--2.5\,$\mu$m} \\
\multicolumn{1}{c}{Ion} & \multicolumn{1}{c}{$\lambda_{\rm air}$ [\AA]} & \multicolumn{1}{c}{$\lambda_{\rm air}$ [$\mu$m]} \\
\hline
O\,{\sc i}     & \phantom{[}7771.94\phantom{$^{\dagger}$}\phantom{]}          & $\cdots$                                                     \\
\\[-1.8ex]
Mg\,{\sc ii}   & \phantom{[}7896.37\phantom{$^{\dagger}$}\phantom{]}          & \phantom{[}1.091\phantom{$^{\dagger}$}\phantom{]}            \\
\\[-1.8ex]
Si\,{\sc ii}   & \phantom{[}6347.11\phantom{$^{\dagger}$}\phantom{]}          & \phantom{[}1.175\phantom{$^{\dagger}$}\phantom{]}            \\
\hline
\end{tabular}
\end{table}

\begin{table}
\caption{Strong lines in the optical and NIR at 22.6\,d past explosion.}\label{tab:stronglines_22.65}
\begin{tabular}{lcc}
\hline
\multicolumn{3}{c}{$t_{\rm exp}=22.6$\,d ($\Delta t_{\rm bol}=+5.1$\,d)} \\
\hline
\multicolumn{1}{c}{} & \multicolumn{1}{c}{3000--10000\,\AA} & \multicolumn{1}{c}{1.0--2.5\,$\mu$m} \\
\multicolumn{1}{c}{Ion} & \multicolumn{1}{c}{$\lambda_{\rm air}$ [\AA]} & \multicolumn{1}{c}{$\lambda_{\rm air}$ [$\mu$m]} \\
\hline
O\,{\sc i}     & \phantom{[}7771.94\phantom{$^{\dagger}$}\phantom{]}          & $\cdots$                                                     \\
\\[-1.8ex]
Mg\,{\sc ii}   & \phantom{[}9218.25\phantom{$^{\dagger}$}\phantom{]}          & \phantom{[}1.091\phantom{$^{\dagger}$}\phantom{]}            \\
\\[-1.8ex]
Si\,{\sc ii}   & \phantom{[}6347.11\phantom{$^{\dagger}$}\phantom{]}          & \phantom{[}1.175\phantom{$^{\dagger}$}\phantom{]}            \\
\hline
\end{tabular}
\end{table}

\begin{table}
\caption{Strong lines in the optical and NIR at 27.4\,d past explosion.}\label{tab:stronglines_27.40}
\begin{tabular}{lcc}
\hline
\multicolumn{3}{c}{$t_{\rm exp}=27.4$\,d ($\Delta t_{\rm bol}=+9.8$\,d)} \\
\hline
\multicolumn{1}{c}{} & \multicolumn{1}{c}{3000--10000\,\AA} & \multicolumn{1}{c}{1.0--2.5\,$\mu$m} \\
\multicolumn{1}{c}{Ion} & \multicolumn{1}{c}{$\lambda_{\rm air}$ [\AA]} & \multicolumn{1}{c}{$\lambda_{\rm air}$ [$\mu$m]} \\
\hline
O\,{\sc i}     & \phantom{[}7771.94\phantom{$^{\dagger}$}\phantom{]}          & $\cdots$                                                     \\
\\[-1.8ex]
Mg\,{\sc ii}   & \phantom{[}9218.25\phantom{$^{\dagger}$}\phantom{]}          & \phantom{[}1.091\phantom{$^{\dagger}$}\phantom{]}            \\
\\[-1.8ex]
Si\,{\sc ii}   & \phantom{[}4130.90\phantom{$^{\dagger}$}\phantom{]}          & \phantom{[}1.175\phantom{$^{\dagger}$}\phantom{]}            \\
\hline
\end{tabular}
\end{table}

\begin{table}
\caption{Strong lines in the optical and NIR at 33.2\,d past explosion.}\label{tab:stronglines_33.15}
\begin{tabular}{lcc}
\hline
\multicolumn{3}{c}{$t_{\rm exp}=33.2$\,d ($\Delta t_{\rm bol}=+15.6$\,d)} \\
\hline
\multicolumn{1}{c}{} & \multicolumn{1}{c}{3000--10000\,\AA} & \multicolumn{1}{c}{1.0--2.5\,$\mu$m} \\
\multicolumn{1}{c}{Ion} & \multicolumn{1}{c}{$\lambda_{\rm air}$ [\AA]} & \multicolumn{1}{c}{$\lambda_{\rm air}$ [$\mu$m]} \\
\hline
Si\,{\sc ii}   & \phantom{[}6347.11\phantom{$^{\dagger}$}\phantom{]}          & \phantom{[}1.175\phantom{$^{\dagger}$}\phantom{]}            \\
\\[-1.8ex]
S\,{\sc ii}    & \phantom{[}5647.02\phantom{$^{\dagger}$}\phantom{]}          & [1.032]\phantom{$^{\dagger}$}                                \\
\\[-1.8ex]
S\,{\sc iii}   & [9530.62]\phantom{$^{\dagger}$}                              & $\cdots$                                                     \\
\hline
\end{tabular}
\end{table}

\begin{table}
\caption{Strong lines in the optical and NIR at 44.1\,d past explosion.}\label{tab:stronglines_44.12}
\begin{tabular}{lcc}
\hline
\multicolumn{3}{c}{$t_{\rm exp}=44.1$\,d ($\Delta t_{\rm bol}=+26.6$\,d)} \\
\hline
\multicolumn{1}{c}{} & \multicolumn{1}{c}{3000--10000\,\AA} & \multicolumn{1}{c}{1.0--2.5\,$\mu$m} \\
\multicolumn{1}{c}{Ion} & \multicolumn{1}{c}{$\lambda_{\rm air}$ [\AA]} & \multicolumn{1}{c}{$\lambda_{\rm air}$ [$\mu$m]} \\
\hline
Si\,{\sc ii}   & \phantom{[}6347.11\phantom{$^{\dagger}$}\phantom{]}          & $\cdots$                                                     \\
\\[-1.8ex]
S\,{\sc ii}    & [4068.60]$^{\dagger}$                                        & [1.032]\phantom{$^{\dagger}$}                                \\
\\[-1.8ex]
S\,{\sc iii}   & [9530.62]\phantom{$^{\dagger}$}                              & $\cdots$                                                     \\
\hline
\end{tabular}
\end{table}

\begin{table}
\caption{Strong lines in the optical and NIR at 53.4\,d past explosion.}\label{tab:stronglines_53.38}
\begin{tabular}{lcc}
\hline
\multicolumn{3}{c}{$t_{\rm exp}=53.4$\,d ($\Delta t_{\rm bol}=+35.8$\,d)} \\
\hline
\multicolumn{1}{c}{} & \multicolumn{1}{c}{3000--10000\,\AA} & \multicolumn{1}{c}{1.0--2.5\,$\mu$m} \\
\multicolumn{1}{c}{Ion} & \multicolumn{1}{c}{$\lambda_{\rm air}$ [\AA]} & \multicolumn{1}{c}{$\lambda_{\rm air}$ [$\mu$m]} \\
\hline
Si\,{\sc ii}   & \phantom{[}6347.11\phantom{$^{\dagger}$}\phantom{]}          & $\cdots$                                                     \\
\\[-1.8ex]
S\,{\sc ii}    & [4068.60]$^{\dagger}$                                        & [1.032]\phantom{$^{\dagger}$}                                \\
\\[-1.8ex]
S\,{\sc iii}   & [9530.62]\phantom{$^{\dagger}$}                              & $\cdots$                                                     \\
\hline
\end{tabular}
\end{table}

\begin{table}
\caption{Strong lines in the optical and NIR at 64.6\,d past explosion.}\label{tab:stronglines_64.59}
\begin{tabular}{lcc}
\hline
\multicolumn{3}{c}{$t_{\rm exp}=64.6$\,d ($\Delta t_{\rm bol}=+47.0$\,d)} \\
\hline
\multicolumn{1}{c}{} & \multicolumn{1}{c}{3000--10000\,\AA} & \multicolumn{1}{c}{1.0--2.5\,$\mu$m} \\
\multicolumn{1}{c}{Ion} & \multicolumn{1}{c}{$\lambda_{\rm air}$ [\AA]} & \multicolumn{1}{c}{$\lambda_{\rm air}$ [$\mu$m]} \\
\hline
Si\,{\sc ii}   & \phantom{[}6347.11\phantom{$^{\dagger}$}\phantom{]}          & $\cdots$                                                     \\
\\[-1.8ex]
S\,{\sc ii}    & [4068.60]$^{\dagger}$                                        & [1.032]\phantom{$^{\dagger}$}                                \\
\\[-1.8ex]
S\,{\sc iii}   & [9068.62]\phantom{$^{\dagger}$}                              & $\cdots$                                                     \\
\hline
\end{tabular}
\end{table}

\begin{table}
\caption{Strong lines in the optical and NIR at 71.1\,d past explosion.}\label{tab:stronglines_71.05}
\begin{tabular}{lcc}
\hline
\multicolumn{3}{c}{$t_{\rm exp}=71.1$\,d ($\Delta t_{\rm bol}=+53.5$\,d)} \\
\hline
\multicolumn{1}{c}{} & \multicolumn{1}{c}{3000--10000\,\AA} & \multicolumn{1}{c}{1.0--2.5\,$\mu$m} \\
\multicolumn{1}{c}{Ion} & \multicolumn{1}{c}{$\lambda_{\rm air}$ [\AA]} & \multicolumn{1}{c}{$\lambda_{\rm air}$ [$\mu$m]} \\
\hline
Si\,{\sc ii}   & \phantom{[}6347.11\phantom{$^{\dagger}$}\phantom{]}          & $\cdots$                                                     \\
\\[-1.8ex]
S\,{\sc ii}    & [4068.60]$^{\dagger}$                                        & [1.032]\phantom{$^{\dagger}$}                                \\
\\[-1.8ex]
S\,{\sc iii}   & [9068.62]\phantom{$^{\dagger}$}                              & $\cdots$                                                     \\
\hline
\end{tabular}
\end{table}

\begin{table}
\caption{Strong lines in the optical and NIR at 86.0\,d past explosion.}\label{tab:stronglines_86.00}
\begin{tabular}{lcc}
\hline
\multicolumn{3}{c}{$t_{\rm exp}=86.0$\,d ($\Delta t_{\rm bol}=+68.4$\,d)} \\
\hline
\multicolumn{1}{c}{} & \multicolumn{1}{c}{3000--10000\,\AA} & \multicolumn{1}{c}{1.0--2.5\,$\mu$m} \\
\multicolumn{1}{c}{Ion} & \multicolumn{1}{c}{$\lambda_{\rm air}$ [\AA]} & \multicolumn{1}{c}{$\lambda_{\rm air}$ [$\mu$m]} \\
\hline
Si\,{\sc ii}   & \phantom{[}6347.11\phantom{$^{\dagger}$}\phantom{]}          & $\cdots$                                                     \\
\\[-1.8ex]
S\,{\sc ii}    & [4068.60]$^{\dagger}$                                        & [1.032]\phantom{$^{\dagger}$}                                \\
\\[-1.8ex]
S\,{\sc iii}   & [9068.62]\phantom{$^{\dagger}$}                              & $\cdots$                                                     \\
\hline
\end{tabular}
\end{table}

\begin{table}
\caption{Strong lines in the optical and NIR at 104.1\,d past explosion.}\label{tab:stronglines_104.10}
\begin{tabular}{lcc}
\hline
\multicolumn{3}{c}{$t_{\rm exp}=104.1$\,d ($\Delta t_{\rm bol}=+86.5$\,d)} \\
\hline
\multicolumn{1}{c}{} & \multicolumn{1}{c}{3000--10000\,\AA} & \multicolumn{1}{c}{1.0--2.5\,$\mu$m} \\
\multicolumn{1}{c}{Ion} & \multicolumn{1}{c}{$\lambda_{\rm air}$ [\AA]} & \multicolumn{1}{c}{$\lambda_{\rm air}$ [$\mu$m]} \\
\hline
S\,{\sc ii}    & [4068.60]$^{\dagger}$                                        & [1.032]\phantom{$^{\dagger}$}                                \\
\\[-1.8ex]
S\,{\sc iii}   & [9068.62]\phantom{$^{\dagger}$}                              & $\cdots$                                                     \\
               & [9530.62]\phantom{$^{\dagger}$}                              &                                                              \\
\\
\hline
\end{tabular}
\end{table}

%%%%%%%%%%%%%%%%%%%%%%%%%%%%%%%%%%%%%%%%%%%%%%%%%%%%%%%%%%%%%%%%%%%%%%

\clearpage

\section{Contribution of individual ions to the total optical and
  near-infrared flux}\label{sect:ladder}

\begin{figure*}
\centering
\includegraphics{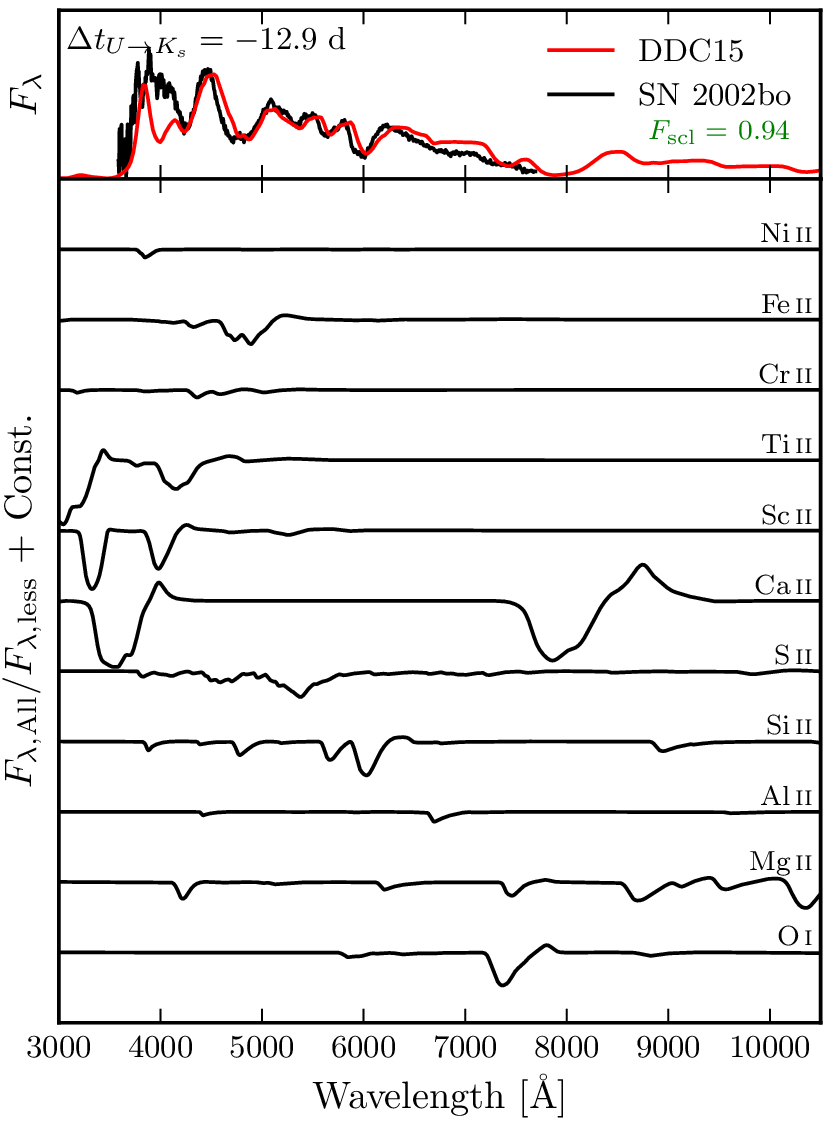}\hspace{.5cm}
\includegraphics{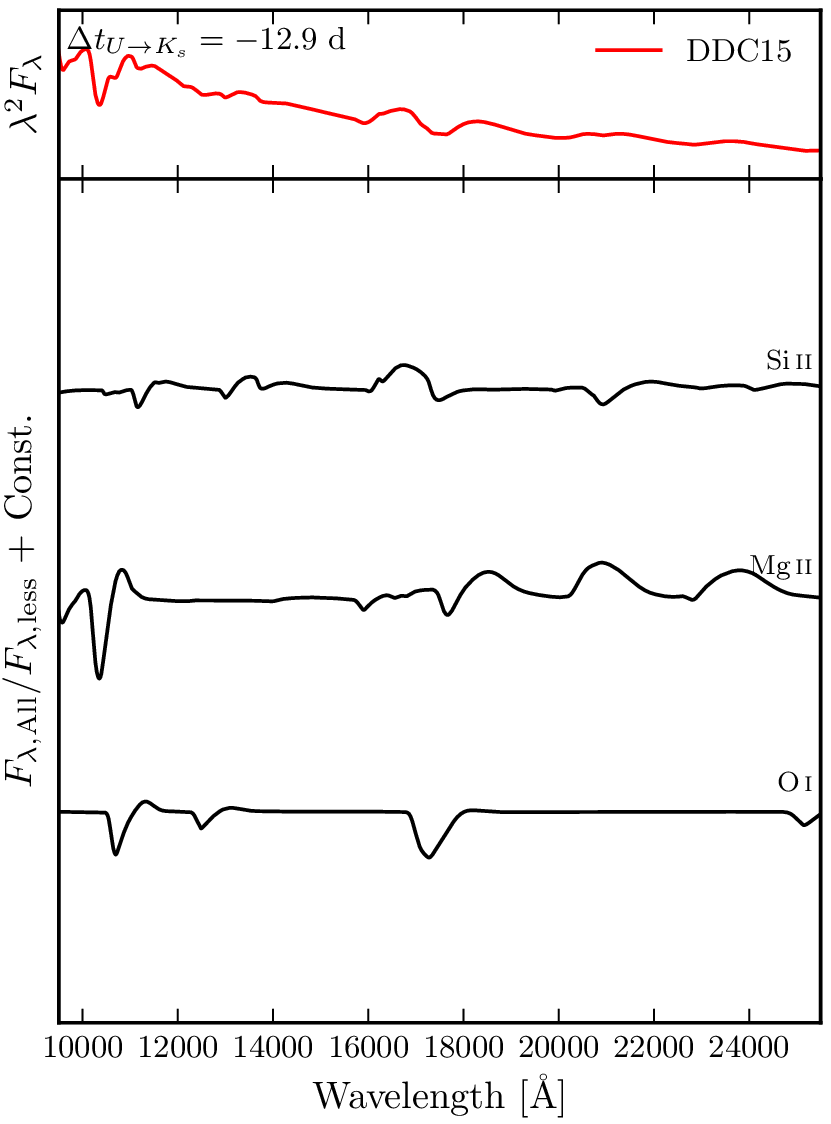}\vspace{.1cm}
\includegraphics{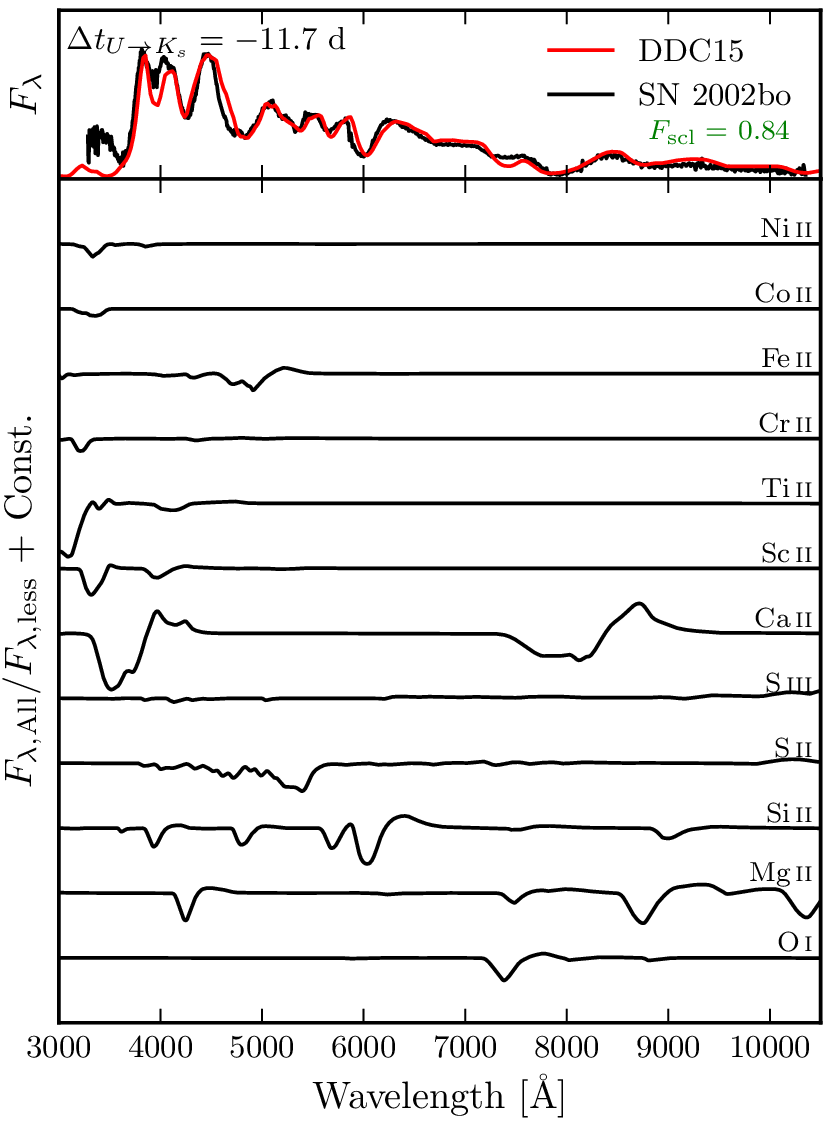}\hspace{.5cm}
\includegraphics{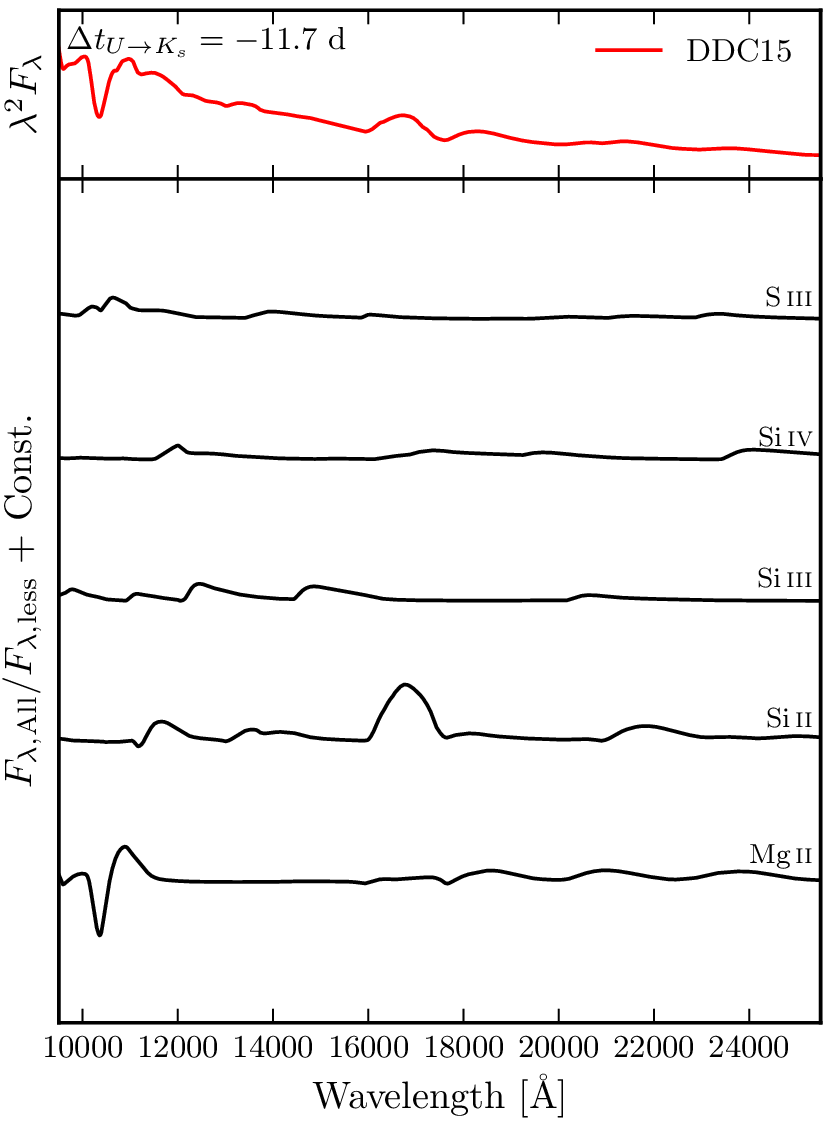}
\caption{\label{fig:ladder_plot1} Contribution of individual ions
  (bottom panels) to the full optical (left) and NIR (right) synthetic
  spectra of DDC15 (top panels, red line), compared to SN~2002bo
  (top panels, black line) at $-12.9$~d and $-11.7$~d from from
  pseudo-bolometric ($U\rightarrow K_s$) maximum.}
\end{figure*}

\begin{figure*}
\centering
\includegraphics{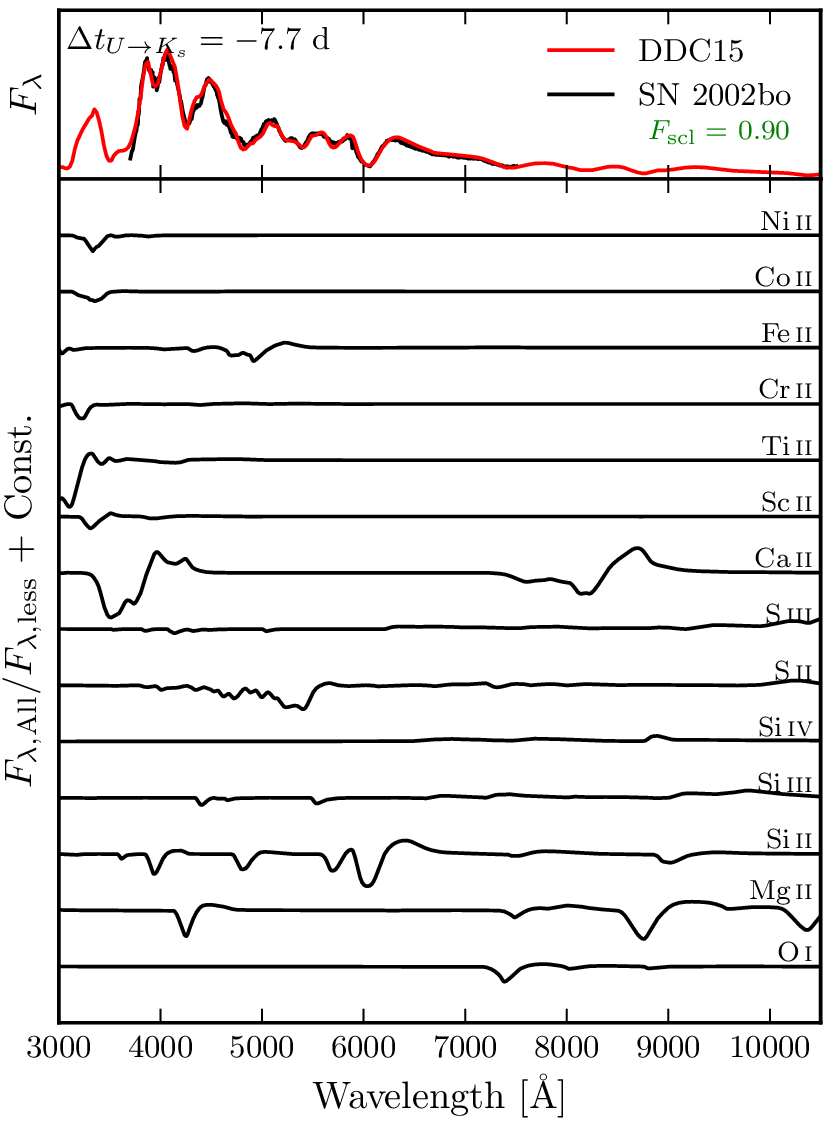}\hspace{.5cm}
\includegraphics{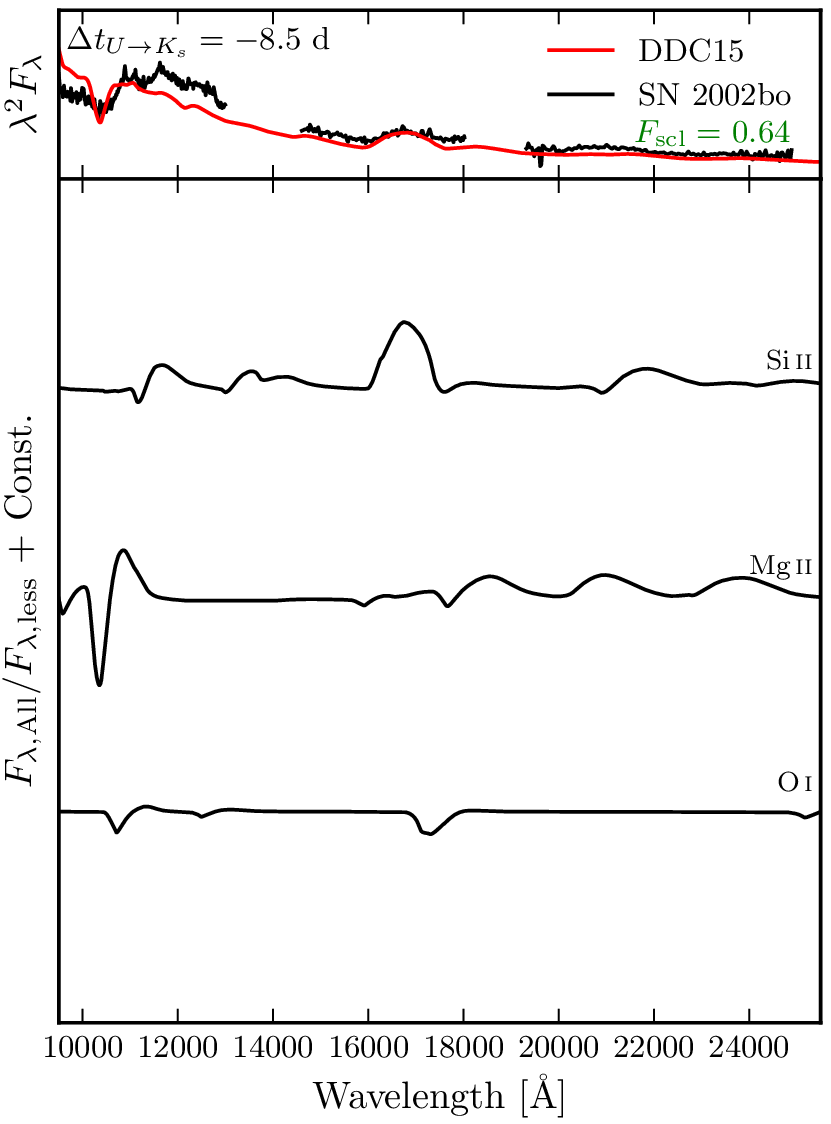}\vspace{.1cm}
\includegraphics{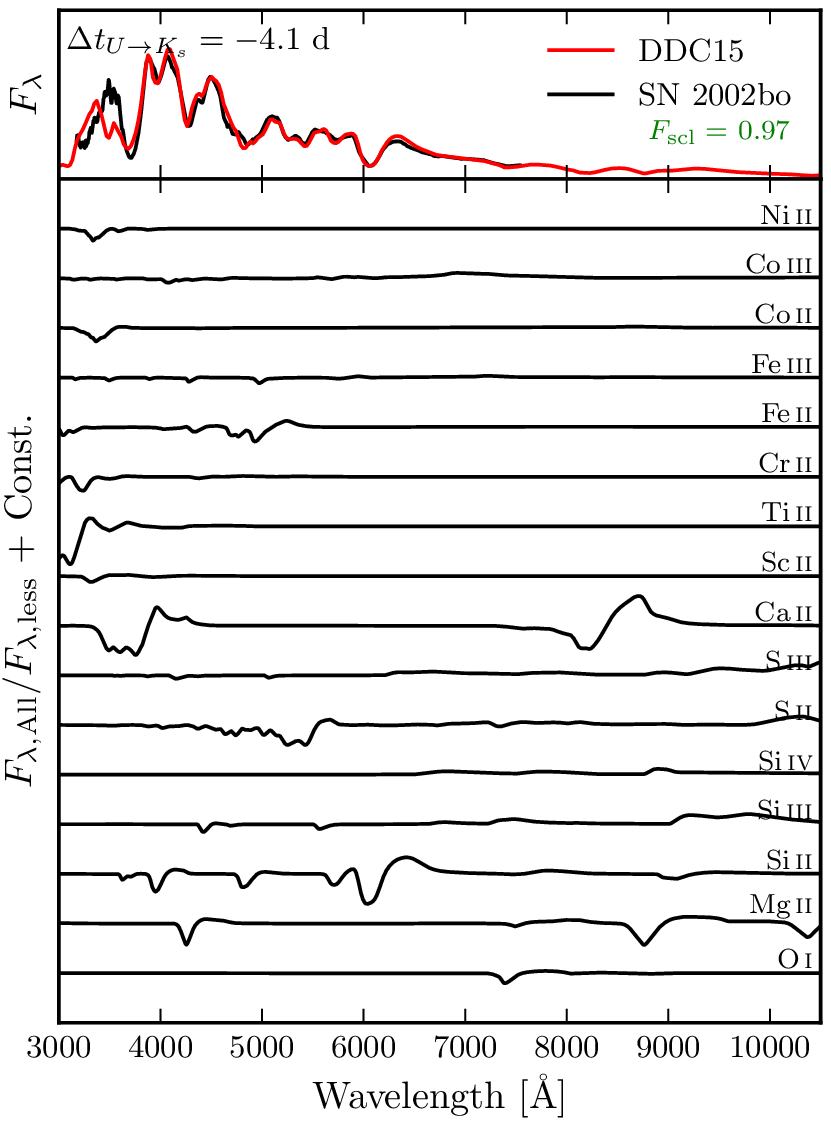}\hspace{.5cm}
\includegraphics{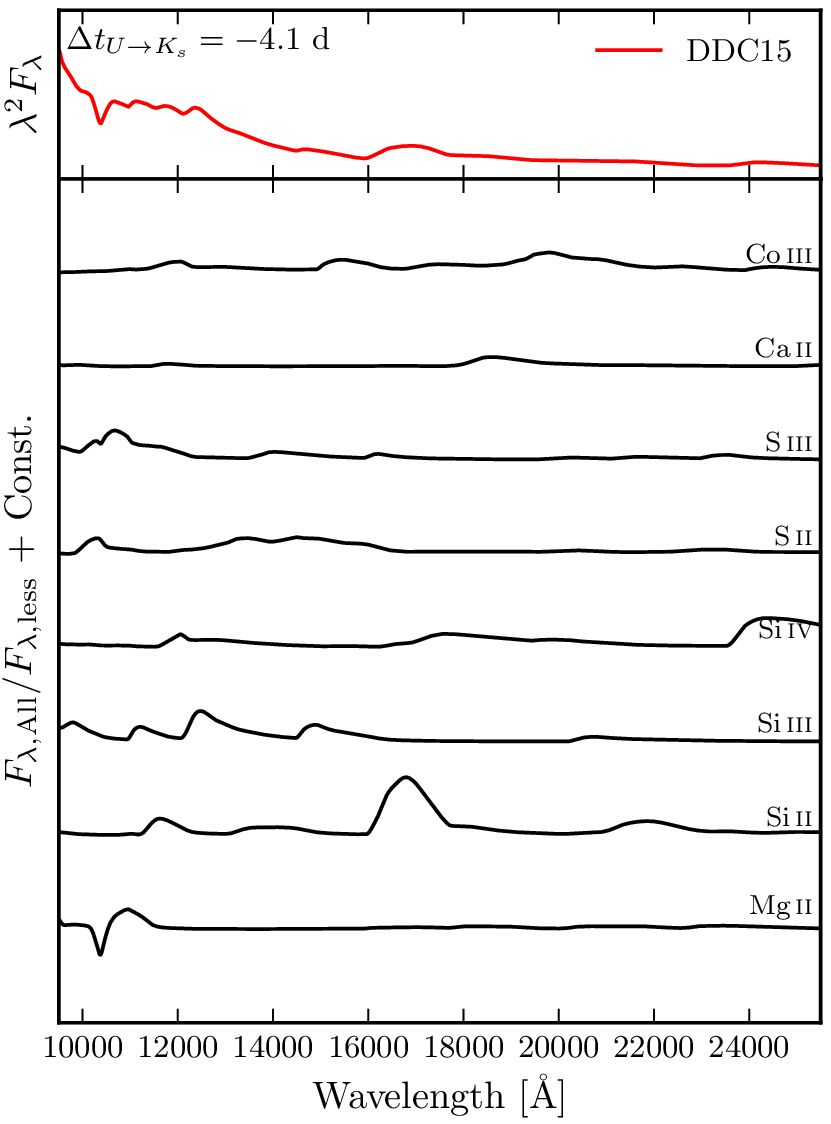}
\caption{\label{fig:ladder_plot2} Contribution of individual ions
  (bottom panels) to the full optical (left) and NIR (right) synthetic
  spectra of DDC15 (top panels, red line), compared to SN~2002bo
  (top panels, black line) at $-8.5$~d (NIR only), $-7.7$~d (optical only), and
  $-4.1$~d from from pseudo-bolometric ($U\rightarrow K_s$) maximum.}
\end{figure*}

\begin{figure*}
\centering
\includegraphics{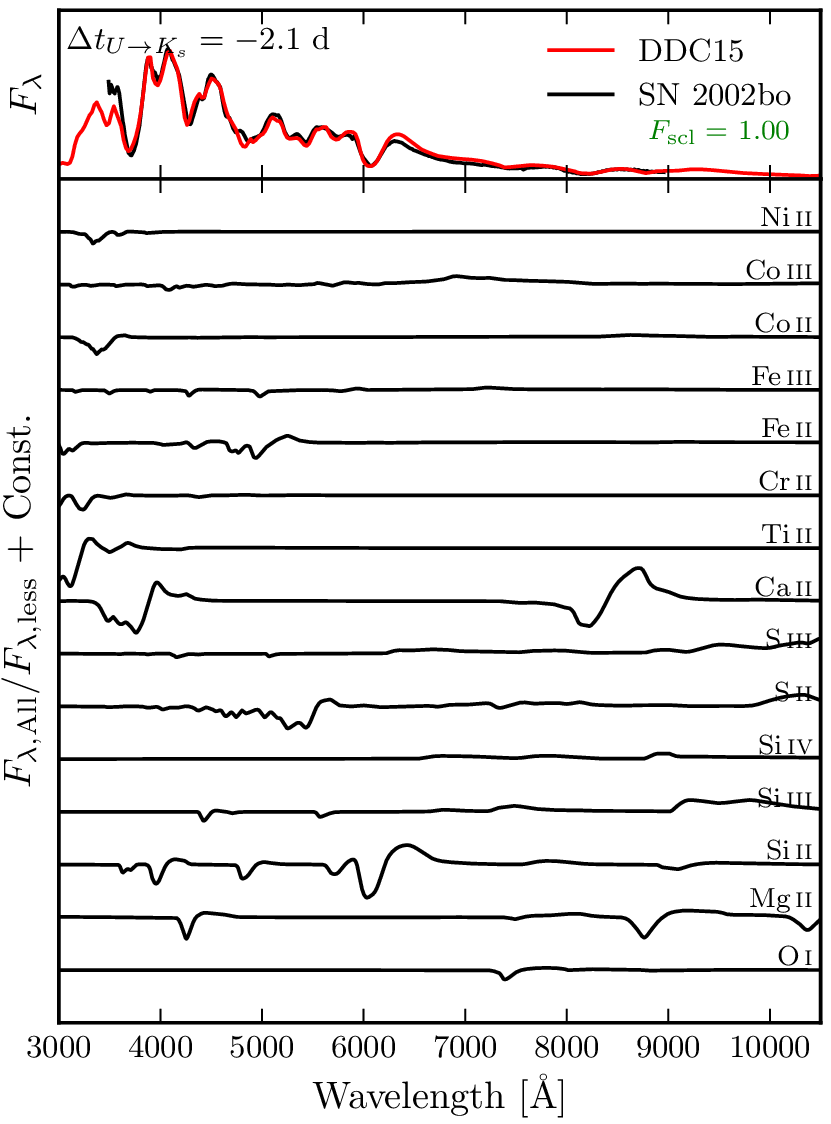}\hspace{.5cm}
\includegraphics{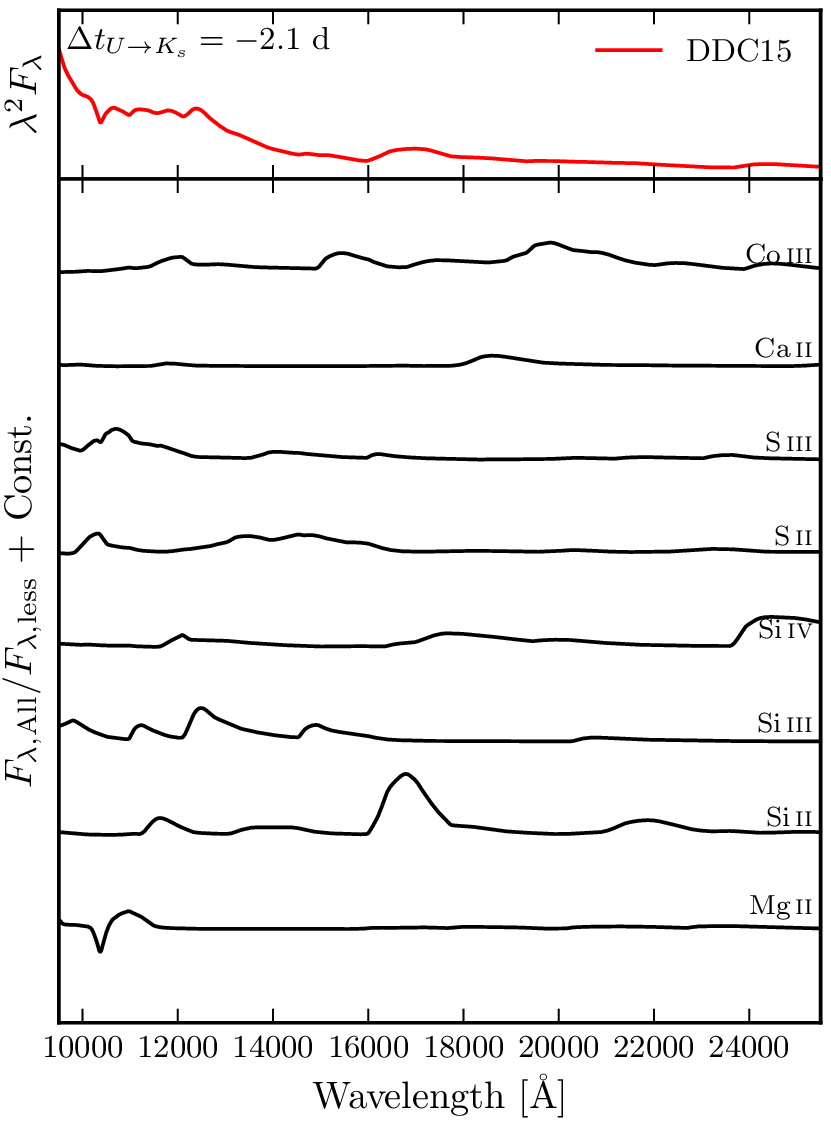}\vspace{.1cm}
\includegraphics{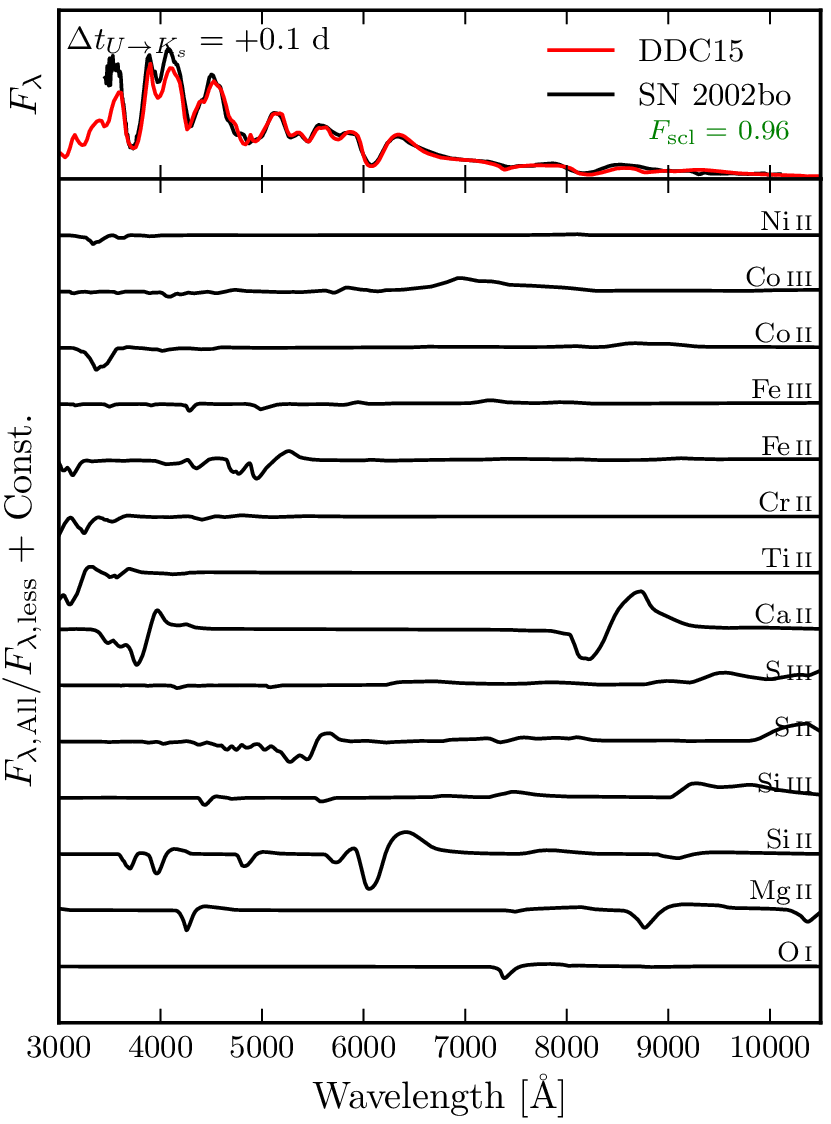}\hspace{.5cm}
\includegraphics{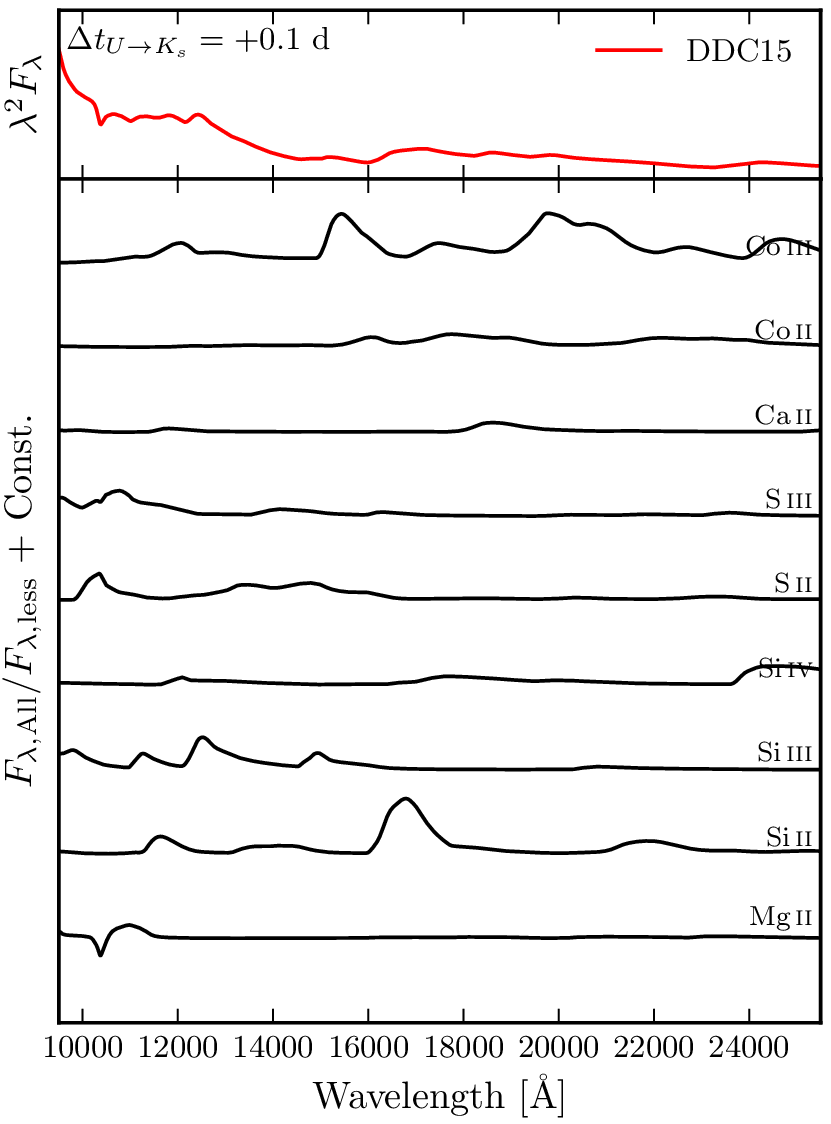}
\caption{\label{fig:ladder_plot3} Contribution of individual ions
  (bottom panels) to the full optical (left) and NIR (right) synthetic
  spectra of DDC15 (top panels, red line), compared to SN~2002bo
  (top panels, black line) at $-2.1$~d and $+0.1$~d from from
  pseudo-bolometric ($U\rightarrow K_s$) maximum.}
\end{figure*}

\begin{figure*}
\centering
\includegraphics{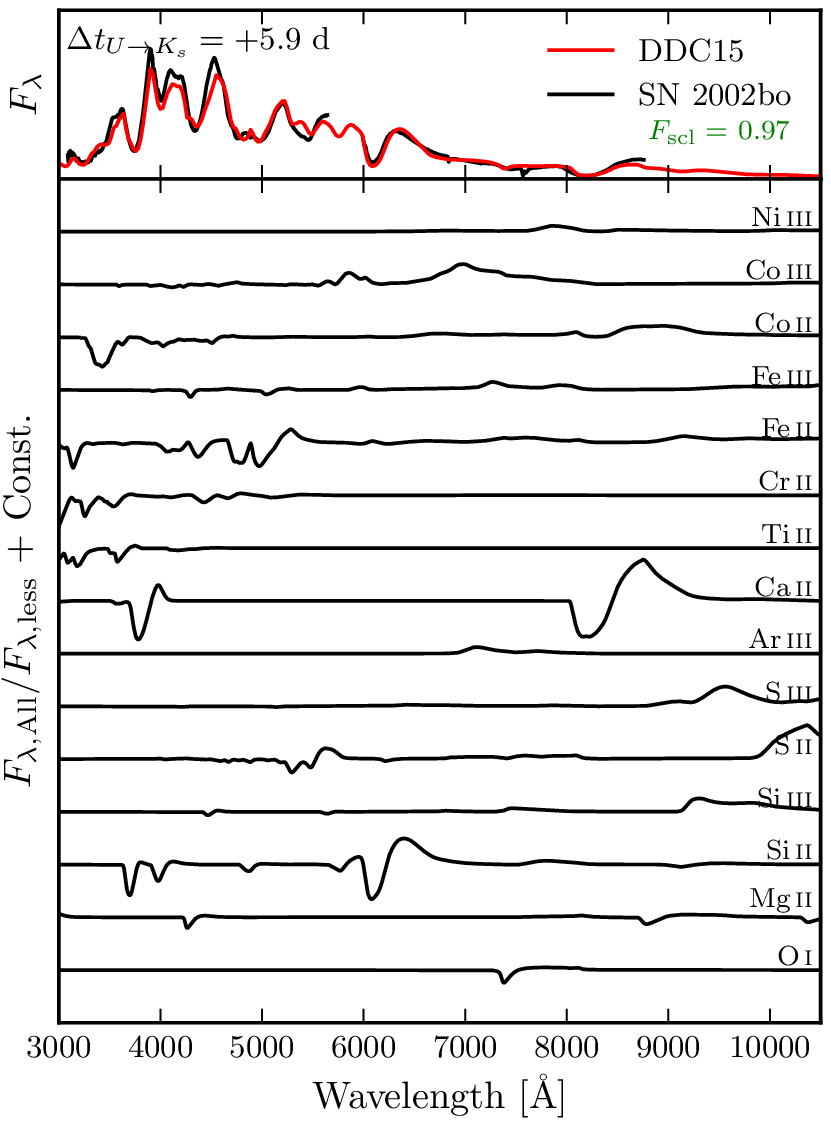}\hspace{.5cm}
\includegraphics{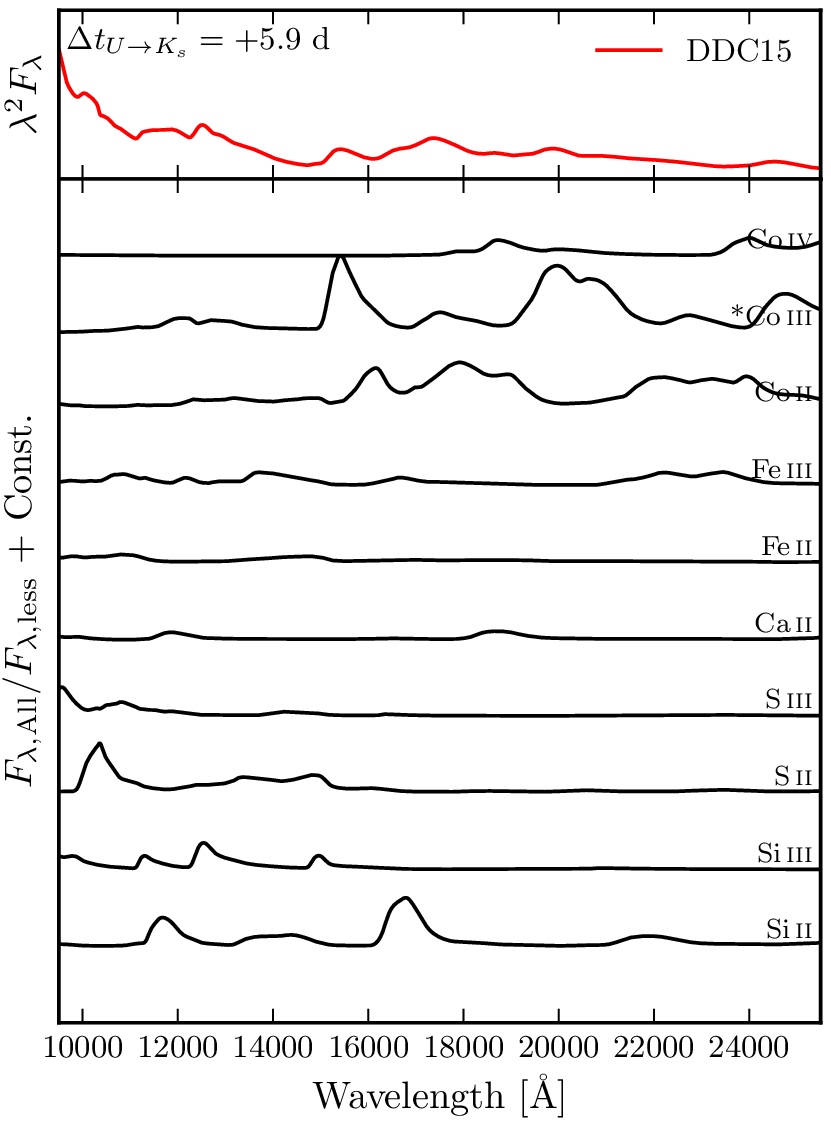}\vspace{.1cm}
\includegraphics{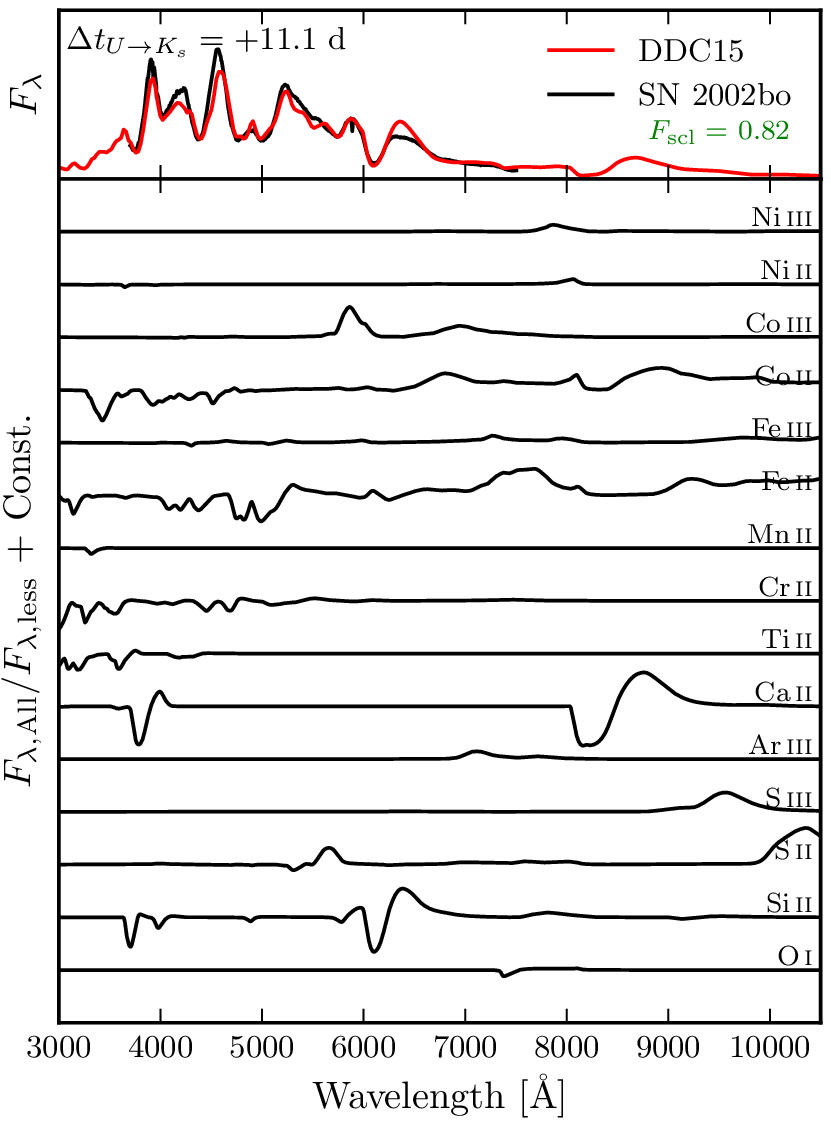}\hspace{.5cm}
\includegraphics{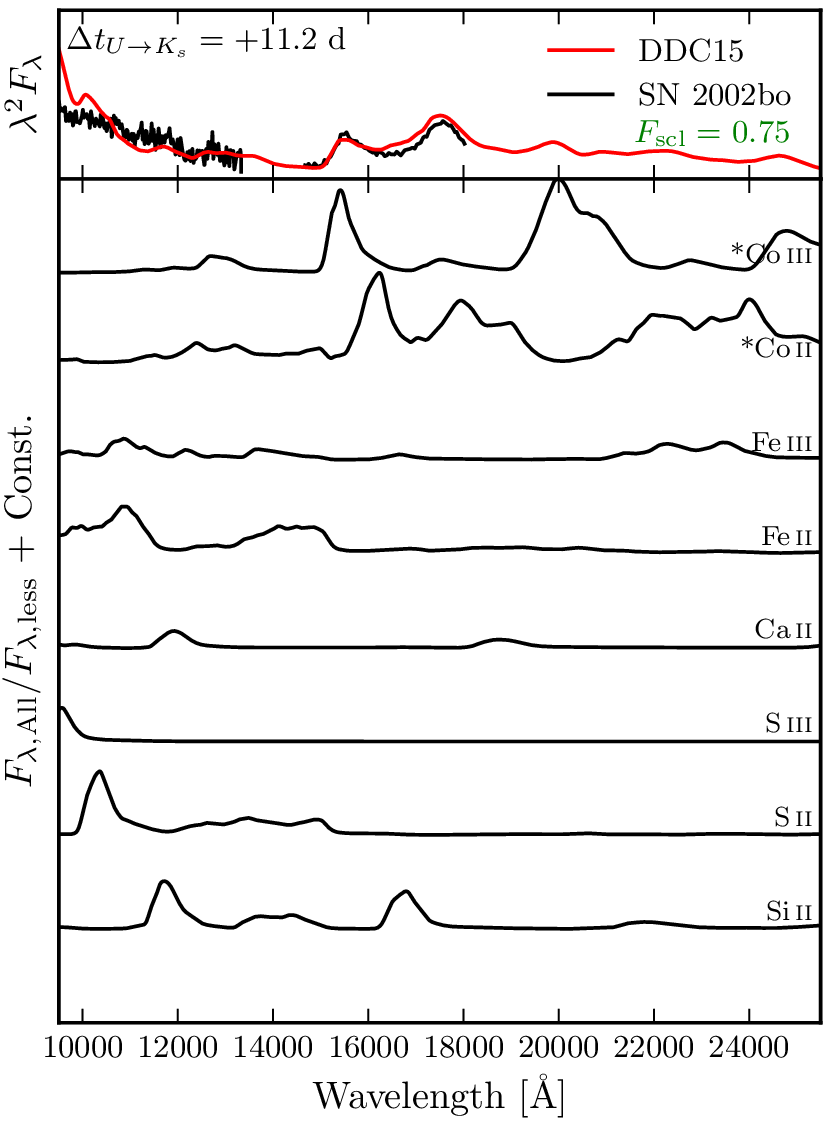}
\caption{\label{fig:ladder_plot4} Contribution of individual ions
  (bottom panels) to the full optical (left) and NIR (right) synthetic
  spectra of DDC15 (top panels, red line), compared to SN~2002bo
  (top panels, black line) at $+5.9$~d, $+11.1$~d  (optical only) and
  $+11.2$~d  (NIR only) from
  from pseudo-bolometric ($U\rightarrow K_s$) maximum. Ion spectra
  marked with a ``*'' have been scaled down for clarity.}
\end{figure*}

\begin{figure*}
\centering
\includegraphics{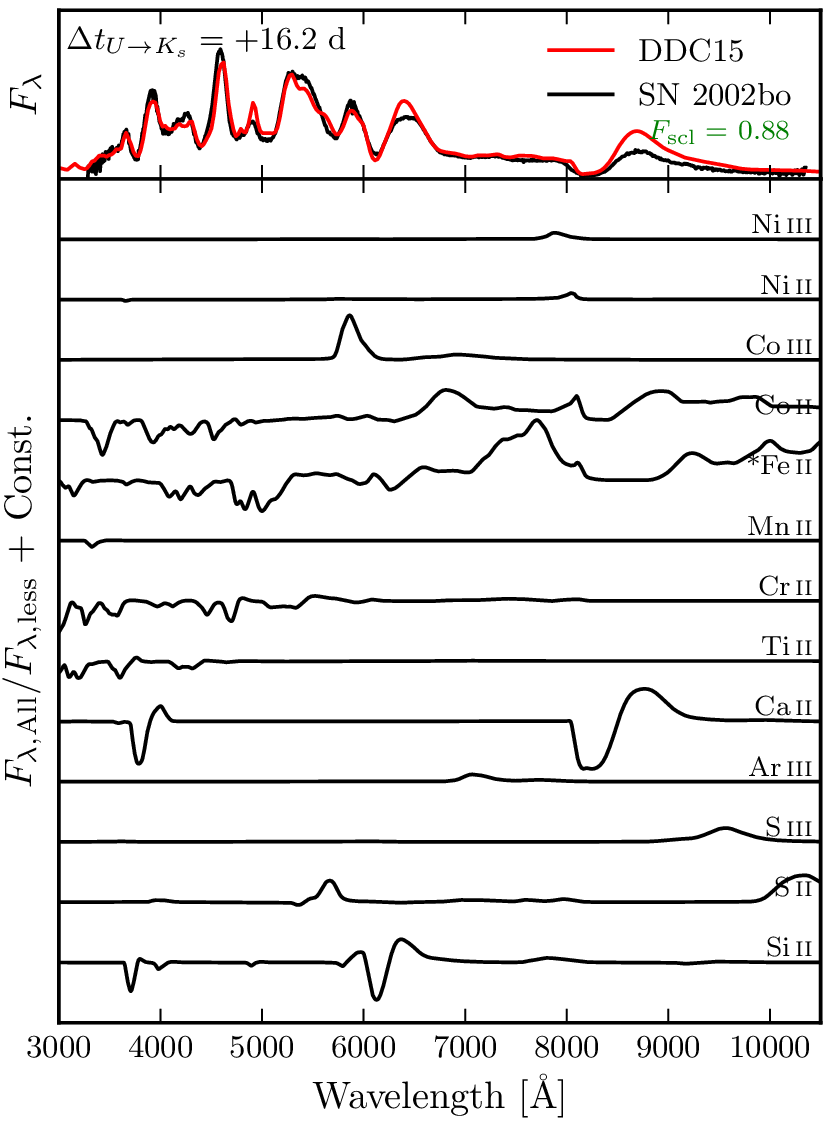}\hspace{.5cm}
\includegraphics{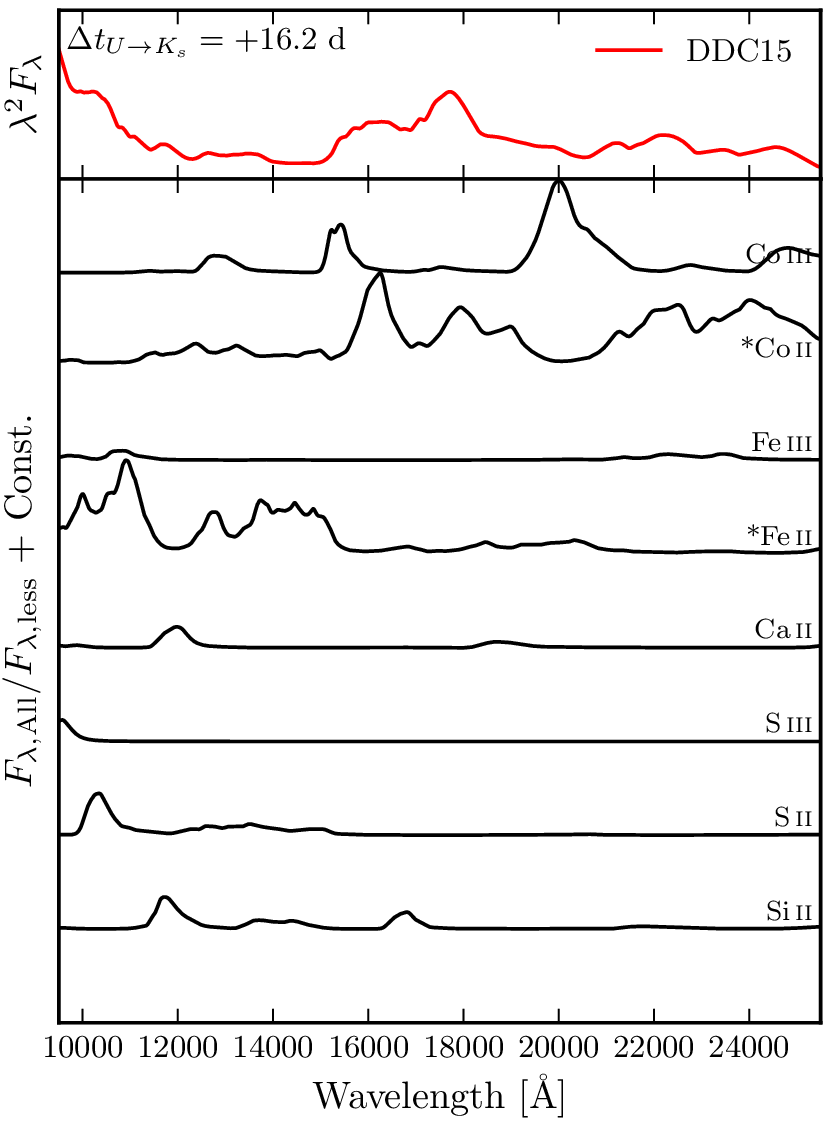}\vspace{.1cm}
\includegraphics{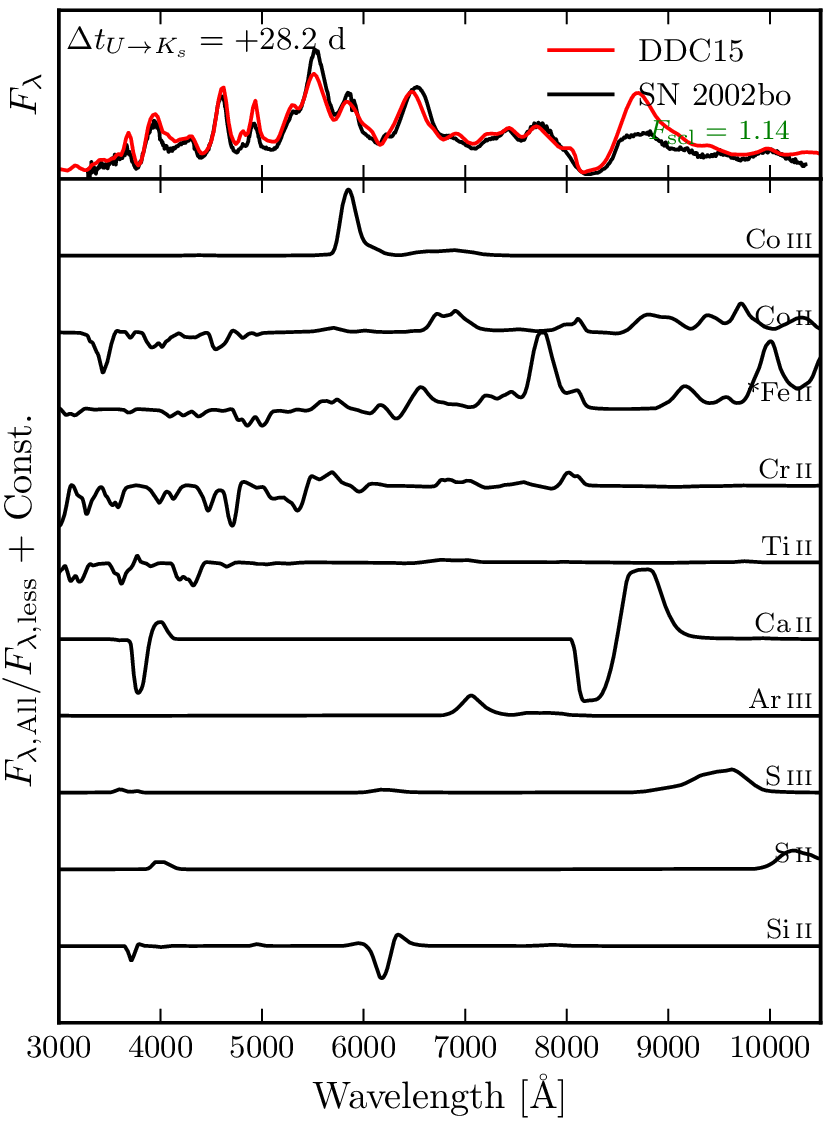}\hspace{.5cm}
\includegraphics{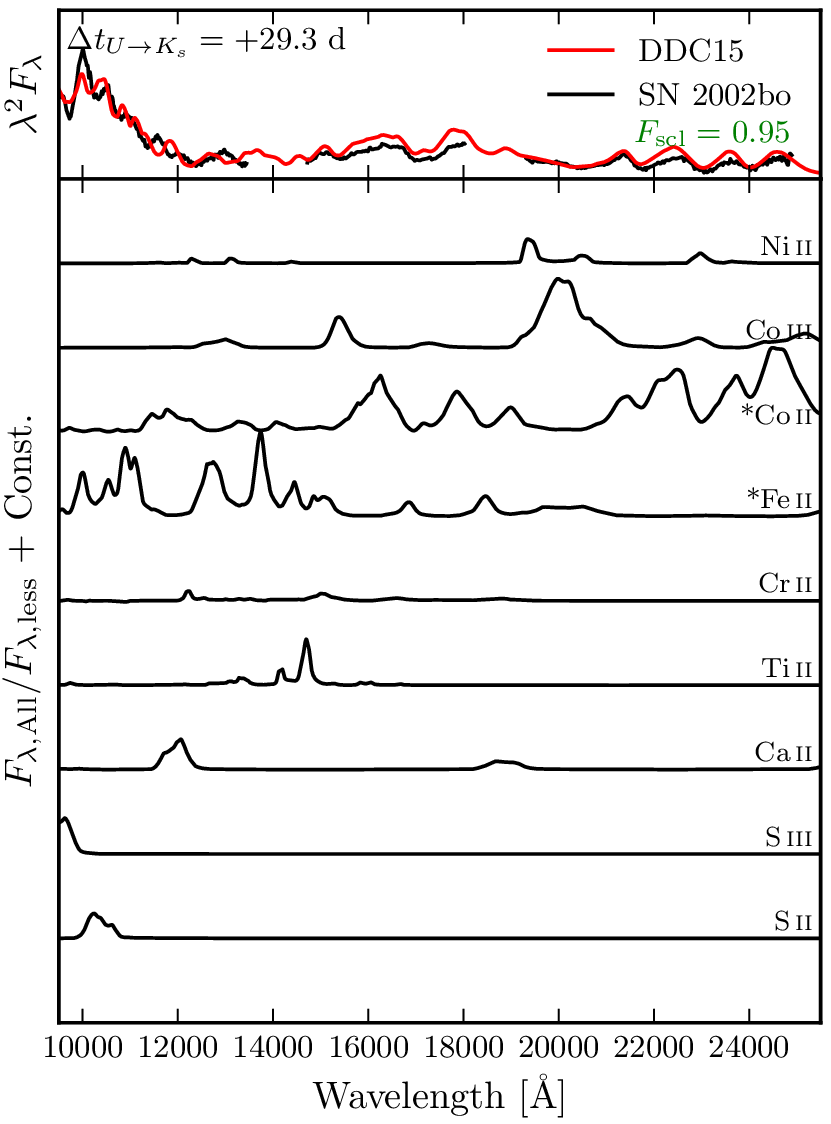}
\caption{\label{fig:ladder_plot5} Contribution of individual ions
  (bottom panels) to the full optical (left) and NIR (right) synthetic
  spectra of DDC15 (top panels, red line), compared to SN~2002bo
  (top panels, black line) at $+16.2$~d, $+28.2$~d (optical only), and
  $+29.3$~d (NIR only) from
  from pseudo-bolometric ($U\rightarrow K_s$) maximum. Ion spectra
  marked with a ``*'' have been scaled down for clarity.}
\end{figure*}

\begin{figure*}
\centering
\includegraphics{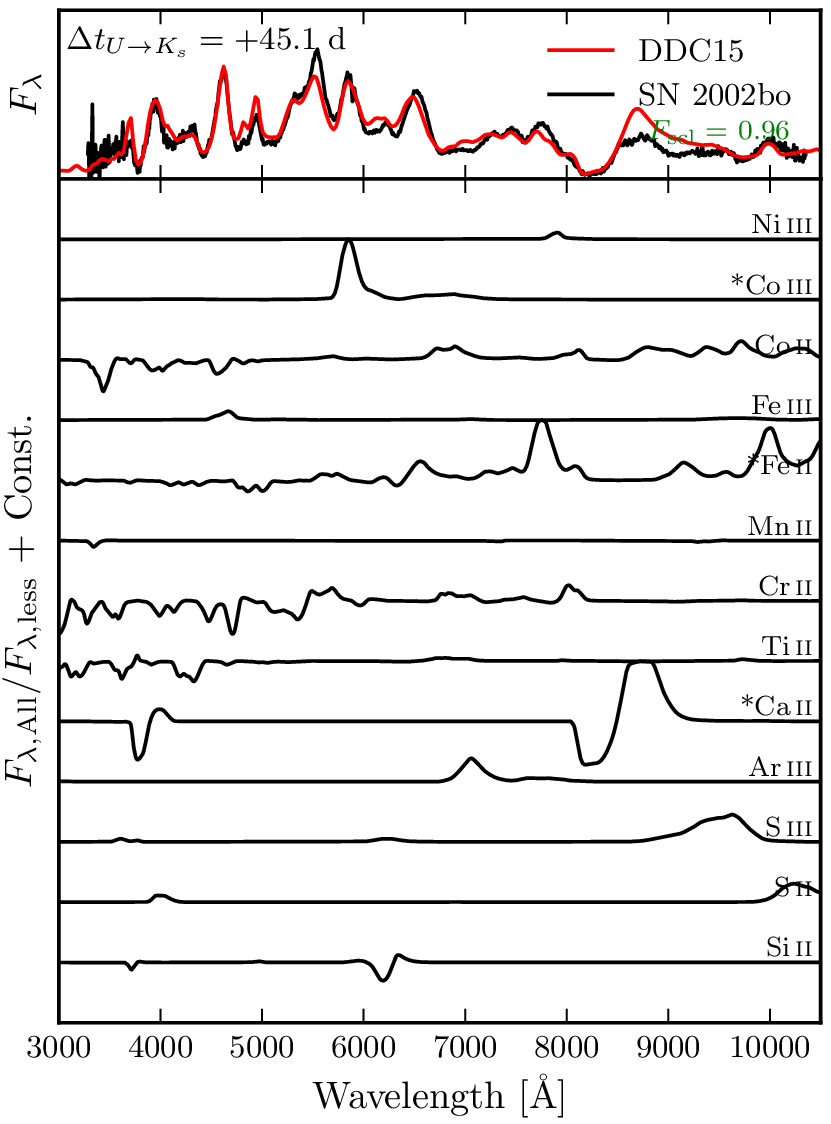}\hspace{.5cm}
\includegraphics{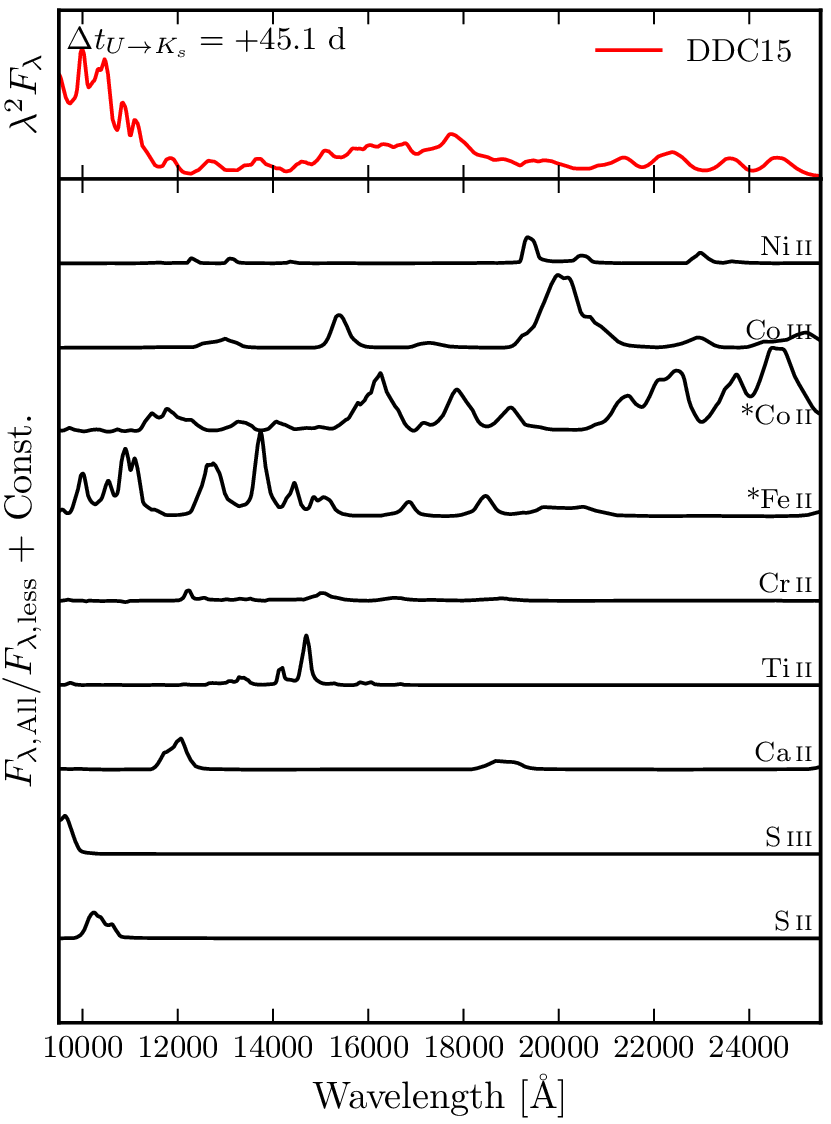}\vspace{.1cm}
\includegraphics{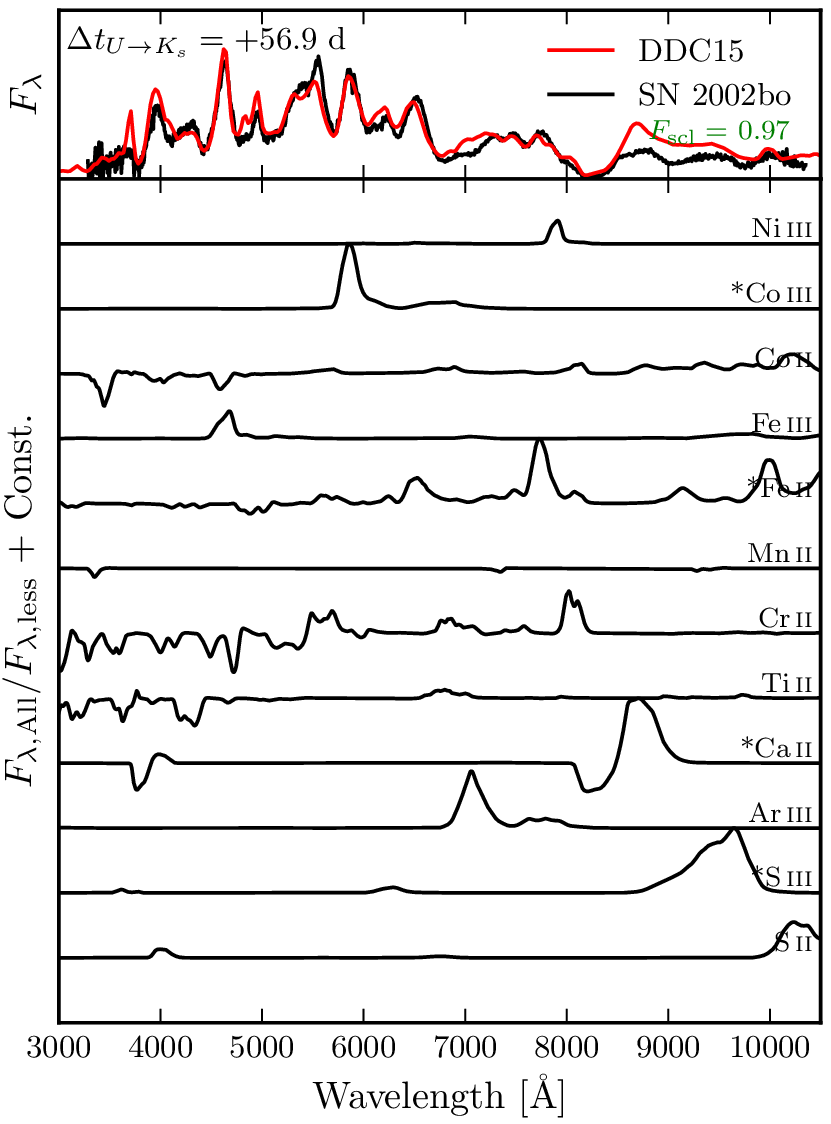}\hspace{.5cm}
\includegraphics{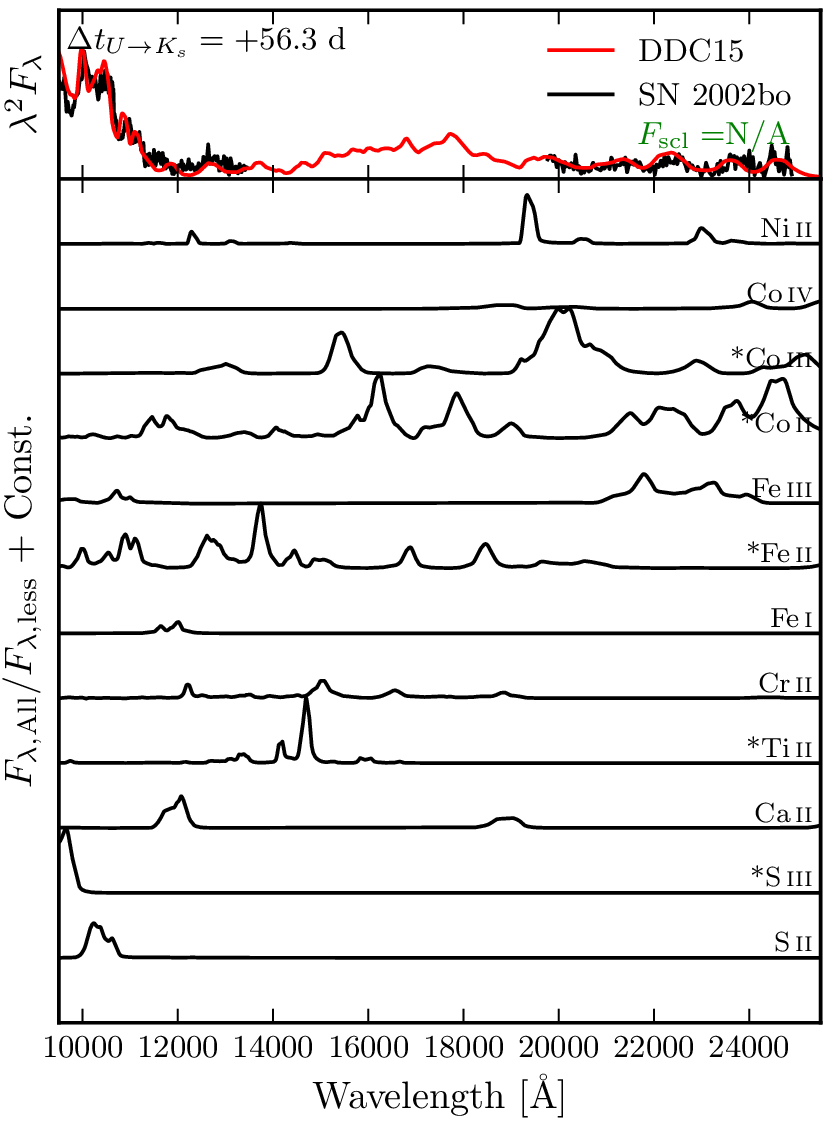}
\caption{\label{fig:ladder_plot6} Contribution of individual ions
  (bottom panels) to the full optical (left) and NIR (right) synthetic
  spectra of DDC15 (top panels, red line), compared to SN~2002bo
  (top panels, black line) at $+45.1$~d, $+56.3$~d (NIR only), and $+56.9$~d
  (optical only) from
  from pseudo-bolometric ($U\rightarrow K_s$) maximum. Ion spectra
  marked with a ``*'' have been scaled down for clarity.}
\end{figure*}

\begin{figure*}
\centering
\includegraphics{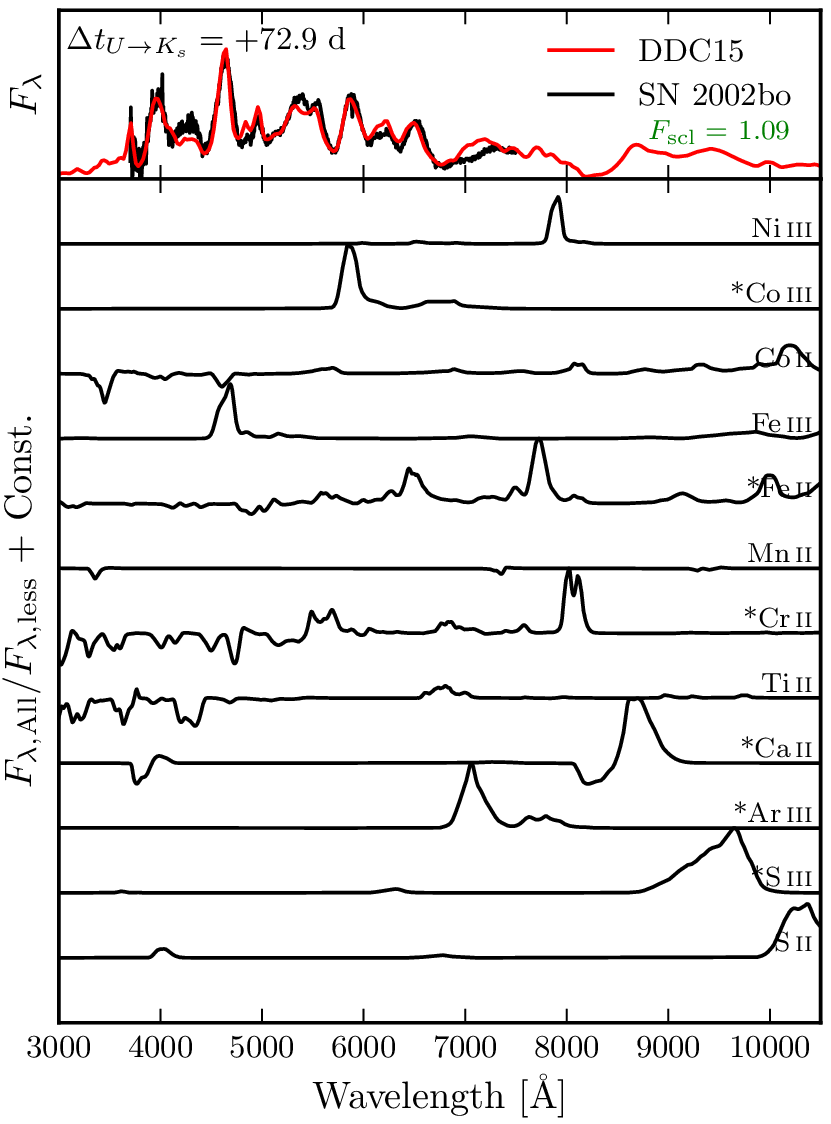}\hspace{.5cm}
\includegraphics{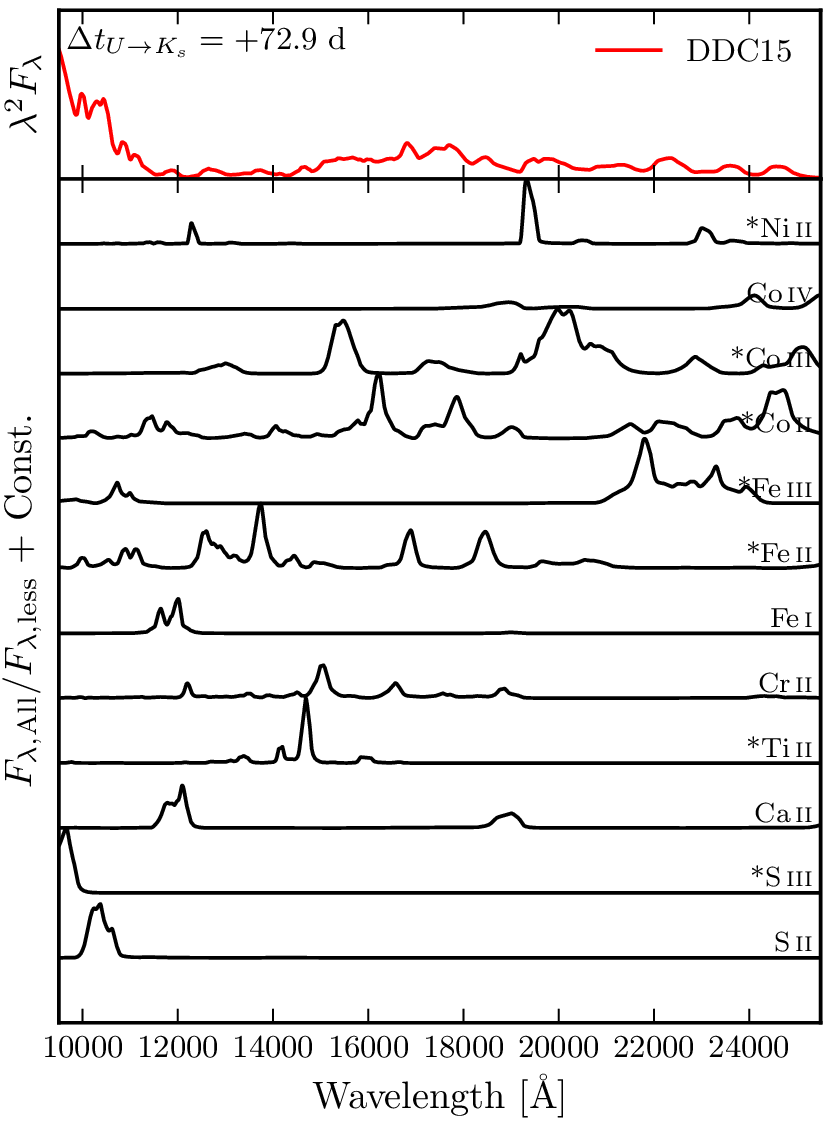}\vspace{.1cm}
\includegraphics{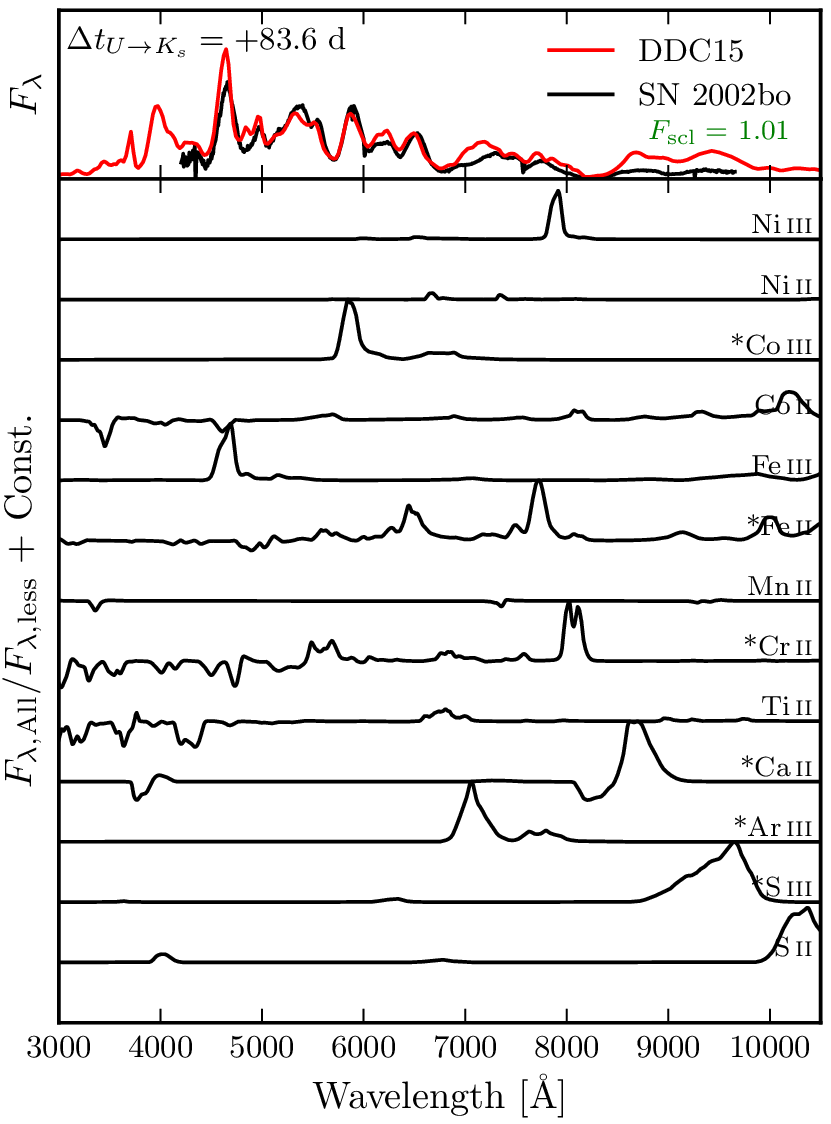}\hspace{.5cm}
\includegraphics{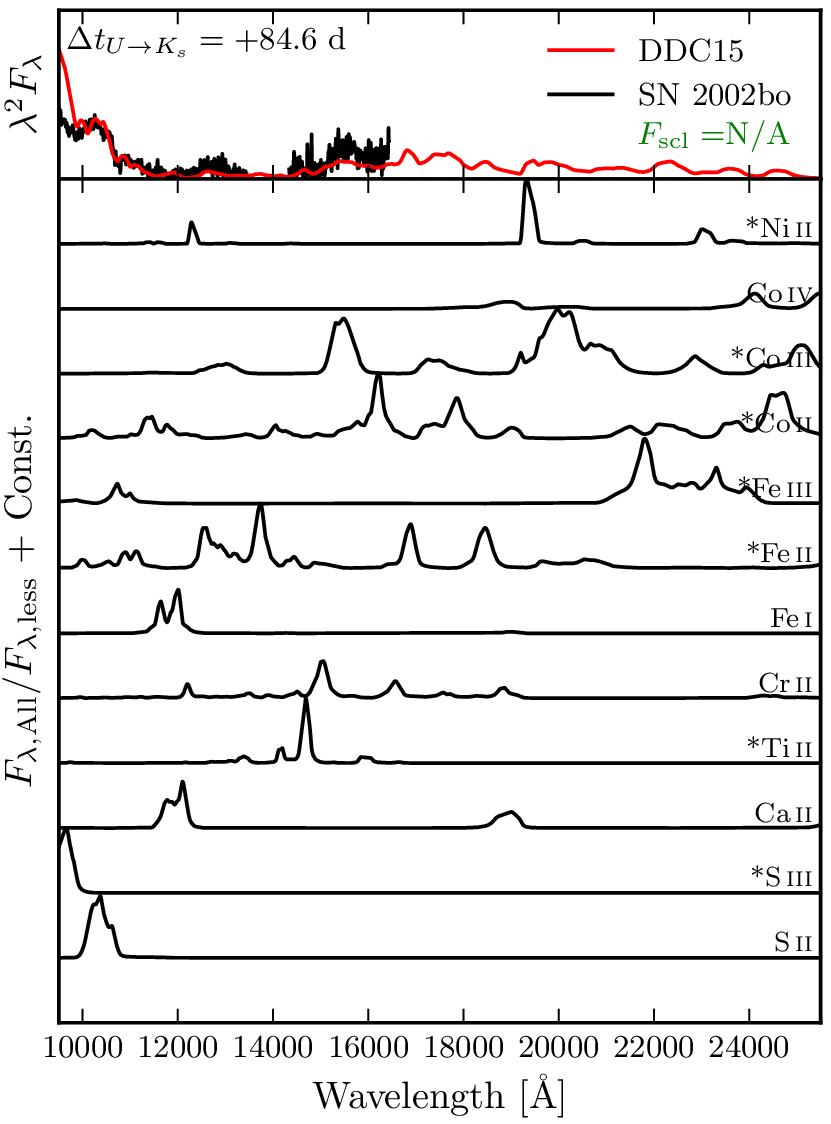}
\caption{\label{fig:ladder_plot7} Contribution of individual ions
  (bottom panels) to the full optical (left) and NIR (right) synthetic
  spectra of DDC15 (top panels, red line), compared to SN~2002bo
  (top panels, black line) at $+72.9$~d, $+83.6$~d (optical only), and
  $+84.6$~d (NIR only) from 
  from pseudo-bolometric ($U\rightarrow K_s$) maximum. Ion spectra
  marked with a ``*'' have been scaled down for clarity.}
\end{figure*}

Figures~\ref{fig:ladder_plot1}--\ref{fig:ladder_plot7} reveal the
contribution of individual ions (bottom panels) to the full optical
(left) and NIR (right) synthetic spectra of DDC15 (top panels, red
line), compared to SN~2002bo (top panels, black line) at all times
presented in Figures~\ref{fig:spec_opt} and \ref{fig:spec_nir}. Only
ions that impact the flux at the $>10$ per cent level are shown.

%%% FIGURE: OPT+NIR ladder plots at all times

\clearpage

\clearpage

\clearpage

\clearpage

\clearpage

\clearpage

%%%%%%%%%%%%%%%%%%%%%%%%%%%%%%%%%%%%%%%%%%%%%%%%%%%%%%%%%%%%%%%%%%%%%%

\bibliographystyle{mn2e} \bibliography{ms}

\label{lastpage}

\end{document}